\documentclass[aps,prd,10pt,twocolumn,superscriptaddress,nofootinbib,dvipsnames]{revtex4-1}
\usepackage[pdftex]{color}
\usepackage[sort&compress]{natbib}
\usepackage[dvipsnames]{xcolor}
\definecolor{my-purple}{RGB}{183,21,153}
\usepackage[colorlinks=true,linkcolor=RubineRed,filecolor=ForestGreen,urlcolor=ForestGreen,citecolor=ForestGreen,pdftex,plainpages=false]{hyperref}
\usepackage{url}
\pdfoutput=1
\usepackage{ifpdf}
\usepackage[utf8]{inputenc}

\usepackage{color,hhline,psfrag,rotating}
\usepackage{dcolumn}
\usepackage{verbatim}
\usepackage{bm}
\usepackage{slashed}
\usepackage{rotating}
\usepackage{amsbsy}
\usepackage{graphicx}
\usepackage{amsfonts,amssymb,amsmath}
\usepackage{booktabs}  

\newcommand{\sea}{\text{sea}}

\begin{document}

\global\let\newpage\relax

\title{Charmonium properties from lattice QCD + QED:
hyperfine splitting, $J/\psi$ leptonic width, charm quark mass and $a_{\mu}^c$}

\author{D.~Hatton}
\email[]{d.hatton.1@research.gla.ac.uk}
\affiliation{SUPA, School of Physics and Astronomy, University of Glasgow, Glasgow, G12 8QQ, UK}
\author{C.~T.~H.~Davies}
\email[]{christine.davies@glasgow.ac.uk}
\author{B. Galloway}
\affiliation{SUPA, School of Physics and Astronomy, University of Glasgow, Glasgow, G12 8QQ, UK}
\author{J.~Koponen}
\affiliation{High Energy Accelerator Research Organisation (KEK), Tsukuba 305-0801, Japan}
\author{G.~P.~Lepage}
\affiliation{Laboratory for Elementary-Particle Physics, Cornell University, Ithaca, New York 14853, USA}
\author{A.~T.~Lytle}
\affiliation{INFN, Sezione di Roma Tor Vergata, Via della Ricerca Scientifica 1, 00133 Roma RM, Italy}
\collaboration{HPQCD collaboration}
\homepage{http://www.physics.gla.ac.uk/HPQCD}
\noaffiliation

\begin{abstract}
We have performed the first $n_f = 2+1+1$ lattice QCD computations 
of the properties (masses and decay constants) 
of ground-state charmonium mesons. 
Our calculation uses the HISQ action to generate
quark-line connected two-point 
correlation functions on MILC gluon field configurations that include 
$u/d$ quark masses going down to the physical point, tuning the 
$c$ quark mass from $M_{J/\psi}$ and including the effect of 
the $c$ quark's electric charge through quenched QED. 
We obtain $M_{J/\psi}-M_{\eta_c}$ (connected) = 120.3(1.1) MeV and interpret the 
difference with experiment as the impact on $M_{\eta_c}$ of 
its decay to gluons, missing from the lattice calculation. This 
allows us to determine $\Delta M_{\eta_c}^{\mathrm{annihiln}}$ =+7.3(1.2) MeV, 
giving its value for the first time. 
Our result of $f_{J/\psi}=$ 0.4104(17) GeV, gives 
$\Gamma(J/\psi \rightarrow e^+e^-)$=5.637(49) keV, in agreement 
with, but now more accurate than experiment. 
At the same time we have improved the determination of 
the $c$ quark mass, including the impact of quenched QED to 
give $\overline{m}_c(3\,\mathrm{GeV})$ = 0.9841(51) GeV. 
We have also used the time-moments of the vector charmonium 
current-current correlators to improve the lattice QCD
result for the $c$ quark HVP contribution to the anomalous magnetic 
moment of the muon. We obtain $a_{\mu}^c =  14.638(47) \times 10^{-10}$, 
which is 2.5$\sigma$ higher than the value derived using
moments extracted from some sets of experimental data on 
$R(e^+e^- \rightarrow \mathrm{hadrons})$. This value for $a_{\mu}^c$ 
includes our determination of the effect of QED on this 
quantity, $\delta a_{\mu}^c = 0.0313(28) \times 10^{-10}$.  
 
\end{abstract}

\maketitle

\section{Introduction}
\label{sec:intro}

The precision of lattice QCD calculations has been steadily improving for some
time and is now approaching, or has surpassed, the 1\% level for multiple
quantities. Good examples are the masses and decay constants 
of ground-state pseudoscalar mesons~\cite{Tanabashi:2018oca}. The meson masses 
can be used to tune, 
and therefore determine, quark masses. 
The decay constants can be combined with experimental annihilation rates 
to leptons to determine elements of the Cabibbo-Kobayashi-Maskawa matrix. 
The accuracy of modern lattice QCD results means that sources of small 
systematic uncertainty that could appear at the percent level need to be 
understood. At this level QED effects, i.e.\ the fact that quarks carry electric 
as well as color charge, come into play. A naive argument that such effects 
could be $\mathcal{O}(\alpha_{\mathrm{QED}})$ would imply a possible 1\% contribution.  
One key driver for the lattice QCD effort to include QED effects has been 
that of calculations of the 
hadronic vacuum polarisation (HVP) contribution to the anomalous
magnetic moment of the muon, $a_{\mu}$. New results are expected soon from the Muon $g-2$
experiment at Fermilab \cite{Grange:2015fou} which aims to clarify the observed tension
between experiment and Standard Model theory seen by the Brookhaven E821 experiment
\cite{Bennett:2004pv}. Current lattice QCD calculations have reached the precision
of a few percent for $a_{\mu}$ and the systematic uncertainties, for example 
from neglecting QED effects, have become a major focus of attention (see, for 
example~\cite{Boyle:2017gzv, Giusti:2019xct, Borsanyi:2020mff}). 

QED effects can have large finite-volume corrections 
 within a lattice QCD calculation 
because the Coulomb interaction is long-range. 
For electrically neutral correlation functions, 
such as the ones 
needed for a calculation of the HVP, and that we study here, 
this is much less of an issue (and we demonstrate this).
Since $\alpha^n_s\alpha_{\mathrm{QED}}$ is not very different in size from 
$\alpha_{\mathrm{QED}}$ at hadronic scales, calculations must be fully nonperturbative 
in QCD. A consistent calculation must also allow for the retuning of quark 
masses needed when QED effects are included. 

Here we examine the properties of ground-state charmonium mesons
more accurately than has been possible in previous lattice QCD 
calculations\footnote{For a different kind of lattice QCD calculation that 
maps out the spectrum more completely, but paying less attention to 
ground-states, see~\cite{Cheung:2016bym}.}.
Since it is possible to obtain statistically very precise results 
for the charmonium system it is a good place to study small systematic 
effects from QED and other sources. 
We are able to see such effects in our results and 
quantify them. 
We include $u$, $d$, $s$ and $c$ quarks in the sea 
for the first time and have results on gluon field configurations 
with physical $u/d$ sea quarks. We also analyse the impact of including 
an electric charge on the valence $c$ quarks (only). 
This approximation, known 
as `quenched QED', should capture the largest QED effects and enable 
us to see how much of a difference QED makes. 
For the vector ($J/\psi$) meson 
properties we improve on earlier results by using an exact method 
to renormalise the lattice vector current (both in QCD and QCD+QED).  
To tune the $c$ quark mass we use the $J/\psi$ meson mass, taking 
into account the retuning that is required when QED is switched on. 
We find the impact of this to be of similar size to the more direct 
QED effects. 

The quantities that we focus on here are the masses and decay constants 
of the ground-state 
$\eta_c$ and $J/\psi$ mesons, 
the $c$ quark mass and the contribution of the 
$c$ vacuum polarisation to $a_{\mu}$. 

The correlation functions that we calculate in our lattice QCD and 
QCD+QED calculations are `connected' correlation functions i.e.\ they 
are constructed from combining charm quark propagators from the 
source to the sink. 
We do not include diagrams in which the $c$ and $\overline{c}$ quarks 
annihilate to multiple gluons and hence hadrons. 
We can gain some insight into the effect this annihilation channel 
has on the meson masses by looking at the meson widths, 
which are twice the imaginary part of the mass and are dominated by 
hadronic channels. 
The $J/\psi$ has a tiny width of 93 keV~\cite{Tanabashi:2018oca} 
but the pseudoscalar $\eta_c$ can annihilate to two gluons, allowing 
it to mix with other flavour-singlet pseudoscalars, and it
has a width of 32 MeV~\cite{Tanabashi:2018oca}. 
The annihilation channel might then be expected to have a larger impact 
on the $\eta_c$ mass and lead lattice QCD 
calculations of the mass from connected correlators to disagree with experiment. 
The only way to achieve the $\mathcal{O}(1\,\mathrm{MeV})$ accuracy required 
to see this is to determine the mass difference between the $J/\psi$ and 
$\eta_c$ (the hyperfine splitting). Any shift in the $\eta_c$ mass 
will have a much larger (by a factor of 30) relative effect in this splitting.  
 Previous lattice QCD calculations of the 
hyperfine splitting from connected correlation functions have not 
been accurate enough to see a significant difference with experiment. 
Here, for the first time, we can see such a difference because we have 
very good control both of discretisation effects and sea-quark mass 
effects and can also determine the impact of QED on this quantity.    

A further place in which QED effects need to be quantified, given our 
accuracy, is that of 
the determination of the $c$ quark mass. We do this by tuning 
our results to the experimental $J/\psi$ meson mass with and without 
electric charge on the valence $c$ quarks and also determine 
the small change in the mass renormalisation factor, $Z_m$, needed 
to convert to the standard $\overline{\text{MS}}$ mass. 

Our analysis of the vector charmonium correlation functions 
with a completely nonperturbative renormalisation of the vector 
current
allows much improved accuracy in the determination of 
the leptonic decay width of the $J/\psi$. 
Using the same correlators, 
we determine the charm quark portion of the hadronic vacuum polarisation  
contribution to the anomalous magnetic of the muon, $a_{\mu}^{\text{HVP,c}}$, 
along with the impact of quenched QED on this quantity.
We can compare this to phenomenological results from 
$R(e^+e^- \rightarrow \text{hadrons})$. 

The paper is laid out as follows: 
\begin{itemize}
\item Section~\ref{sec:setup} describes the 
lattice QCD calculation and the inclusion of quenched QED; 
\item Section~\ref{sec:hyperfine} describes our determination of the hyperfine splitting; 
\item Section~\ref{sec:mass}, 
the $c$ quark mass; 
\item Section~\ref{sec:jpsidecay}, the $J/\psi$ and $\eta_c$ 
decay constants;
\item Section~\ref{sec:amuc} the time-moments of $c$ vector-vector 
correlators and the $c$ quark hadronic vacuum 
polarisation contribution to $a_{\mu}$; 
\item Section~\ref{sec:conclusions} gives our conclusions. 
\end{itemize}
Each section of results includes a description of the pure QCD calculation followed 
by a determination of 
the impact of quenched QED on the result and then a discussion subsection 
including comparison to both experiment and 
previous lattice QCD calculations, where applicable. Finally 
Section~\ref{sec:conclusions} collects up all of our results 
and summarises our conclusions.

\section{Lattice setup} \label{sec:setup}

We perform calculations on a total of 17 gluon field ensembles, concentrating our 
analysis on 15 of these.  16 sets include the effects of
light, 
strange and charm quarks in the sea with up and down quarks having the 
same mass ($n_f = 2+1+1$), one set has up and down quarks set separately 
to their physical values ($n_f = 1+1+1+1$). All gluon field ensembles include 
sea quarks 
using the HISQ action~\cite{Follana:2006rc} and
were generated by the MILC collaboration~\cite{Bazavov:2012xda,Bazavov:2017lyh}. 
Parameters for the ensembles are
given in Table~\ref{tab:ensembles}.
Our sets include lattices at 6 different
$\beta$ (the bare QCD coupling) values corresponding to 6 different sets of 
lattice spacing values with the finest
lattice reaching a spacing of $\sim 0.03$ fm. Although `topology freezing' 
has been seen on the very finest lattices used here, we do not expect this 
to have significant impact on the charmonium quantities we study here because 
no valence light quarks are involved in the calculations~\cite{Bernard:2017npd}.  
We use ensembles with sea $u/d=l$ masses at the physical point on all but
the finest two lattice spacings. We also employ three ensembles (sets 5, 6 and 7)
with shared
parameters except for their spatial extent in units of the lattice spacing $L_s$. These ensembles allow us to
investigate finite volume effects in our QED analysis. 
We test the impact of strong-isospin breaking effects in the sea by using 
two ensembles (sets 3A and 3B) with all parameters the same except that one ensemble has 
$m_u=m_d=m_l$ and one has $(m_u+m_d)/2=m_l$ and $m_d/m_u=2.18$.  
The
gluon action on these ensembles is improved so that discretisation errors through
$\mathcal{O}(\alpha_sa^2)$ are removed~\cite{Hart:2008sq}. 

\begin{table*}
  \caption{Details of the lattice gluon field ensembles and calculation parameters used.
  The lattice spacing is determined from the Wilson flow parameter, $w_0$~\cite{Borsanyi:2012zs}, with 
$w_0/a$ values given in column 2. The lattice spacing can be determined in fm 
by using $w_0=0.1715(9)$ fm~\cite{fkpi} (fixed from $f_{\pi}$). $L_s$
  and $L_t$ are the lattice spatial and temporal extents in lattice units.
Columns 6, 7 and 8 give the sea quark masses in lattice units. Note that all 
of the configuration sets are $n_f=2+1+1$, i.e. with equal mass $u$ and $d$ quarks 
(denoted $l$) {\it{except}} for set 3B which is $n_f=1+1+1+1$. For set 3B $am^{\sea}_u$ 
and $m^{\sea}_d$ are listed separately with $am^{\sea}_u$ on top.   
  $am_c^{\mathrm{val}}$ is the valence $c$ quark mass with $\epsilon_{\mathrm{Naik}}$ the corresponding 
Naik parameter (see text). The column headed
  $N_{\mathrm{cfg}} \times N_t$ shows both the number of configurations used in the pure QCD calculation
  and the number of time sources for propagators per configurations. 
$N_{\mathrm{cfg,QED}}$ refers to the number of
  configurations (and time sources) used in QCD+QED calculations. Sets 1-3 will
  be referred to as very coarse ($a \approx$ 0.15 fm), sets 4-8 as coarse ($a \approx$ 0.12fm), sets 9-11 as fine ($a \approx$ 0.09 fm), 12 and
  13 as superfine ($a \approx$ 0.06 fm), 14 as
  ultrafine ($a \approx 0.045$ fm) and 15 as exafine ($a \approx$ 0.03 fm).}
  \label{tab:ensembles}
\begin{ruledtabular}
\begin{tabular}{llllllllllll}
Set & $\beta$ & $w_0/a$ & $L_s$ & $L_t$ & $am_l^{\sea}$ & $am_s^{\sea}$ & $am_c^{\sea}$ & $am_c^{\mathrm{val}}$ & $\epsilon_{\mathrm{Naik}}$ & $N_{\mathrm{cfg}} \times N_t$ & $N_{\mathrm{cfg,QED}} \times N_t$ \\
\hline
1 & 5.80 & 1.1119(10) & 16 & 48 & 0.013 & 0.065 & 0.838 & 0.888 & -0.3820 & 1020$\times$8 & - \\
2 & 5.80 & 1.1272(7) & 24 & 48 & 0.0064 & 0.064 & 0.828 & 0.873 & -0.3730 & 1000$\times$8 & 340$\times$16 \\
3 & 5.80 & 1.1367(5) & 32 & 48 & 0.00235 & 0.0647 & 0.831 & 0.863 & -0.3670 & 1000$\times$8 & - \\
3A & 5.80 & 1.13215(35) & 32 & 48 & 0.002426 & 0.0673 & 0.8447 & 0.863 & -0.3670 & 1762$\times$16 & - \\
3B & 5.80 & 1.13259(38) & 32 & 48 & 0.001524 & 0.0673 & 0.8447 & 0.863 & -0.3670 & 1035$\times$16 & - \\
 &  &  &  &  & 0.003328 &  &  &  &  &  & - \\
\hline
4 & 6.00 & 1.3826(11) & 24 & 64 & 0.0102 & 0.0509 & 0.635 & 0.664 & -0.2460 & 1053$\times$8 & - \\
5 & 6.00 & 1.4029(9) & 24 & 64 & 0.00507 & 0.0507 & 0.628 & 0.650 & -0.2378 & - & 340$\times$16 \\
6 & 6.00 & 1.4029(9) & 32 & 64 & 0.00507 & 0.0507 & 0.628 & 0.650 & -0.2378 & 1000$\times$8 & 220$\times$16 \\
7 & 6.00 & 1.4029(9) & 40 & 64 & 0.00507 & 0.0507 & 0.628 & 0.650 & -0.2378 & - & 220$\times$16 \\
8 & 6.00 & 1.4149(6) & 48 & 64 & 0.00184 & 0.0507 & 0.628 & 0.643 & -0.2336 & 1000$\times$8 & - \\
\hline
9 & 6.30 & 1.9006(20) & 32 & 96 & 0.0074 & 0.037 & 0.440 & 0.450 & -0.1250 & 300$\times$8 & -  \\
10 & 6.30 & 1.9330(20) & 48 & 96 & 0.00363 & 0.0363 & 0.430 & 0.439 & -0.1197 & 300$\times$8 & 371$\times$16 \\
11 & 6.30 & 1.9518(7) & 64 & 96 & 0.00120 & 0.0363 & 0.432 & 0.433 & -0.1167 & 565$\times$8 & - \\
\hline
12 & 6.72 & 2.8941(48) & 48 & 144 & 0.00480 & 0.0240 & 0.286 & 0.274 & -0.0491 & 1019$\times$8 & 265$\times$16 \\
13 & 6.72 & 3.0170(23) & 96 & 192 & 0.0008 & 0.022 & 0.260 & 0.260 & -0.0443 & 100$\times$8 & - \\
\hline
14 & 7.00 & 3.892(12) & 64 & 192 & 0.00316 & 0.0158 & 0.188 & 0.194 & -0.0250 & 200$\times$8 & -  \\
\hline
15 & 7.28 & 5.243(16) & 96 & 288 & 0.00223 & 0.01115 & 0.1316 & 0.138 & -0.0127 & 100$\times$4 & -  \\
\end{tabular}
\end{ruledtabular}
\end{table*}

On these gluon field configurations we calculate propagators for valence $c$ 
quarks by solving the Dirac equation for a source consisting of a set of 
Gaussian random numbers across a timeslice (a random wall source). 
We use multiple time sources per configuration to improve statistical accuracy. 
The number of configurations used and the number of time sources is given 
in Table~\ref{tab:ensembles}. The table also gives the valence $c$ quark masses in lattice 
units, which may differ from those in the sea because they are tuned more 
accurately. 
This will be discussed 
further below. The HISQ action~\cite{Follana:2006rc} includes an improved discretisation of 
the covariant derivative in the Dirac equation. 
This removes tree-level $a^2$ discretisation errors 
by the addition of a 3-link `Naik' term to the symmetric 1-link 
difference. 
For heavy quarks the coefficient of the Naik term is adjusted from 1 to 
$1+\epsilon_{\mathrm{Naik}}$ to 
remove $(am)^4$ errors at tree-level~\cite{Follana:2006rc}. 
A closed-form expression for 
$\epsilon$ in terms of the tree-level quark pole mass is given 
in~\cite{Monahan:2012dq} along with the formula for the tree-level 
quark pole mass in terms of the 
bare mass. Table~\ref{tab:ensembles} gives the values of $\epsilon$ that we use.  

We combine charm quark and antiquark propagators to 
calculate two types of quark-line connected 
two-point correlation functions: pseudoscalar and
vector. The ground state of the pseudoscalar correlation function 
corresponds to the $\eta_c$ meson and the
vector correlation function, to the $J/\psi$. 
When using staggered quarks, as here,
the different spin structures are implemented using position
dependent phases in the operators at source and sink. The
two-point `Goldstone' pseudoscalar 
($\gamma_5 \otimes \gamma_5$ in spin-taste notation) 
correlation functions are simply constructed from quark propagators $S(x,0)$ from the origin to $x$ 
as
\begin{equation}
\label{eq:pscorr}
  C(t) = \frac{1}{4} \sum_{x} \langle \mathrm{Tr}(S(x,0)S^{\dagger}(x,0)) \rangle ,
\end{equation}
where the factor of 4 accounts for the taste multiplicity with staggered quarks. 
For the vector correlation functions we use a local vector operator 
(spin-taste $\gamma_i \otimes \gamma_i$). The correlation functions 
then combine $S(x,0)$ with a propagator made from patterning the source with 
a phase $(-1)^{x_i}$ and inserting $(-1)^{x_i}$ at the sink timeslice as 
we tie the propagators together and sum over spatial sites. Our vector correlation 
functions average over all spatial polarisations, $i$, for improved 
statistical precision. 
Note that we do not calculate any quark-line disconnected correlation functions. 

The HISQ local vector current is not conserved and requires
renormalisation. For this purpose we use the RI-SMOM momentum subtraction scheme
implemented on the lattice as discussed in~\cite{Hatton:2019gha}. 
In~\cite{Hatton:2019gha}
it was shown that, because of the Ward-Takahashi identity, these
renormalisation factors do not suffer any contamination by nonperturbative 
artefacts (condensates) 
and can therefore be safely used
in calculations such as those presented here. 
The quenched QED correction to the RI-SMOM
vector current renormalisation was also given in~\cite{Hatton:2019gha} 
and shown to be tiny
($\sim 0.01\%$) for the HISQ action (as expected since the pure QCD $Z_V$ values 
only differ from 1 at the 1\% level and quenched QED provides a small correction 
to this difference from 1). 
Here we use the $Z_V$ values from \cite{Hatton:2019gha} at a scale $\mu$ of 2 GeV. 
We will also demonstrate (see Section~\ref{sec:jpsidecay}) 
that using $\mu$ = 3 GeV gives the same result 
as it must for a $Z_V$ that correctly matches the lattice to continuum physics.

Since we make use of an ensemble (set 14) with a finer lattice spacing than those
studied in \cite{Hatton:2019gha} we have directly calculated the 
value of $Z_V$ on set 14 at
$\mu = 2$ GeV in addition. 
We have, however, only used a small number of configurations (6)
in that calculation due to the computational limitation of the stringent Landau
gauge fixing required. We therefore double the statistical uncertainty
for $Z_V$ on that ensemble. This has very little impact on our final
results as the $Z_V$ uncertainty is small. See Appendix~\ref{appendix:zv} 
for a discussion of our $Z_V$ values, where we also derive a $Z_V$ 
value for set 15. 

In order to tune the mass of the valence charm quark we use bare
charm mass values on each lattice that produce a $J/\psi$ mass equal to the
experimental value (both in pure QCD and in QCD+QED). We
choose the $J/\psi$ here rather than the $\eta_c$ because the relatively large width
of the $\eta_c$ means that annihilation effects that we are not including 
could lead to small (order 0.1\% ) uncertainties in the mass. This is 
mentioned in 
Section~\ref{sec:intro} and will be discussed 
further in Section~\ref{sec:hyperfine}.  
We measure our valence $c$ mass mistunings as the
difference between our lattice $J/\psi$ mass and the experimental average value 
of 3.0969 GeV (with negligible uncertainty)~\cite{Tanabashi:2018oca}. 
The two panels of Fig.~\ref{fig:jpsi-masses}, where the horizontal line is the
experimental value, show that our mistunings are well below 0.5\%. 
These mistunings are allowed for in our final fits. 

\subsection{Two-point correlator fits}

We fit the two-point correlation functions described above 
as a function of the time separation, $t$, between source and sink. 
The aim is to extract the energies (masses) and amplitudes 
(giving decay constants) of the ground-state mesons in each channel. 
However it is important to allow for the systematic effect of excited 
states that are present in the correlation functions and can affect 
the ground-state values if they are not taken into account. 
We do this by fitting the correlators
to sums of exponentials associated with each energy eigenvalue and  
using Bayesian priors to
constrain the (ordered) excited states in the standard way~\cite{Lepage:2001ym}. 
The pseudoscalar correlators are fit to
\begin{eqnarray}
\label{eq:psfit}
  C_P(t) &=& \sum_i A^P_i f(E^P_i, t) , \\ 
f(E,t) &=& e^{-E t} + e^{-E(T-L_t)} \, . \nonumber
\end{eqnarray}
The vector correlators require a more complicated form because of the 
presence of opposite
parity states as a result of the use of staggered quarks:
\begin{equation}
\label{eq:vfit}
  C_V(t) = \sum_i \left(A^V_i f(E^V_i, t) - (-1)^t A^{V,o}_i f(E^{V,o}_i, t) \right) .
\end{equation}

We cut out the correlator values 
at low values of $t$, below some value
$t_{\mathrm{min}}$ ($5 - 10$) where excited state contamination is most pronounced. 
We also use a standard procedure (see Appendix D of~\cite{Dowdall:2019bea}) 
to avoid underestimating the low eigenvalues of the correlation matrix 
and hence the uncertainty. 

The fit parameters that we need from Eq.~\ref{eq:psfit} and~\ref{eq:vfit} are 
the mass of the ground-state ($E^P_0$ and $E^V_0$) and the amplitude ($A^P_0$ 
and $A^V_0$). 
From the amplitude we determine the decay constant, see Section~\ref{sec:jpsidecay}. 

\subsection{QED formalism}
\label{sec:incqed}

We perform calculations in both lattice QCD and in lattice QCD with 
quenched QED. By quenched QED we mean 
that we include effects from the valence quarks having electric charge but 
we neglect effects from the electric charge of the sea quarks. 
We will first describe how we include QED and then 
discuss the expected impact on our 
results of not including 
the QED effects from the sea quarks.

To include quenched QED effects 
we generate a random momentum space photon
field $A_{\mu}(k)$ in Feynman gauge for each QCD gluon field configuration. 
This choice of gauge simplifies the
generation of the photon field as the QED path integral weight takes the form of
a Gaussian with variance $1/\hat{k}^2$ where $\hat{k}=2\mathrm{sin}(k_{\mu}/2)$. 
The results presented here do not depend
on this gauge choice. Once the momentum space field is generated zero modes
are set to zero using the QED$_L$ formulation~\cite{Hayakawa:2008an}. 
$A_{\mu}$ is then Fourier
transformed into position space. We have checked that these Feynman gauge $A_{\mu}$
fields produce the plaquette and average link expected from
$\mathcal{O}(\alpha_{\mathrm{QED}})$ perturbation theory (readily obtained 
from $\mathrm{O}(\alpha_s)$ calculations in lattice QCD 
with Wilson glue~\cite{weisz:1982zw, Hart:2004jn}). 
These gauge fields are exponentiated as $\mathrm{exp}(ieQA_{\mu})$
to give a U(1) field which is then multiplied into the QCD gauge links before
HISQ smearing. $Q$ is the quark electric charge in units of the proton charge $e$.

This approach is known as the stochastic approach to 
quenched QED~\cite{Duncan:1996xy}, in contrast to the 
perturbative approach of~\cite{deDivitiis:2013xla}. Since 
$\mathcal{O}(\alpha_{\mathrm{QED}})$ is already a very small effect, 
we are seeking here only to pin down the linear term in $\alpha_{\mathrm{QED}}$, 
fully nonperturbatively in $\alpha_s$. The two approaches should then 
give the same result to our level of accuracy; we use the stochastic 
approach because it is more straightforward (in the quenched case) 
and gives very precise results 
for the $c$ quarks we are interested in here. 

\subsection{Impact of quenching QED}
\label{sec:quenching}

Quenched QED affects only the valence quarks; the sea quarks remain uncharged. 
We expect the valence quark contribution to be by far 
the largest QED effect (although already very small as we will see) 
and discuss here the small systematic error that 
remains from ignoring sea quark QED effects. 
The impact of having $m_u=m_d$ in the sea for most of our results is 
at the same level and we also discuss that.   

We first discuss the determination of the lattice spacing in 
QCD with quenched QED. 
In such a calculation there is no coupling 
of QED effects to purely gluonic quantities. 
This means that the Wilson flow parameter $w_0/a$, measured 
on each ensemble, is unchanged from pure QCD. 
The physical value of $w_0$ that is 
used to determine the lattice spacing on each ensemble was determined 
in~\cite{fkpi} by matching the decay constant of 
the $\pi$ meson, $f_{\pi}$, in lattice QCD to that obtained from 
experiment. 
The experimental value of $f_{\pi}$ is obtained from 
measurement of the rate for $\pi \rightarrow \ell \nu [\gamma]$ decay where 
$[\gamma]$ indicates that the rate is fully inclusive of additional 
photons. The rate 
obtained is then adjusted 
to remove electromagnetic and electroweak 
corrections and to give a `purely leptonic 
rate' corresponding to weak annihilation at lowest order in the absence 
of QED~\cite{freview-pdg}. Combining this with  
a determination of $|V_{ud}|$ from nuclear $\beta$ 
decay~\cite{Tanabashi:2018oca} gives an experimental value of 
$f_{\pi} \equiv f_{\pi}^{\mathrm{expt}}$ which is a `pure QCD' 
value, albeit that for a physical $\pi^+$ meson. 
The dominant uncertainty in $f_{\pi}^{\mathrm{expt}}$ is that 
from the remaining uncertainty in the electromagnetic corrections to 
the experimental rate, 
mainly from the hadronic-structure dependent contributions to 
the emission of additional photons. This is set at
0.1\% in~\cite{freview-pdg}. 

Because $f_{\pi}^{\mathrm{expt}}$ is a pure QCD quantity it 
can be used to set the lattice spacing in lattice QCD in a way 
that should be minimally different for lattice QCD+QED\footnote{Indeed, 
$f_{\pi}$ cannot readily be calculated in lattice QCD+QED because 
of infrared QED effects from an electrically charged $\pi^+$. 
Calculations have been done that confirm the size of radiative 
corrections to $f_{\pi}$, however~\cite{DiCarlo:2019thl}.}.  
Small differences might still be expected between the lattice 
QCD $f_{\pi}$ and the experimental value from the way that 
the quark masses are tuned in a pure QCD scenario. 
The lattice QCD calculation in~\cite{fkpi}
used $m_u=m_d$ and tuned the average mass, $m_l$, to the 
experimental mass of the $\pi^0$, which is the mass that both 
neutral and charged $\pi$ mesons have in the absence of QED, 
up to quadratic corrections in the $u-d$ mass difference. 
An uncertainty was included in the $\pi^0$ mass to allow 
for these corrections, taking an estimate from chiral 
perturbation theory~\cite{Amoros:2001cp}. We expect the impact of 
such effects to be tiny, well below 0.1\%. They are at the same level 
as potential effects from QED in the sea and would therefore be only 
possible to pin down with a calculation that included the impact 
of having electrically charged quarks in the sea.  

These expectations are backed up by recent lattice QCD+QED 
results~\cite{Borsanyi:2020mff} that used 
the $\Omega$ baryon mass to determine $w_0$. The impact 
of QED for the sea quarks was included to first order in 
$\alpha_{\mathrm{QED}}$. No effects linear in $m_u-m_d$ are 
expected in $M_{\Omega}$ because, like $f_{\pi}$ and $M_{\pi}$ 
above, it is symmetric 
under $u \leftrightarrow d$ interchange. Strong-isospin 
breaking effects were therefore ignored. The impact of QED 
in the sea on $w_0M_{\Omega}$ was found to be $\mathcal{O}(0.01\%)$, 
whereas the effect of QED for the valence quarks 
(already allowed for in the $f_{\pi}$ analysis) 
was $\mathcal{O}(0.05\%)$. The final value of $w_0$ using $M_{\Omega}$ 
from~\cite{Borsanyi:2020mff} 
agrees well with the result using $f_{\pi}$ from~\cite{fkpi}, 
although the uncertainties in both cases are completely dominated 
by those from the pure QCD, isospin-symmetric part of the calculation. 

From this we conclude that, at the sub-0.1\% level, 
we can compare 
lattice QCD plus quenched QED with pure lattice QCD using the same 
value of the lattice spacing, determined from $f_{\pi}$, in 
both calculations. 

The impact of quenching QED on charmonium quantities follows 
a similar discussion to that for the lattice spacing 
because interaction with the electric 
charge of the sea quarks is suppressed by sea quark mass effects 
and by powers of $\alpha_s$. Since the sum of electric charges 
of $u$, $d$ and $s$ sea quarks is zero, 
QED interaction between valence $c$ quarks 
and light sea quarks will be suppressed by sea quark mass differences. 
The impact of $c$ quarks in the sea is already small and so 
we can safely neglect the even smaller QED effects from valence/sea 
$c$ quark interactions. The leading sea-quark 
QED effect will then come from photon exchange across a sea-quark 
bubble at $\mathcal{O}(\alpha_s^2\alpha_{\mathrm{QED}})$~\cite{Hatton:2019gha}
in perturbative language. 
The expected size is then 10\% of that of the QED effects from valence $c$ quark 
interactions, which we will see are themselves typically a small fraction of 1\%.. 

\subsection{Impact of having $m_u=m_d$ in the sea}
\label{sec:seaeffects}

As discussed above, we expect the effects of having 
$m_u=m_d$ in the sea, i.e. not including strong-isospin breaking 
effects, to be negligible. For both the scale-setting determination and for 
the charmonium quantities themselves, pure strong-isospin breaking effects 
are quadratic in $(m_u-m_d)/\Lambda$. Effects linear in the sea 
quark masses are already small, for example 
an $\mathcal{O}(5\,\mathrm{MeV})$ shift in the average $u/d$ 
quark mass produces a 1\% effect 
on $w_0/a$~\cite{Chakraborty:2014aca}. We might then expect quadratic  
effects to be $\mathcal{O}(0.01\%)$. 

We can provide a test of these expectations from results on gluon 
ensemble sets 3A and 3B
that differ only in the values of $m_u$ and $m_d$ in the sea for the 
same average (which has its physical value). Set 3B has $m_d/m_u$ 
set to the expected ratio~\cite{Basak:2018yzz}. 
We see from Table~\ref{tab:ensembles} 
that the determinations of $w_0/a$ agree at the level of their 0.03\% 
uncertainties. In contrast,
the $w_0/a$ value on set 3A differs 
by a clearly visible 0.40(5)\% from that on set 3, which has sea 
quark masses that are slightly mistuned from the physical point  
by an amount (summing over $u$, $d$ and $s$) equal 
to 5\%~\cite{Chakraborty:2014aca} of $m_s$. 

\begin{figure}
  \includegraphics[width=0.47\textwidth]{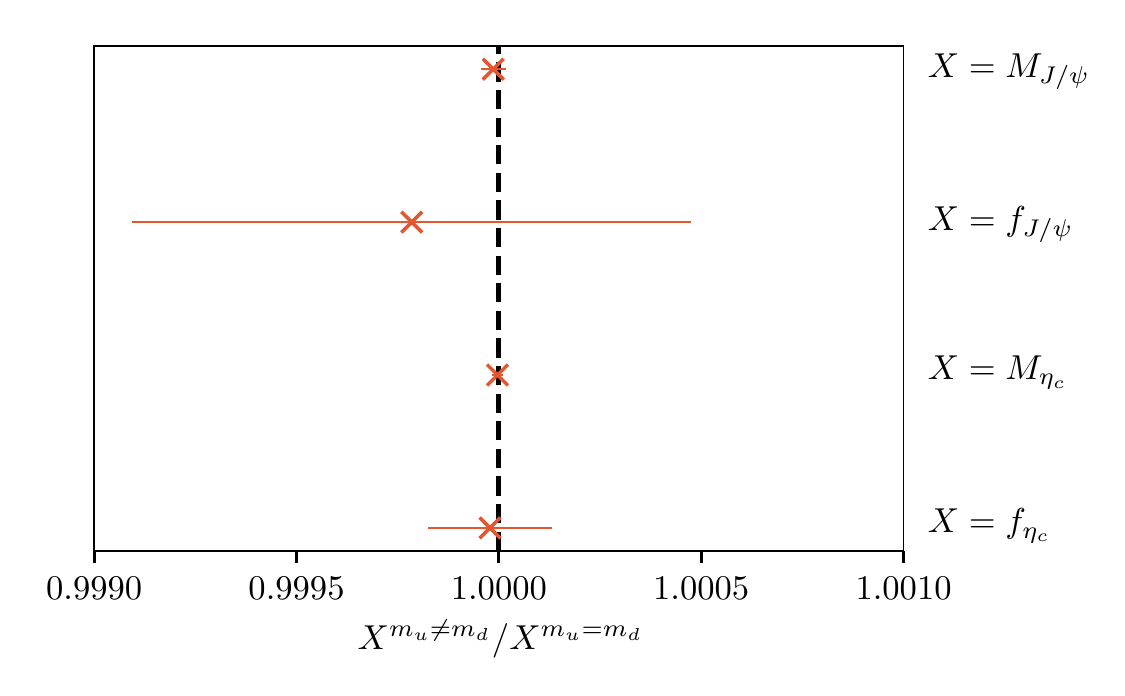}
  \caption{ A test of the impact of strong-isospin breaking in the sea 
through a comparison of results for the charmonium quantities 
given in Table~\ref{tab:3AB} for sets 3A and 3B (Table~\ref{tab:ensembles}).
These sets differ only in the $u/d$ quark masses in the sea. Set 3A has 
$m_u=m_d$ and set 3B has $m_d/m_u=2.18$ with the same average value. 
The plot above gives the ratio of results in lattice units, 
for the same $am_c^{\mathrm{val}}$, for the $J/\psi$ 
and $\eta_c$ masses and decay constants. 
The masses agree to within their 0.001\% statistical errors and the 
decay constants agree to within their statistical errors (0.05\% for 
the $J/\psi$). 
 }
  \label{fig:isospincomp}
\end{figure}

Table~\ref{tab:3AB} compares values for the $J/\psi$ and $\eta_c$ 
masses in lattice units on these two ensembles for the same 
$am_c^{\mathrm{val}}$ and the ratio of the two results is plotted 
in Figure~\ref{fig:isospincomp}. The meson masses in lattice units agree on the two 
ensembles to within their statistical errors which are at the level of 
0.001\%. We also 
tabulate results for the decay constants and plot the ratio 
of these values. Again agreement is seen between results on set 3A 
and set 3B. They provide a weaker constraint because of much 
larger statistical errors, but nevertheless they agree within 0.05\%. 
Notice that this comparison does not allow for possible changes in 
$w_0/a$ in the two cases. As discussed above, this could be at the 
0.01\% level. Again we can contrast the agreement between sets 
3A and 3B with the results on set 
3 where, for the same $am_c^{\text{val}}$, the $\eta_c$ mass differs 
by 0.02\%.  

We conclude that we can safely neglect strong-isospin breaking 
in the sea and proceed with calculations on gluon field ensembles 
with $m_u^{\mathrm{sea}}=m_d^{\mathrm{sea}}$. 

\begin{table}
  \caption{Results for the $\eta_c$ and $J/\psi$ masses 
and decay constants (see Section~\ref{sec:jpsidecay} 
for how these are calculated) 
in lattice units on gluon field ensembles sets 3A 
and 3B (Table~\ref{tab:ensembles}). Both 
sets have the average $u/d$ quark mass in the sea set to its physical value 
but set 3B has $m_d/m_u$ also set to the expected ratio.  
The valence $c$ quark mass was set to 0.863 in lattice units. 
  }
  \label{tab:3AB}
\begin{ruledtabular}
\begin{tabular}{lll}
 & 3A ($n_f$=2+1+1) & 3B ($n_f$=1+1+1+1) \\
\hline
$aM_{\eta_c}$ & 2.288139(19) & 2.288131(25)  \\
$aM_{J/\psi}$ & 2.375618(44) & 2.375585(60)  \\
$af_{\eta_c}$ & 0.366596(38) & 0.366588(41)  \\
$af_{J/\psi}/Z_V$ & 0.41799(16) & 0.41790(24)  \\
\end{tabular}
\end{ruledtabular}
\end{table}

\begin{figure}
  \includegraphics[width=0.47\textwidth]{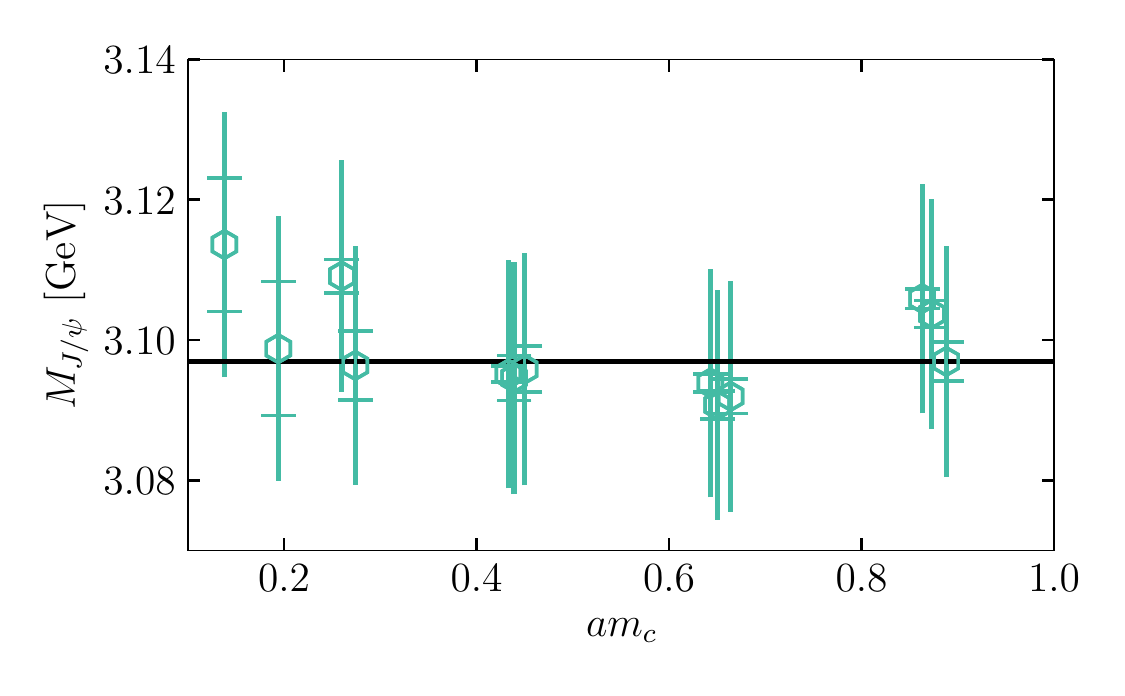}
  \includegraphics[width=0.47\textwidth]{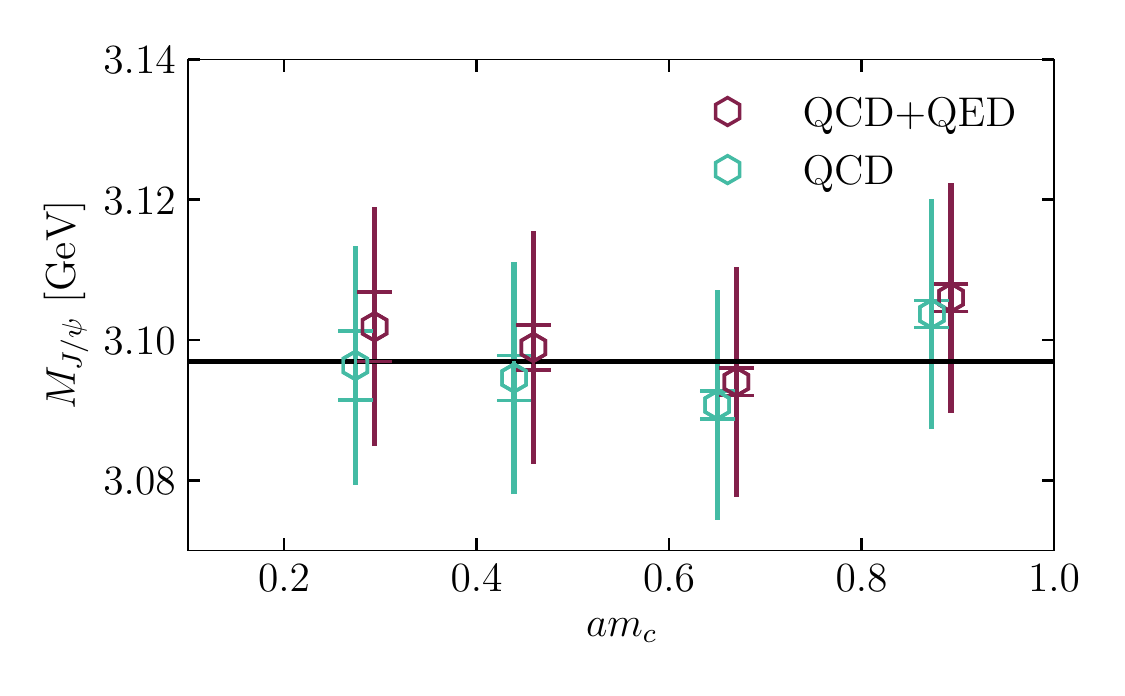}
  \caption{Top panel: Pure QCD $J/\psi$ masses on all 15 sets from
  Table~\ref{tab:ensembles} using the valence masses in that Table. Two error
  bars are shown. The error can be broken into parts that are uncorrelated
  between different sets and the contribution from fixing the lattice spacing 
from the physical value of $w_0$ which is correlated.
  The outer error bar shows the full uncertainty and the inner bar the
  uncertainty without the contribution from $w_0$.
  Bottom panel: The $J/\psi$ masses on sets 2, 6, 10 and 12 with and without the
  inclusion of quenched QED. On each set the same valence mass (and lattice spacing) 
was used for both
  pure QCD and QCD+QED but the points have been separated on the $x$-axis for
  clarity. The error bars are the same as for the top panel, but note that here
  there is a correlation of the uncertainty from $w_0/a$ for the QCD and QCD+QED
  results on each set. The QCD+QED results are above the pure QCD results in every case. }
  \label{fig:jpsi-masses}
\end{figure}

\subsection{A first look at quenched QED effects}
\label{sec:firstlook}

An estimate of the size of corrections from quenched 
QED (often simply referred to as QED in what follows) in 
charmonium systems can be obtained
by studying the effect on the $J/\psi$ mass. 
The bottom panel of Fig.~\ref{fig:jpsi-masses}
shows the $J/\psi$ mass for the same valence $c$ mass 
for both QCD+QED and pure QCD calculations 
on sets 2, 6, 10 and 12. 
The
QCD+QED and QCD results at the same lattice spacing are separated on the
$x$-axis for clarity. 
All points share a correlated uncertainty (the outer error bar) from $w_0$ and 
this dominates the uncertainty. The
uncorrelated error is shown by the smaller inner error bar. 
Note that the points at the same lattice
spacing are also correlated through their $w_0/a$ value. 
The shift of the mass in QCD+QED compared to pure QCD is 
very small, at the level 
of 0.1\%, and is upwards.  

When discussing QCD+QED and pure QCD calculations of some quantity $X$ we will
use the notation $X[\mathrm{QCD}+\mathrm{QED}]$ and $X[\mathrm{QCD}]$
respectively. We will often consider the ratio of the two
for which we
will use the shortened notation $R^{(0)}_{\mathrm{QED}}[X]$.
$R^{0}_{\mathrm{QED}}$ will refer to the `bare' ratio defined using the 
same bare quark mass $am_c$ in both QCD+QED and pure QCD calculations. 
$R_{\mathrm{QED}}$  will refer to the final QED-renormalised 
ratio which includes the impact 
of retuning the $c$ quark mass to give the experimental 
$J/\psi$ mass in both the QCD+QED and pure QCD cases. 
So
\begin{eqnarray}
\label{eq:Rdef}
R^{0}_{\mathrm{QED}}[X]  &\equiv& \left. \frac{X[\mathrm{QCD+QED}]}{X[\mathrm{QCD}]}\right|_{{\mathrm{fixed}}\, am_c}  \\
R_{\mathrm{QED}}[X]  &\equiv&  \left.\frac{X[\mathrm{QCD+QED}]}{X[\mathrm{QCD}]}\right|_{{\mathrm{fixed}}\, M_{J/\psi}} \, . \nonumber
\end{eqnarray}
As shown in Fig.~\ref{fig:jpsi-masses}
the bare $c$ quark mass has to be re-adjusted 
downwards for QCD+QED relative to pure QCD.   
 
\subsection{Fitting strategy}
\label{sec:fitstrategy}

We have results in pure QCD for all of the sets in 
Table~\ref{tab:ensembles} and QCD+QED results on a subset 
of ensembles. To be able
to simultaneously account both for the `direct' effects of QED and 
for the effects of valence $c$ mass mistuning,
which may be similarly sized, we choose to fit all of this data 
in a single fit
for each quantity we consider. 
The generic form of the fit we use for a quantity
$X$ is
\begin{eqnarray} \label{eq:X-fit}
  X(a^2,Q) &=& x \bigg [1 + \sum_{i=1}^{5}c_{a}^{(i)} (am_c)^{2i} + \\  
&& \hspace{-2.0em} c_{m,{\mathrm{sea}}} \delta_m^{\mathrm{sea},uds} \{ 1+ c_{a^2,\mathrm{sea}}(\Lambda a)^2 + c_{a^4,\mathrm{sea}}(\Lambda a)^4 \} + \nonumber \\ 
&&  c_{c,{\mathrm{sea}}}\delta_m^{\mathrm{sea},c} + c_{c,\mathrm{val}}\delta_m^{\mathrm{val},c}+ \nonumber \\
&& \hspace{-5.0em} \alpha_{\mathrm{QED}} Q^2
  \{ c_{\mathrm{QED}} + \sum_{p=1}^{3} c_{aQ}^{(p)} (am_c)^{2p} +
   c_{\mathrm{val},Q} \delta_m^{\mathrm{val},c} \} \bigg ] . \nonumber
\end{eqnarray}

Here $Q$ is the valence quark electric charge (in units of $e$) 
used in the calculation and is therefore 0 in
pure QCD. The pure QCD value of this fit at the physical point 
(the continuum limit with quark masses set to their physical values)
is $x$. The value including quenched QED
corrections is $x[1+\alpha_{\mathrm{QED}}Q^2 c_{\mathrm{QED}}]$. 
Note that the factor of $\alpha_{\mathrm{QED}}$ multiplying the QED part of the fit
function is there so that the fit parameters are order 1. 
The stochastic method that we use includes 
in principle all orders of $\alpha_{\mathrm{QED}}$
but we expect to see only linear terms, 
including $\alpha_{\mathrm{QED}}\alpha_s^n$ pieces. 

Our fits are typically to 15 pure QCD data points 
and 4 QCD+QED points. The pure QCD points do not include those on 
sets 6 and 7 which are used to test finite-volume effects or on sets 
3A and 3B that are used to test strong-isospin breaking effects, but they 
do include 
additional results at mistuned $c$ quark masses to test mistuning effects. 

We now describe each of the terms in Eq.~(\ref{eq:X-fit}) in turn. 
The $(am_c)^i$ terms on the first line account for discretisation 
effects. Because we are dealing with heavy quarks here, the scale of 
discretisation effects can be set by $m_c$ and will typically be larger 
than those for light-quark quantities. Since $m_c > \Lambda_{\mathrm{QCD}}$ any 
discretisation effects set by scale $\Lambda_{\mathrm{QCD}}$ will simply 
appear as $m_c$-scale discretisation effects with a small coefficient.   

The terms on the second line allow for mistuning of the sea 
$u/d$ and $s$ masses and discretisation effects in that mistuning 
(we shall see that those are important for the hyperfine 
splitting). The total of the mistuning of the sea masses is defined as 
in~\cite{Chakraborty:2014aca}: 
\begin{equation} \label{eq:deltam}
   \delta_m^{\mathrm{sea},uds} =  \frac{2m_l^{\mathrm{sea}} + m_s^{\mathrm{sea}} - 2m_l^{\mathrm{phys}} - m_s^{\mathrm{phys}}}{10 m_s^{\mathrm{phys}}} .
\end{equation}
$m_s^{\mathrm{phys}}$ is taken from
\cite{Chakraborty:2014aca} or, where not available on the finest lattices, 
calculated from the tuned $c$ quark mass and the $m_c/m_s$
ratio given in \cite{Chakraborty:2014aca}.
The value of $\Lambda$ in the discretisation effects multiplying the sea-quark 
mistuning is taken as 1 GeV ($\sim m_c$). 

The effect of mistuning the
charm mass in the sea is included in the third line of Eq.~(\ref{eq:X-fit}) using
\begin{equation}
  \delta_m^{\mathrm{sea},c} = \frac{m_c^{\mathrm{sea}} - m_c^{\mathrm{phys}}}{m_c^{\mathrm{phys}}} .
\end{equation}
The values of $m_c^{\mathrm{phys}}$ are taken from~\cite{Chakraborty:2014aca}. 
Although this used a slightly different tuning method the differences 
are negligible 
for this purpose. We have tested that including discretisation effects for 
this term has no effect on the fit.  

Mistuning in the valence mass is accounted for through
$\delta_m^{\mathrm{val},c}$ on the third and fourth lines of Eq.~(\ref{eq:X-fit}). 
We define this as 
\begin{equation}
  \delta_m^{\mathrm{val},c} = \frac{M_{J/\psi}-M^{\mathrm{expt}}_{J/\psi}}{M^{\mathrm{expt}}_{J/\psi}} .
\end{equation}
where $M_{J/\psi}$ is our lattice result for that ensemble in either the 
QCD or the QCD+QED case. 
Thus $\delta_m^{\mathrm{val},c}$ is zero (the valence $c$ quark mass is tuned) 
when the $J/\psi$ mass takes its experimental value on 
each ensemble (and with or without QED). 
The fit parameters $c_{c,\mathrm{val}}$ and $c_{\mathrm{val},Q}$ then 
determine the dependence on the valence $c$ mass of the quantity being fit, and 
the QED corrections to that dependence, respectively. 
The experimental value of the $J/\psi$ mass is 3.0969 GeV~\cite{Tanabashi:2018oca} with 
negligible uncertainty. 

In order to make use of the correlations between our QCD+QED and pure QCD
results on the same gluon field configurations 
we perform simultaneous
fits to the correlators in each case. The fits then 
capture the correlations and we can propagate them to the fit 
of Eq.~(\ref{eq:X-fit}).
At the same time it allows us to determine the ratio of QCD+QED 
to pure QCD for the quantities that we will study. We will give 
results for these ratios in the sections that follow. 

The fit form of Eq.~(\ref{eq:X-fit}) has been constructed 
such that the coefficients (apart from $x$)
are expected to be of order 1. We therefore use priors of $0\pm 1$ for all fit
parameters except $x$ for which we take a prior width on its expected value
of 20\% (the prior mean for $x$ depends on the quantity being fitted).

\section{Hyperfine splitting} \label{sec:hyperfine}

\subsection{Pure QCD} \label{sec:hypQCD}

\begin{table}
  \caption{The $\eta_c$ and $J/\psi$ masses and their difference
  ($a\Delta M_{\mathrm{hyp}}$) in pure QCD on each set in lattice units.
The values of $am_c^{\mathrm{val}}$ are those given in Table~\ref{tab:ensembles} 
except for two cases with a deliberately mistuned $c$ quark mass: 
set 6 denoted by a * where $am_c=0.643$ and set 14 denoted by a $\dag$ where 
$am_c$=0.188. 
  The pseudoscalar and vector correlator fits have been performed separately and
  the correlations between $aM_{\eta_c}$ and $aM_{J/\psi}$ have therefore been
  ignored because they have little impact.}
  \label{tab:masses}
\begin{ruledtabular}
\begin{tabular}{llll}
Set & $aM_{\eta_c}$ & $aM_{J/\psi}$ & $a\Delta M_{\mathrm{hyp}}$ \\
\hline
1 & 2.331899(72) & 2.42072(19) & 0.08883(20) \\
2 & 2.305364(39) & 2.39308(14) & 0.08772(14) \\
3 & 2.287707(26) & 2.37476(21) & 0.08705(21) \\
\hline
4 & 1.876536(48) & 1.94364(10) & 0.06710(11) \\
6 & 1.848041(35) & 1.914749(67) & 0.066708(76) \\
6$^*$ & 1.834454(34) & 1.901479(66) & 0.067025(74) \\
8 & 1.833950(18) & 1.900441(39) & 0.066491(43) \\
\hline
9 & 1.366839(72) & 1.41568(16) & 0.04884(17) \\
10 & 1.342455(21) & 1.391390(43) & 0.048935(48) \\
11 & 1.329313(18) & 1.378237(51) & 0.048924(54) \\
\hline
12 & 0.896675(24) & 0.929860(54) & 0.033185(59) \\
13 & 0.862689(22) & 0.895650(37) & 0.032961(43) \\
\hline
14 & 0.666818(39) & 0.691981(54) & 0.025163(67) \\
14$^\dag$ & 0.652439(56) & 0.67798(14) & 0.02554(15) \\
\hline
15 & 0.496991(47) & 0.516126(68) & 0.019135(82) \\
\end{tabular}
\end{ruledtabular}
\end{table}

\begin{figure}
  \includegraphics[width=0.47\textwidth]{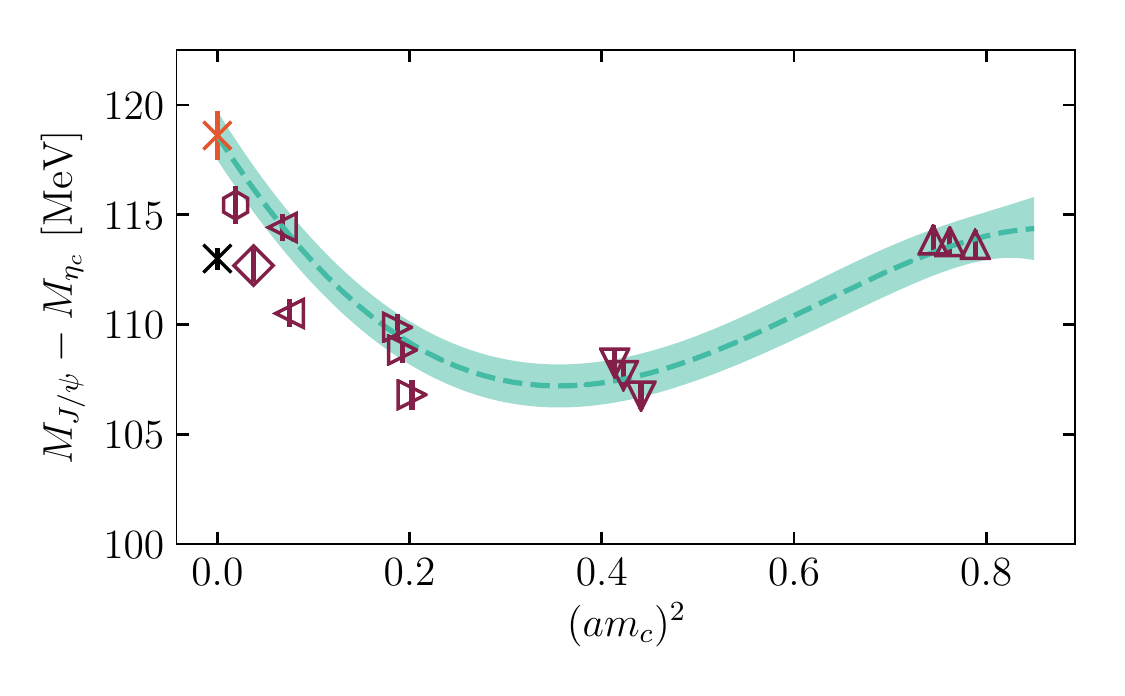}
  \caption{The charmonium hyperfine splitting as a function of lattice spacing on pure QCD ensembles.
Our results are from ensembles including $u$, $d$, $s$ and $c$ quarks in the sea with 
varying $u/d$ average quark mass. 
   The raw lattice results, for well-tuned $c$ quark masses, 
are given by symbols with error bars. The error bars include 
both statistical uncertainties and those from determining the lattice spacing from $w_0$, 
which are correlated between values.   
The results on each group of ensembles with approximately the same lattice spacing are 
given the same symbol. Within these groups, the results go from right to left 
as the $u/d$ quark mass changes from $m_s/5$ to the physical value. 
Notice the small range on the $y$-axis; this is the results of discretisation 
effects being so small for the 
HISQ action. Results at mistuned $c$ masses are not plotted but are 
included in the fit.  
  The fit line is the output of the fit from Eq.~(\ref{eq:X-fit}) at physical 
quark masses and with $Q=0$. 
The orange cross gives our result in the continuum limit 
for physical quark masses. The black cross gives the experimental average result~\cite{Tanabashi:2018oca}. 
 }
  \label{fig:hyperfine}
\end{figure}

\begin{table}
  \caption{QCD+QED $\eta_c$ and $J/\psi$ masses and hyperfine splitting 
  presented as the ratio of the QCD+QED
  result to the pure QCD one on that set. Correlations between the calculations 
in the QCD+QED and pure QCD cases are used in the determination of the 
ratio and result in the very high statistical accuracy obtained. 
Note that the ratio is calculated for the same $am_c$ value in the two cases i.e.\ 
the ratio given here does {\it{not}} include the impact of retuning the $c$ quark mass. 
  }
  \label{tab:masses-qed}
\begin{ruledtabular}
\begin{tabular}{llll}
Set & $R^0_{\mathrm{QED}}[M_{\eta_c}]$ & $R^0_{\mathrm{QED}}[M_{J/\psi}]$ & $R^0_{\mathrm{QED}}[\Delta M_{\mathrm{hyp}}]$ \\
\hline
2 & 1.000450(26) & 1.000750(27) & 1.0086(10) \\
6 & 1.0008335(59) & 1.0010742(81) & 1.00774(28) \\
10 & 1.0011861(54) & 1.0014044(76) & 1.00739(26) \\
12 & 1.0015755(48) & 1.001787(11) & 1.00750(33) \\
\end{tabular}
\end{ruledtabular}
\end{table}

\begin{figure}
  \includegraphics[width=0.47\textwidth]{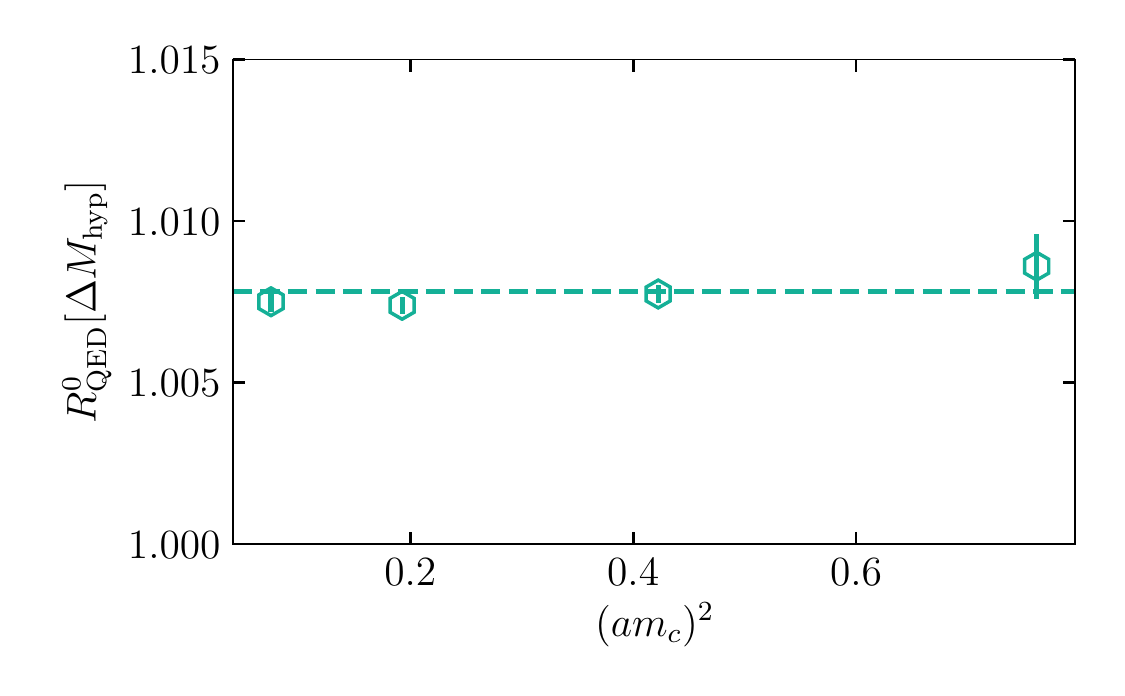}
  \caption{The fractional effect of quenched QED 
on the charmonium hyperfine splitting plotted against $(am_c)^2$. 
The fractional effect is determined at the same $am_c$ value, i.e.\ it does 
not include $c$ quark mass retuning effects. The dashed line is horizontal and 
shows the weighted average value. 
The results show the precision that can be obtained by capitalising on the 
correlations between QCD+QED and QCD. This enables a clear demonstration that 
the impact of quenched QED here is not dependent on the lattice spacing. 
}
  \label{fig:hyp-qed}
\end{figure}

The hyperfine splitting, $\Delta M_{\mathrm{hyp}}$, is calculated on each ensemble as the difference
of the vector and pseudoscalar ground state masses, in lattice units, divided
by the lattice spacing. 
The results for
$aM_{\eta_c}$ and $aM_{J/\psi}$ and their difference 
are given in Table~\ref{tab:masses} for the pure QCD case. 
Although the pseudoscalar and vector
correlators on each configuration are correlated the fit outputs for the vector
correlator dominate the uncertainties and so the correlations have 
very little effect
as a result.

The pure QCD results are plotted in Fig.~\ref{fig:hyperfine} along
with the fit form of Eq.~(\ref{eq:X-fit}) for the pure QCD case (i.e.\ $Q=0$).
Note the small range of the $y$-axis - this is possible for our results 
because we have a highly-improved quark action with small discretisation 
errors. Since all tree-level $a^2$ errors have been removed, the shape of
the curve reflects the fact that higher-order $a^4$ and $a^6$ errors are 
visible; discretisation errors of this kind are present in all formalisms 
of course, but often they are hidden below much larger $a^2$ effects and 
consequently overlooked.  
Note also the clear dependence on the light sea quark mass seen on the
finest lattices. 
To pin down the value of the valence mass mistuning parameter, $c_{\mathrm{val}}$, 
we include results at deliberately mistuned $c$ quark masses 
(see Table~\ref{tab:masses}). These are not 
shown in the Figure but are included in the fit. 
The result for the hyperfine splitting in the pure QCD case in the 
continuum limit and for physical quark masses is 118.6(1.1) MeV, which is higher 
than the experimental average value, as is clear in Fig.~\ref{fig:hyperfine}. 
In order to understand what this means, we need to quantify all possible 
sources of small systematic effects in our calculation, including those from QED.  

\subsection{Impact of Quenched QED} \label{sec:hypQED}

The fractional direct effect of quenched QED on the $\eta_c$ and $J/\psi$ masses and the
hyperfine splitting 
are given in Table~\ref{tab:masses-qed}. The correlation between 
the QCD+QED and the pure QCD results enables very high statistical accuracy to be obtained 
in the ratio. The inclusion of
quenched QED shifts both the $\eta_c$ and $J/\psi$ masses up by
$\mathcal{O}(0.1\%)$, depending on lattice spacing, 
at a given $am_c$ value. 
Although these mass shifts are small, there is a difference 
between the shift for the $J/\psi$ and that for $\eta_c$ and so the inclusion of 
quenched QED also changes the hyperfine splitting. 
The impact here is more substantial, ~0.7\%, because the hyperfine splitting is 
so much smaller.
The size of the direct QED effect on the hyperfine splitting can be simply estimated by 
replacing $C_F\alpha_s$ by $Q^2\alpha_{\mathrm{QED}}$ in a potential model 
estimate of the splitting. This gives a fractional effect 
of $\alpha_{\mathrm{QED}}/(3\alpha_s)$, consistent with what we find. 

The values of $R^0_{\mathrm{QED}}[\Delta M_{\mathrm{hyp}}]$ are plotted against
$(am_c)^2$ in Fig.~\ref{fig:hyp-qed}. This shows that the results are 
consistent across
all lattice spacings and thus discretisation effects in this ratio are smaller than 
for the masses themselves. 

\begin{figure}
  \includegraphics[width=0.47\textwidth]{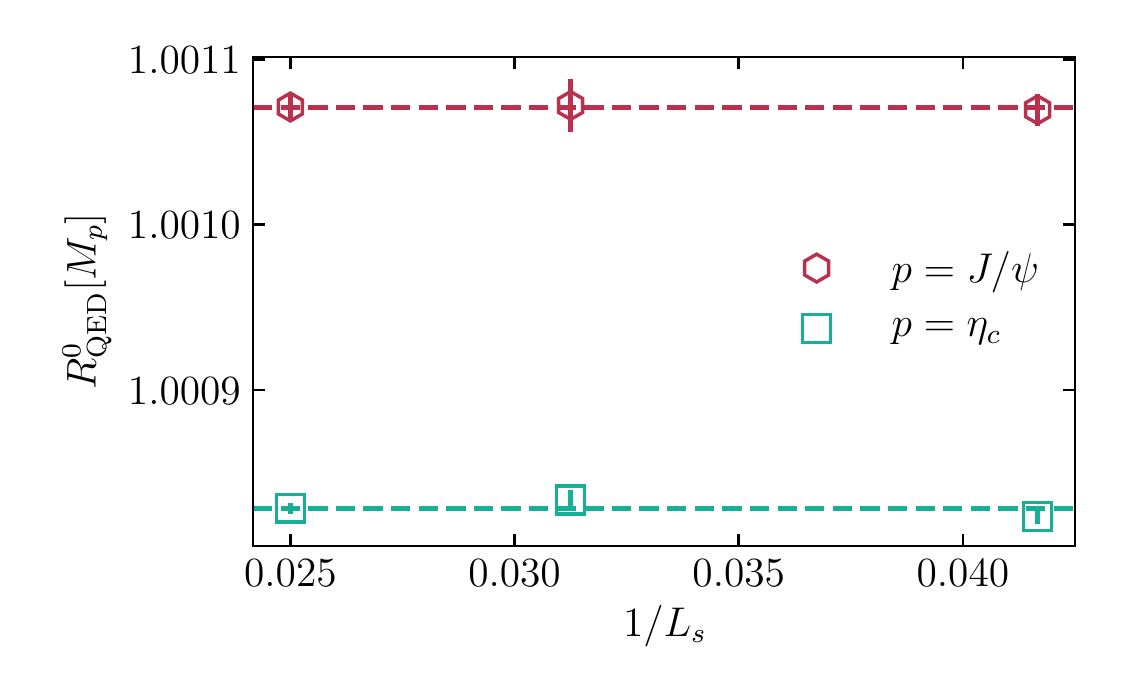}
  \caption{The fractional shift in the $J/\psi$ and $\eta_c$ masses 
from the inclusion of quenched
  QED plotted as a function of $1/L_s$ on sets 5, 6 and 7.  
The dashed lines are horizontal lines at the weighted average values. 
}
  \label{fig:mass-qed-vol-dep}
\end{figure}

Finite-volume effects are an issue in general for QED 
corrections to meson masses but we expect them to be small for the 
electrically neutral and spatially small charmonium mesons that we study here. 
In~\cite{Davoudi:2014qua} it is shown that the finite volume
expansion for electrically neutral mesons starts at $\mathcal{O}(1/L_s^4)$. 
In Fig.~\ref{fig:mass-qed-vol-dep} we compare results for the fractional effect 
of QED on the $J/\psi$ and $\eta_c$ as a function of $1/L_s$. 
This calculation is done on sets 5, 6 and 7 (see Table~\ref{tab:ensembles}) 
which differ only in their spatial extent. We see no finite-volume effects  
to well within 0.01\%, and we therefore ignore such effects. 

\begin{figure}
  \includegraphics[width=0.47\textwidth]{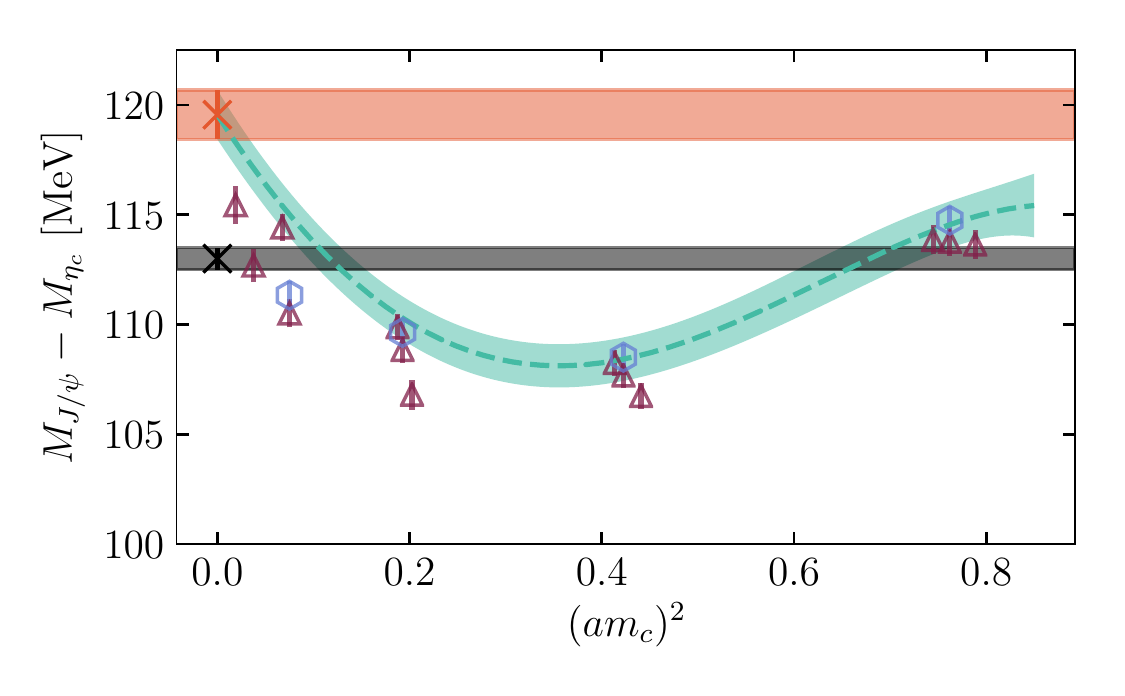}
  \caption{The charmonium hyperfine splitting as a function of lattice spacing, 
including both QCD+QED and pure QCD points. The red open triangles are 
the same lattice results 
as in Fig.~\ref{fig:hyperfine}. The additional QCD+QED points are 
given as blue open hexagons.  
  The green fit band is the output of the fit from Eq.~(\ref{eq:X-fit}), but now with the 
$c$ quark electric charge, $Q=2/3$. 
The orange cross and orange band gives our result in the continuum limit 
for physical quark masses. The black cross and black band 
gives the experimental average result~\cite{Tanabashi:2018oca}. 
 }
  \label{fig:hyp-all}
\end{figure}

\begin{figure}
  \includegraphics[width=0.47\textwidth]{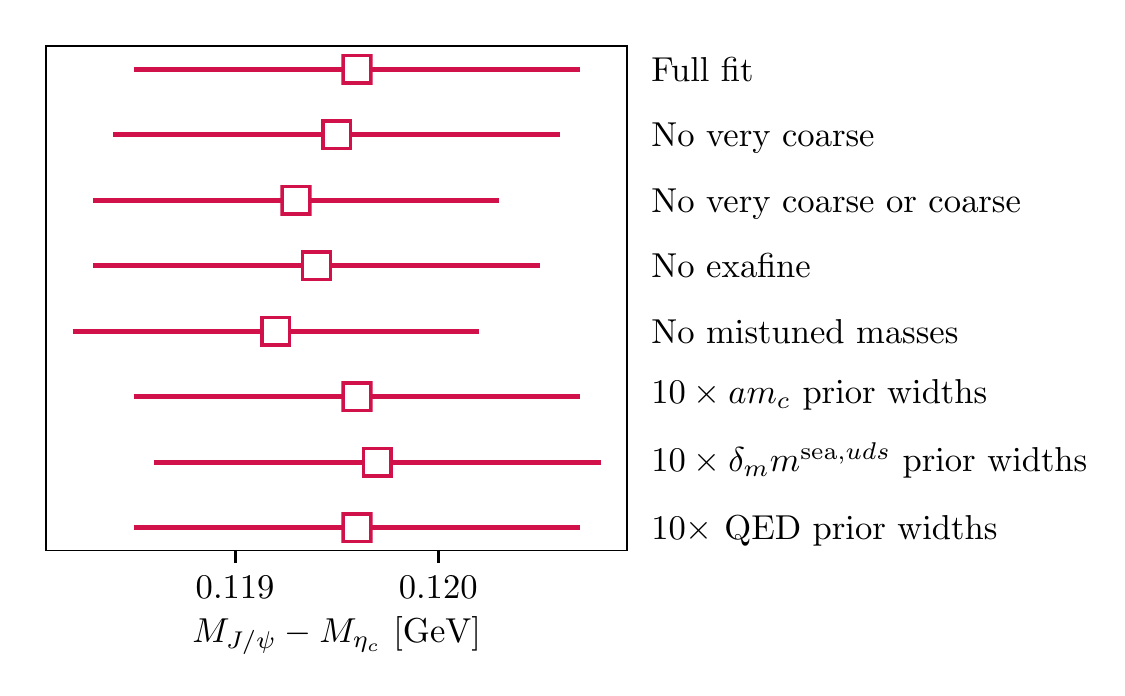}
  \caption{A stability analysis for our fit (see text) to the charmonium hyperfine 
splitting. 
Our full fit result is shown at the top. The lines with error bars below 
this show the result for different modifications of this fit. From 
the top, these include 
missing out results on particular sets of configurations, then missing 
out the results using mistuned $c$ quark masses and then changing the priors 
by a factor of ten on different sets of parameters in the fit. 
 }
  \label{fig:hyp-stability}
\end{figure}

Our results including both pure QCD and QCD+QED are shown in 
Fig.~\ref{fig:hyp-all}, plotted against $(am_c)^2$. The fit curve 
from Eq.~(\ref{eq:X-fit}) at physical quark masses is also shown. 
The fit has a $\chi^2/\mathrm{dof}$ of 0.59 and gives a hyperfine splitting 
in the continuum limit at physical quark masses of 119.6(1.1) MeV. 
In Figure~\ref{fig:hyp-stability} we show the results of a stability analysis 
for this fit. The figure shows that the fit is very robust.  

Taking the
(correlated) ratio between the physical value of the full QCD+QED fit 
and the physical value from the fit at $Q=0$ (i.e.\ the pure QCD result) 
we obtain $R_{\mathrm{QED}}[\Delta M_{\mathrm{hyp}}] =$ 1.00804(43). This ratio now does include the effect of retuning 
the $c$ quark mass to obtain the experimental $J/\psi$ mass when 
quenched QED is included. This retuning requires a reduction of the bare $c$ quark mass 
by $\mathcal{O}(0.1\%)$ (see Table~\ref{tab:masses-qed}) and 
this further increases the hyperfine splitting, but only by $\mathcal{O}(0.1\%)$. 
The impact of QED here is therefore dominated by the `direct' quenched 
QED effect. 

There is an additional pure QED contribution to the $J/\psi$ mass that has not 
been included yet since it is quark-line disconnected. 
This comes from a diagram in which
the $c\overline{c}$ annihilate into a photon which converts back into $c\overline{c}$.
The contribution of this diagram is
\begin{equation}
  \frac{8\pi \alpha_{\mathrm{QED}}Q^2}{M_{J/\psi}^2} \vert \psi(0) \vert^2 ,
\end{equation}
where $\psi(0)$ is the nonrelativistic $J/\psi$ wavefunction equal (in the
normalisation being used here) to 
$f_{J/\psi}\sqrt{M_{J/\psi}}/\sqrt{6}$~\cite{Gray:2005ur}. The
contribution evaluates to +0.7 MeV, 
which is a tiny, and completely negligible, 
effect for the $J/\psi$ meson 
mass (0.03\%). It has some impact on the hyperfine splitting because 
that is 30 times smaller and so we include it here. 
We add this contribution to our hyperfine splitting
result with a 30\% uncertainty from possible QCD corrections 
to give a final result of 
\begin{equation}
\label{eq:finalhyp}
M_{J/\psi} - M_{\eta_c} (\mathrm{connected}) = 120.3(1.1) \ \mathrm{MeV}.
\end{equation}

\begin{table}
  \caption{ Error budget for our final result for the charmonium hyperfine
  splitting including quenched QED corrections. The uncertainties shown are
  given as a percentage of the final result. The largest uncertainties are
  clearly from the determination of the lattice spacing. 
  }
  \label{tab:hf-errbudg}
\begin{ruledtabular}
\begin{tabular}{ll}
 & $\Delta M_{\mathrm{hyp}}$ \\
\hline
$a^2 \to 0$ & 0.13 \\
Pure QCD statistics & 0.24 \\
QCD+QED statistics & 0.08 \\
$w_0/a$ & 0.24 \\
$w_0$ & 0.87 \\
Valence mistuning & 0.02 \\
Sea mistuning & 0.06 \\
\hline
Total & 0.96 \\
\end{tabular}
\end{ruledtabular}
\end{table}

The error budget for our hyperfine splitting result is given in
Table~\ref{tab:hf-errbudg}. We follow
Appendix A of~\cite{Bouchard:2014ypa} for the meaning of the uncertainties 
contributing to the error budget. The majority of the uncertainty is associated with
the lattice spacing determination, either from the correlated $w_0$ uncertainty
or the individual $w_0/a$ uncertainties. This is not surprising because the 
hyperfine splitting is sensitive to uncertainties in the determination of the 
lattice spacing for the reasons discussed in~\cite{Donald:2012ga}. 
We have separated out the uncertainty
arising from the pure QCD data and the 
$R^0_{\mathrm{QED}}[\Delta M_{\mathrm{hyp}}]$ values from
Table~\ref{tab:masses-qed} which we label `Pure QCD Statistics' and `QCD+QED
Statistics' in Table~\ref{tab:hf-errbudg}. The sea mistuning uncertainty comes
from the $c_m$ coefficients in Eq.~\ref{eq:X-fit} and the valence mistuning
uncertainty from the $c_{\mathrm{val}}$ and $c_{\mathrm{val},Q}$ coefficients.
The $a^2 \to 0$ uncertainty is from the $c_a$ and $c_{aQ}$ coefficients.

Our final result is for the charmonium hyperfine splitting determined 
from quark-line connected correlation functions in QCD and including 
the impact of QED effects, through explicit calculation in quenched 
QED. We expect the effect of further QED effects in the sea to be negligible 
compared to our 1\% uncertainty.  
The only significant Standard Model effect then missing is that of quark-line disconnected
diagrams in which the $c\overline{c}$ annihilate to gluons. 
We expect this effect to be much larger for the $\eta_c$ than for the $J/\psi$ 
so a comparison of our result for the hyperfine splitting to experiment can 
yield information on the size and sign of the annihilation contribution 
to the $\eta_c$ mass. This is discussed in the next subsection. 

\subsection{Discussion: Hyperfine Splitting}

\begin{figure}
  \includegraphics[width=0.47\textwidth]{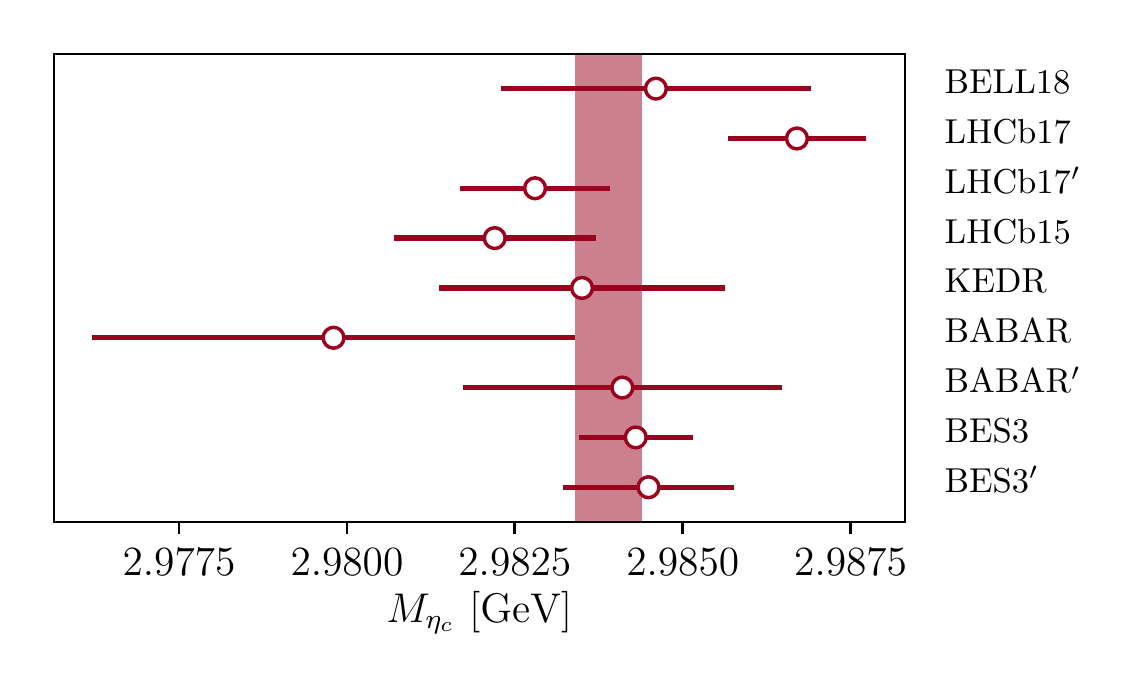}
  \includegraphics[width=0.47\textwidth]{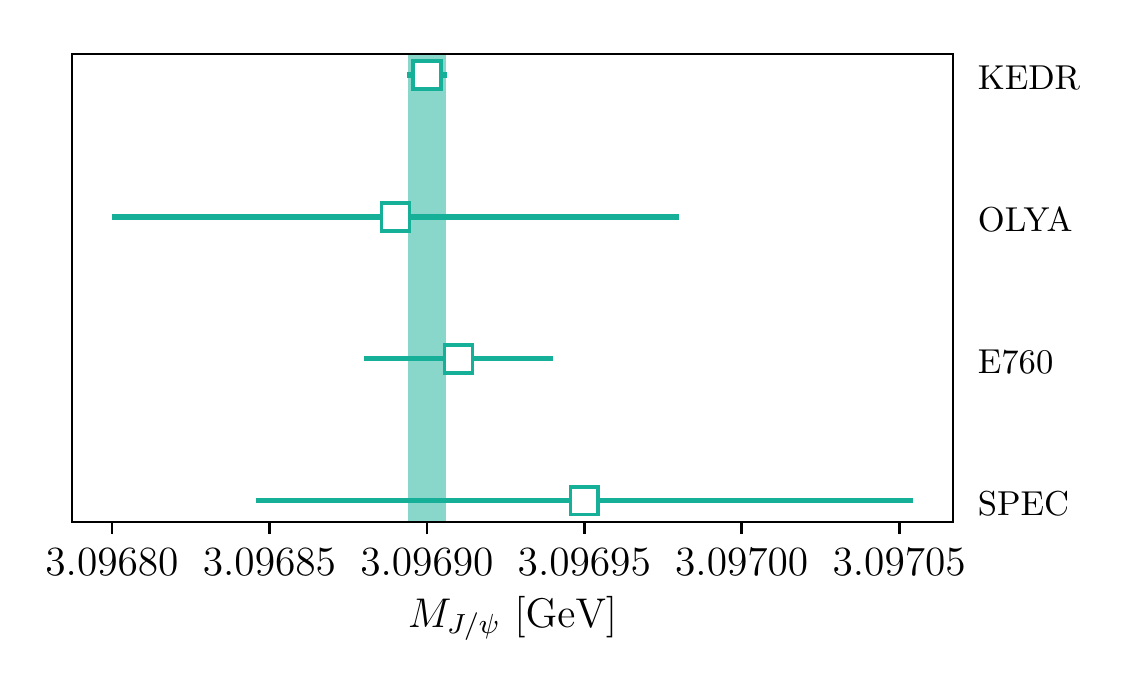}
  \caption{
Comparison of different experimental results for the $J/\psi$ and 
$\eta_c$ masses along with the PDG average values. 
  The $\eta_c$ results represent a recent subset of those used in the
  PDG average. The most recent result is from BELLE 
(denoted BELL18)~\cite{Xu:2018uye}. There are
  three different determinations from LHCb \cite{Aaij:2014bga, Aaij:2016kxn,
  Aaij:2017tzn}, two of which also measured the hyperfine splitting. We include
  a KEDR measurement \cite{Anashin:2014wva}, two from different BaBar analyses
  \cite{Lees:2014iua} and two from BESIII \cite{BESIII:2011ab,Ablikim:2012ur}.}
  \label{fig:exp-masses}
\end{figure}

\begin{figure}
  \includegraphics[width=0.47\textwidth]{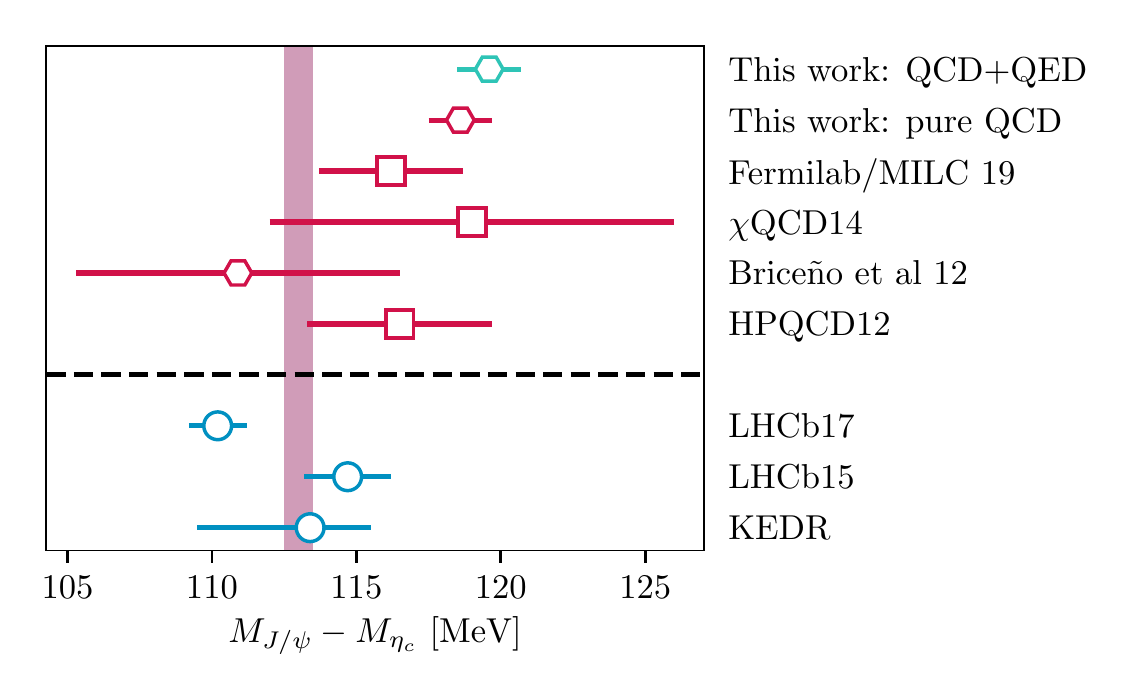}
  \caption{
Comparison of different lattice results for the charmonium hyperfine
  splitting and separate experimental results that measure this difference, 
as well as the PDG average (pink band). The
  PDG average is obtained from taking the differences of the PDG $J/\psi$ and
  $\eta_c$ masses (see Fig.~\ref{fig:exp-masses}) 
rather than only from experiments that directly measure the
  splitting. 
The squares give lattice QCD results from calculations that include $n_f=2+1$ 
flavours of quarks in the sea. The hexagons give results 
that include $n_f=2+1+1$ flavours of sea quarks, including the results we present 
here at the top of the plot. 
  All lattice QCD results have had uncertainties from neglecting
  $\eta_c$ annihilation removed so that we might expect some difference
  between them and experiment (see text), but previous lattice QCD results have not been 
accurate enough to see this. The Fermilab/MILC result used Fermilab $c$ quarks 
on gluon field configurations with asqtad sea quarks~\cite{DeTar:2018uko} and 
the previous HPQCD result~\cite{Donald:2012ga} used HISQ quarks on the same 
asqtad-sea ensembles. Brice\~{n}o et al~\cite{Briceno:2012wt} used a modification of 
the Fermilab approach known as relativistic heavy quarks on the $n_f=2+1+1$ HISQ 
sea quark ensembles that we use here. The $\chi$QCD result~\cite{Yang:2014sea} used overlap 
quarks on gluon field configurations including $n_f=2+1$ domain-wall sea quarks. }
  \label{fig:hf-comp}
\end{figure}

The experimental average value of the hyperfine splitting (113.0(5) MeV) 
from the Particle Data 
Group (PDG)~\cite{Tanabashi:2018oca} is calculated as the difference of the separate
averages for the $J/\psi$ and $\eta_c$ masses. The different experimental
results contributing to the PDG average of the two masses are shown in
Fig.~\ref{fig:exp-masses}. For the $J/\psi$ mass the average is dominated
by the most recent result from KEDR~\cite{Anashin:2015rca}. There are only three experimental
results used in these analyses that can independently produce values for the
hyperfine splitting. These are the KEDR experiment~\cite{Anashin:2015rca,Anashin:2014wva}
and two LHCb analyses in different channels \cite{Aaij:2014bga,
Aaij:2016kxn}. The LHCb result in~\cite{Aaij:2016kxn} used the
$\eta_c(2\mathrm{S}) \to p\overline{p}$ decay
while the analysis of~\cite{Aaij:2014bga} used $\eta_c(1\mathrm{S}) \to p\overline{p}$.
In the comparison plot of Fig.~\ref{fig:hf-comp}
\cite{Aaij:2014bga} is referred to as LHCb15 and \cite{Aaij:2016kxn} as LHCb17.

Fig.~\ref{fig:hf-comp} shows a comparison of lattice QCD results for the
charmonium hyperfine splitting along with the PDG average value 
and separate experimental values that measured this splitting.
Previous calculations on gluon field configurations that included 
$n_f=2+1$ flavours of sea quarks by HPQCD~\cite{Donald:2012ga} and by Fermilab/MILC~\cite{DeTar:2018uko} both obtained values above 
the experimental average, although only by just over one standard deviation. 

The
result we present here is substantially more precise than these earlier studies and for
the first time displays a significant, 6$\sigma$, difference from the
experimental average, clearly showing that the lattice result lies above the
experimental one. We interpret this as the effect of ignoring annihilation
to gluons in the calculation of the $\eta_c$ mass. From the comparison of our 
results to experiment we conclude that these annihilation effects increase the 
$\eta_c$ mass by 7.3(1.2) MeV, where the uncertainty is dominated by that from 
the lattice calculation. 

Previous analyses of this issue have given mixed results. In NRQCD perturbation 
theory~\cite{Follana:2006rc} we can relate the shift in the $\eta_c$ mass to 
its total (hadronic) width through the perturbative amplitude 
for $c\overline{c} \rightarrow gg \rightarrow c\overline{c}$ 
at threshold~\cite{Bodwin:1994jh}. Then 
\begin{eqnarray}
\Delta M_{\eta_c} &=& \frac{\Gamma_{\eta_c}}{2}\left(\frac{2(\ln 2 -1)}{\pi} + \mathcal{O}(\alpha_s,v^2/c^2)\right)  \\
&=& \frac{31.9(7)}{2} \,\mathrm{MeV} \times \left( -0.195 + \mathcal{O}(\alpha_s,v^2/c^2) \right) . \nonumber
\end{eqnarray} 
The leading term here gives $-$3.1 MeV, but sub-leading corrections 
could easily change the sign. 
An alternative way to think about the effect is non-perturbatively and then 
the gluon annihilation allows mixing between the $\eta_c$ and other flavour-singlet 
pseudoscalar mesons. Since these are lighter than the $\eta_c$ this mixing 
could give a positive correction to the $\eta_c$ mass.  
Direct lattice QCD determination of the effect, by calculating the appropriate 
quark-line disconnected correlation functions, has so far not proved possible. 
This is because the lighter states that are introduced by the 
 mixing make it very hard to pin down a small effect on the mass of 
a particle, the $\eta_c$, which is so much further up the spectrum in this 
channel. An estimate of the mass shift of 
+1--4 MeV was obtained in the quenched approximation 
in which this mixing is not possible but where mixing with glueballs could happen 
instead~\cite{Levkova:2010ft}. 

Our result for the hyperfine splitting, by its accuracy, 
provides for the first time a clear 
indication of the size of the impact of the $\eta_c$ annihilation to gluons 
on its mass: 
\begin{equation}
\label{eq:final-deltametac}
\Delta M^{\mathrm{annihln}}_{\eta_c} = +7.3(1.2) \,\mathrm{MeV}.
\end{equation}

\section{Determination of $m_c$}
\label{sec:mass}

\subsection{Pure QCD} 
\label{sec:massQCD}

In~\cite{Lytle:2018evc} we showed that it is possible to determine 
the strange and charm quark masses accurately in lattice QCD using an 
intermediate momentum-subtraction scheme. By intermediate we mean that 
the mass renormalisation factor to convert the tuned bare lattice quark mass 
to the momentum-subtraction scheme is calculated on the lattice. The conversion 
from the momentum-subtraction to the final preferred $\overline{\text{MS}}$ 
scheme is carried out using QCD perturbation theory in the continuum.  
To do this accurately it is important to use a momentum-subtraction scheme that 
has only one momentum scale, $\mu$. This means that the squared 4-momentum on 
each leg of the vertex diagram, from which the mass renormalisation 
factor is calculated, is $\mu^2$. The RI-SMOM scheme~\cite{Sturm:2009kb} 
used in~\cite{Lytle:2018evc} 
is such a scheme. A further important point is that the mass renormalisation 
factor will be contaminated by nonperturbative (condensate) artefacts through 
its nonperturbative calculation on the lattice. To identify and remove these 
artefacts (that appear as inverse powers of $\mu$) 
requires calculations at multiple values of $\mu$ and a fit to the results, 
as discussed in~\cite{Lytle:2018evc}.  

Below we briefly summarise the procedure followed in~\cite{Lytle:2018evc}:
\begin{enumerate}
  \item Determine the tuned bare quark mass and lattice spacing at physical 
sea quark masses for each set of gluon field configurations at a fixed $\beta$ value. 
We do this following Appendix A of~\cite{Chakraborty:2014aca}. 
  \item Calculate the mass renormalisation factor, $Z_m^{\mathrm{SMOM}}$, that converts the lattice 
quark masses to the RI-SMOM scheme for each $\beta$ value at multiple values of $\mu$. 
We thereby obtain the quark mass in the RI-SMOM scheme at scale $\mu$. 
  \item Convert the mass to the $\overline{\mathrm{MS}}$ scheme at scale 
$\mu$ using a perturbative 
  continuum matching calculation. We denote this conversion factor by
  $Z_m^{\overline{\mathrm{MS}}/\mathrm{SMOM}}(\mu)$.
  \item Run all the $\overline{\text{MS}}$ quark masses at a range of scales $\mu$ 
to a reference scale of 3 GeV using the four loop QCD $\overline{\text{MS}}$
  $\beta$ function. We denote these running factors $r(3\ \mathrm{GeV}, \mu)$.
  \item Fit all of the results for the $\overline{\text{MS}}$ mass at 3 GeV 
to a function that allows for discretisation effects and 
  condensate contamination, which begins at $1/\mu^2$ with the nongaugeinvariant
  $\langle A^2 \rangle$ condensate. 
  \item Obtain from the fit the physical value for the quark mass 
in the $\overline{\text{MS}}$ 
scheme at 3 GeV with condensate contamination removed. 
\end{enumerate}

\begin{table}
  \caption{Lattice spacing values and tuned $c$ quark masses 
at physical 
sea quark masses for each group of ensembles at a fixed $\beta$ 
value, denoted by their name in column 1 (see Table~\ref{tab:ensembles}). 
The lattice spacing value is given in units of $w_0$ in column 2 
and the $c$ quark mass, fixed from the $J/\psi$ meson mass, is given 
in GeV units in column 3. The first uncertainty on the mass is 
uncorrelated between lattice spacing values, and the second is 
correlated. 
  }
  \label{tab:mc-vals}
\begin{ruledtabular}
\begin{tabular}{lll}
 & $w_0/a$ & $m_c^{\mathrm{tuned}}$ [GeV] \\
\hline
coarse  & 1.4055(33)  & 1.0524(10)(30)\\
fine  & 1.9484(33) & 0.9736(10)(30) \\
superfine & 3.0130(56) & 0.8973(10)(30) \\
ultrafine & 3.972(19) & 0.8592(20)(30)\\
\end{tabular}
\end{ruledtabular}
\end{table}

\begin{figure}
  \includegraphics[width=0.47\textwidth]{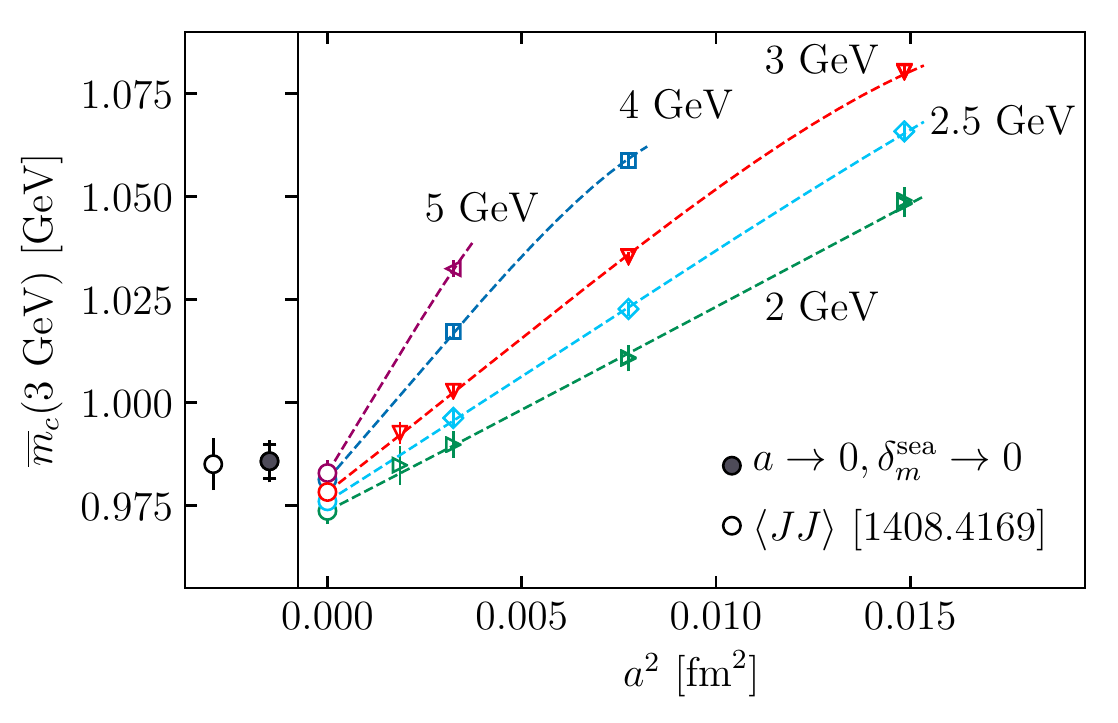}
  \caption{$\overline{m}_c(3\ \mathrm{GeV})$ extrapolated to the continuum with
  a fit form that allows for condensate terms that behave like inverse powers of
  the renormalisation scale $\mu$. This plot is an updated version of the upper section of
  Fig. 10 in~\cite{Lytle:2018evc}, with added data points on ultrafine lattices (set 14) 
and a retuned bare $c$ quark mass fixed from the $J/\psi$, rather than $\eta_c$, meson.
The
  ultrafine points have slightly mistuned $\mu$ values compared to 
the corresponding lines (see text). 
}
  \label{fig:mc}
\end{figure}

Here we provide three small updates to~\cite{Lytle:2018evc}. We first list 
them and then discuss them in more detail below. 
The three updates are: 
\begin{enumerate}
\item  we improve the uncertainty in the tuning of the bare lattice $c$ quark mass by using the $J/\psi$ 
mass rather than the $\eta_c$;
\item we include results from a finer ensemble of lattices (set 14) to provide 
even better control of the continuum limit;
\item we use the new 3-loop-accurate SMOM to $\overline{\mathrm{MS}}$ matching factors, 
$Z_m^{\overline{\mathrm{MS}}/\mathrm{SMOM}}$, 
calculated in~\cite{Kniehl:2020sgo, Bednyakov:2020ugu} to reduce the perturbative matching uncertainty. 
\end{enumerate}

The first update is to change how the tuning of the bare charm quark 
mass is done. In~\cite{Lytle:2018evc} bare charm masses were used that had
been tuned to the experimental $\eta_c$ mass, adjusted to allow for estimates 
of missing QED (from a Coulomb potential) and gluon 
annihilation effects (from perturbation theory). A 100\% uncertainty 
was included on the adjustment~\cite{Chakraborty:2014aca}. 
Now that we are explicitly including quenched 
QED it makes more sense to have a tuning process that uses an 
experimental meson mass with no adjustments. We also 
want to use the same tuning process 
for both the pure QCD case and the QCD+QED case to allow for a 
clear comparison and one that reflects the procedures that would be 
followed in a complete QCD+QED calculation. 
This means that we should use the $J/\psi$ meson mass, 
as we have done in Section~\ref{sec:hyperfine}.  
The $J/\psi$ mass is more accurate experimentally than that of 
the $\eta_c$ and the $J/\psi$ has a much smaller width, implying
little effect on the mass from its 3-gluon annihilation mode.  
The impact of $J/\psi$ annihilation to a single photon is a sub-1 MeV 
shift to the mass which is a 0.02\% effect, so negligible. 

Using our $J/\psi$ meson masses and 
following the procedure of~\cite{Chakraborty:2014aca}
we obtain 
tuned bare $c$ masses for each $\beta$ value. These are given in  
Table~\ref{tab:mc-vals} along with the $w_0/a$ values corresponding 
to physical sea quark masses at that $\beta$ value, which 
are also updated slightly from~\cite{Lytle:2018evc}. 
These slight changes in $w_0/a$ lead to small adjustments in the $\mu$ values 
relative to those given in Table IV of~\cite{Lytle:2018evc}.
This is accounted for when we run 
the $Z_m$ to the correct reference scale 
in $\overline{\mathrm{MS}}$. 

The second update is to
include results from the ultrafine $\beta=7.0$ ensemble (set 14). 
The appropriate tuned mass and $w_0/a$ value for physical sea 
quark masses is given in Table~\ref{tab:mc-vals}. 
We have also calculated new $Z_m^{\mathrm{SMOM}}(\mu)$ values on set 14. 
These are given in Appendix~\ref{appendix:zv}. 

The third update is to add the $\alpha_s^3$ correction to 
the SMOM to $\overline{\mathrm{MS}}$ conversion 
factor, $Z_m^{\overline{\mathrm{MS}}/\mathrm{SMOM}}$, for the mass renormalisation. 
This correction was recently calculated in continuum perturbative 
QCD~\cite{Kniehl:2020sgo, Bednyakov:2020ugu}. For $n_f=4$, as here, the 
$\alpha_s^3$ correction is a small effect (0.2\%), 
continuing the picture seen at $\mathcal{O}(\alpha_s)$ 
and $\mathcal{O}(\alpha_s^2)$ and consistent with the uncertainty 
taken from missing it in~\cite{Lytle:2018evc}.   

Once we have determined results for $\overline{m}_c$(3 GeV) 
at a variety of lattice spacing values using the SMOM intermediate 
scheme at a variety of $\mu$ values, we need to fit the results 
to determine $\overline{m}_c$(3 GeV) in the continuum limit. 
We do this following our previous calculation~\cite{Lytle:2018evc},  
where the fit function is given in Eq.~(26). The fit includes 
discretisation effects and condensate artefacts in the lattice 
calculation of $Z_m^{\mathrm{SMOM}}$. In~\cite{Lytle:2018evc} 
we included a term in the fit, $c_{\alpha}\alpha_s^3(\mu)$ (with $\alpha_s$ 
in the $\overline{\mathrm{MS}}$ scheme) to allow for the then-missing 
$\alpha_s^3$ term in the SMOM to $\overline{\mathrm{MS}}$ conversion. 
Here we replace that term with $c_{\alpha}\alpha_s^4(\mu)$ since the conversion 
is now calculated through $\alpha_s^3$ and included in our results. 
We take a prior value on 
$c_{\alpha}$ of $0.0 \pm 0.5$. This allows for the coefficient of 
the $\alpha_s^4$ term in the conversion factor to be 4 times as large
as the $\alpha_s^3$ coefficient.

\begin{table}
  \caption{Error budget (in \%) for the calculation of the charm quark mass in the
  $\overline{\mathrm{MS}}$ scheme at a scale of 3 GeV using RI-SMOM as an
  intermediate scheme. The listed contributions have the same meaning as those in
  \cite{Lytle:2018evc} except that we use $r$ here for the running factor rather than 
$R$ and we have an additional one labelled `QED effects' which comes from the
  continuum extrapolation shown in Fig.~\ref{fig:Zm}.}
  \label{tab:mc-errbudg}
\begin{ruledtabular}
\begin{tabular}{ll}
 & $\overline{m}_c$(3 GeV) \\
\hline
$a^2 \to 0$ & 0.23 \\
Missing $\alpha_s^4$ term & 0.10 \\
Condensate & 0.21 \\
$m_{\mathrm{sea}}$ effects & 0.00 \\
$Z_m^{\overline{\mathrm{MS}}/\mathrm{SMOM}}$ and $r$ & 0.07 \\
$Z_m^{\mathrm{SMOM}}$ & 0.12 \\
Uncorrelated $m^{\mathrm{tuned}}$ & 0.15 \\
Correlated $m^{\mathrm{tuned}}$ & 0.30 \\
Gauge fixing & 0.09 \\
$\mu$ error from $w_0$ & 0.12 \\
QED effects & 0.02 \\
\hline
Total & 0.52 \\
\end{tabular}
\end{ruledtabular}
\end{table}

The updated fit to our results for $\overline{m}_c$ in the $\overline{\text{MS}}$ 
scheme at 3 GeV is shown in Fig.~\ref{fig:mc}. 
The fit has a $\chi^2/\mathrm{dof}$ of 0.71. The error budget for this
calculation is shown in Table~\ref{tab:mc-errbudg}. Most of the
entries are very similar to those in \cite{Lytle:2018evc}. 
The contribution due to the continuum
extrapolation has, unsurprisingly, dropped a little, as has 
the uncertainty from the missing higher order terms (here $\alpha_s^4$) in the 
SMOM to $\overline{\mathrm{MS}}$ conversion. 
The correlated tuning uncertainty comes largely from the 
uncertainty in the physical value of $w_0$ used to fix the lattice 
spacing. 
We will be able to reduce that uncertainty 
in future by improving the determination of the value of $w_0$. 

Our updated pure QCD result is 
\begin{equation}
\label{eq:mcval-qcd}
\overline{m}_c(3\,\mathrm{GeV})_{\mathrm{QCD}} = 0.9858(51) \, \mathrm{GeV} . 
\end{equation} 
Running this down to a scale equal to the mass gives 
\begin{equation}
\label{eq:mcmc-qcd}
\overline{m}_c(\overline{m}_c)_{\mathrm{QCD}} = 1.2723(78) \, \mathrm{GeV} . 
\end{equation} 
These results improve on and supersede the value in~\cite{Lytle:2018evc}. 

\subsection{Impact of Quenched QED}
\label{sec:QEDmass}

To include quenched QED effects in the determination of the $c$ quark mass 
we must determine both the bare quark mass and the mass renormalisation factor 
with quenched QED switched on. 

We include quenched QED
effects for $Z_m$ in the RI-SMOM scheme in the same way as that 
described for the vector current
renormalisation in~\cite{Hatton:2019gha}. 
This involves the generation of U(1) fields in Landau
gauge to remain consistent with the Landau gauge QCD configurations used in 
the pure QCD calculation. 
When converting from the RI-SMOM scheme for $Z_m$ 
to the $\overline{\mathrm{MS}}$ scheme it is also necessary to include QED
effects in the perturbative matching factor.
We can evaluate the QED contribution to $Z_m^{\overline{\text{MS}}/\text{SMOM}}$ 
at order $\alpha_{\mathrm{QED}}$ by multiplying the known 
coefficient of $\alpha_s$~\cite{Sturm:2009kb} by (3/4)$Q^2$ to give the coefficient of 
$\alpha_{\mathrm{QED}}$.  
The impact is very small ($<0.1\%$) and we
therefore safely neglect higher order terms. The numerical values of the 
$\mathcal{O}(\alpha_{\mathrm{QED}})$ term 
we do include for the RI-SMOM to $\overline{\mathrm{MS}}$ matching are given in
Table~\ref{tab:Zm-qed}. 

\begin{table}
  \caption{Table giving factors needed for the determination of the quark mass in 
a calculation including quenched QED for different $\mu$ values and 
lattice spacings (denoted by set numbers). The fractional QED correction 
to $Z_m^{\mathrm{SMOM}}$ is given in the third column, the QED 
  component of the RI-SMOM to $\overline{\mathrm{MS}}$ matching for
  each $\mu$ in the fourth column and the factor giving the QED mass running 
from $\mu$ to a reference scale of 3 GeV in the fifth column.}
  \label{tab:Zm-qed}
\begin{ruledtabular}
\begin{tabular}{lllll}
Set & $\mu$ [GeV] & $R_{\mathrm{QED}}[Z_m^{\text{SMOM}}]$ & $Z_{m,\mathrm{QED}}^{\overline{\mathrm{MS}}/\mathrm{SMOM}}$ & $r^{\mathrm{QED}}$(3 GeV, $\mu$) \\
\hline
5 & 2 & 1.001200(83) & 0.999872 & 0.999372 \\
10 & 2 & 1.001516(35) & 0.999872 & 0.999372 \\
12 & 2 & 1.001853(83) & 0.999872 & 0.999372 \\
\hline
5 & 2.5 & 1.000827(31) & 0.999873 & 0.999718 \\
\hline
5 & 3 & 1.000540(15) & 0.999873 & - \\
10 & 3 & 1.000851(11) & 0.999873 & - \\
12 & 3 & 1.001308(18) & 0.999873 & - \\
\hline
10 & 4 & 1.0005001(21) & 0.999873 & 1.000446 \\
12 & 4 & 1.0009331(34) & 0.999873 & 1.000446 \\
\end{tabular}
\end{ruledtabular}
\end{table}

\begin{figure}
  \includegraphics[width=0.47\textwidth]{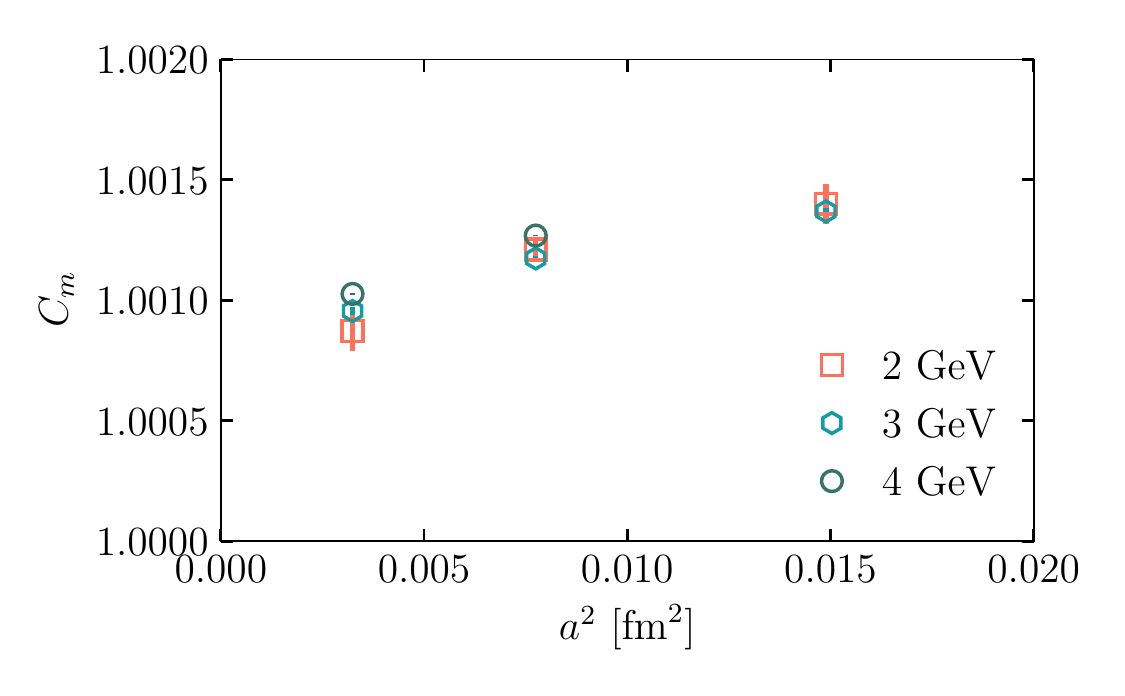}
  \caption{The factor $C_m$ that forms part of the QED effect in the 
lattice to $\overline{\text{MS}}$ mass renormalisation, as 
defined in Eq.~(\ref{eq:Cmdef}), plotted against the square of the lattice 
spacing. The fact that $C_m$, as shown here, has almost no $\mu$ or $a$ dependence
demonstrates that the $\mu a$ dependence seen 
in $R_{\mathrm{QED}}[Z^{\overline{\text{MS}}}_m]$ from columns 3 and 
4 of Table~\ref{tab:Zm-qed} is 
simply that expected from perturbation theory. 
  }
  \label{fig:pert}
\end{figure}

\begin{figure}
  \includegraphics[width=0.47\textwidth]{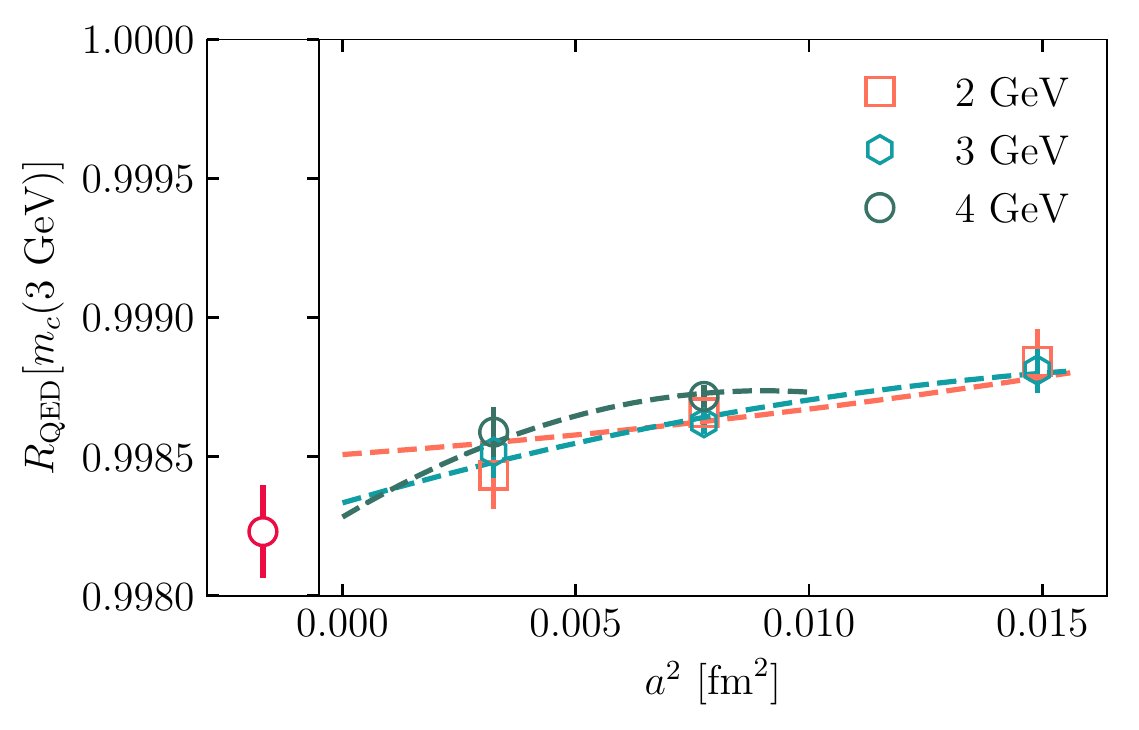}
  \caption{QED correction to the charm quark mass in the
  $\overline{\mathrm{MS}}$ scheme at a scale of 3 GeV. The different $\mu$
  values, shown as different colours and shapes, have all been run to 3 GeV and
  only differ by discretisation and condensate effects. 
The red point on the left is the result for $R_{\mathrm{QED}}[\overline{m}_c(3\,\mathrm{GeV})]$
returned by the fit of Eq.~(\ref{eq:fitmc}).
  }
  \label{fig:Zm}
\end{figure}

To assess the QED impact on the tuned bare $c$ mass we use the QCD+QED $J/\psi$ masses
given in Table~\ref{tab:masses-qed} and shown in Fig.~\ref{fig:jpsi-masses}. 
As we have corrected the pure QCD determination of $m_c$  to be
tuned to the experimental $J/\psi$ mass this is the tuning we will 
use for the QCD+QED case as well. The fractional shift in $am_c$ required to obtain the
correct $J/\psi$ mass after QED has been included (which we denote
$R_{\mathrm{QED}}[am_c]$) can be evaluated from the fractional change in the $J/\psi$ mass. 
$R_{\mathrm{QED}}[am_c]$ is the fractional change in $am_c$ required to return 
the $J/\psi$ mass to the value it had in pure QCD (i.e.\ the experimental 
value) once QED is switched on. Thus the increase in $J/\psi$ mass seen 
with QED must be compensated by a corresponding reduction in $am_c$. 
From the deliberately mistuned $am_c$ values in 
Table~\ref{tab:masses} we see that the fractional change in the 
$c$ mass is 1.5 times larger than 
the change seen in the meson mass. We can use this factor, 
along with $R^0_{\text{QED}}[M_{J/\psi}]$ values from 
Table~\ref{tab:masses-qed} and the required change of sign in the 
shift, 
to determine the retuning of 
the quark masses in QCD+QED. 
 We therefore take
$R_{\mathrm{QED}}[am_c]$ on coarse, fine and superfine lattices (sets 6, 10 and 12) to be 
0.99840(8), 0.99790(4) and 0.99734(9) respectively. We increase
the uncertainty on $R_{\mathrm{QED}}[am_c]$ compared to that 
for $R^0_{\mathrm{QED}}[M_{J/\psi}]$ by a factor of 5 to allow for 
the uncertainty in our conversion factor of 1.5 above. 

We calculate the ratio of $Z_m$ in the SMOM scheme in 
QCD+QED to pure QCD ($R_{\mathrm{QED}}[Z_m^{\text{SMOM}}]$) 
following the methods of~\cite{Lytle:2018evc}. The calculations are 
carried out at multiple values of the
renormalisation scale $\mu$ and extrapolated to zero valence 
quark mass. Results are given in Table~\ref{tab:Zm-qed}.  
Notice that these values are larger than 1 and so compensate to a large 
extent for the changes in the tuning of the bare lattice $am_c$ induced by QED. 
This reflects the fact that most of the shift is an unphysical ultraviolet self-energy 
effect. 
We then convert from the SMOM scheme 
to $\overline{\mathrm{MS}}$ by making use of the 
ratio of the perturbative conversion factor for QCD+QED to pure QCD, i.e.\ 
the $\mathcal{O}(\alpha_{\text{QED}})$ piece of $Z_m^{\overline{\text{MS}}/\text{SMOM}}$ 
also given in Table~\ref{tab:Zm-qed}. 
This is less than 1 but only by a very small amount. 

Multiplying these two factors together gives the ratio of the 
lattice to $\overline{\mathrm{MS}}$ mass renormalisation factors 
for QCD+QED to that for pure QCD, i.e.\ $R_{\mathrm{QED}}[Z_m^{\overline{\text{MS}}}]$. 
From Table~\ref{tab:Zm-qed}, multiplying columns 3 and 4, we can see that 
$R_{\mathrm{QED}}[Z_m^{\overline{\text{MS}}}]$ varies with $\mu$ and with 
lattice spacing over a range of about 0.001. In perturbation theory 
we expect $R_{\mathrm{QED}}[Z_m^{\overline{\text{MS}}}]$ to consist of a power 
series in $\alpha_{\text{QED}}$ and $\alpha_s$ multiplied by constants 
and powers of logarithms of $a\mu$. The leading logarithm at each order 
can be derived from the anomalous dimensions of the mass, allowing 
us to write~\cite{Bednyakov:2016onn} 
\begin{equation}
\label{eq:Cmdef}
R_{\mathrm{QED}}[Z_m^{\overline{\text{MS}}}] = 1 + C_m - \frac{3\alpha_{\mathrm{QED}}Q^2}{4\pi}\log(\mu^2a^2) .
\end{equation}
Here $C_m$ is a constant, up to discretisation effects and higher order terms
multiplying powers of $\log(a\mu)$. 
Figure~\ref{fig:pert} plots our results for $C_m$. These show very little variation 
with $a$ and $\mu$, confirming that the dependence on $a$ and $\mu$ of 
$R_{\mathrm{QED}}[Z_m^{\overline{\text{MS}}}]$ is almost entirely captured by Eq.~(\ref{eq:Cmdef}). 

Once the impact of QED on the 
$c$ mass in the $\overline{\text{MS}}$ scheme at scale $\mu$ is obtained, 
as above, we then need to allow for QED effects in the running of the masses from 
$\mu$ to the reference scale of 3 GeV. 
This is done by multiplication by a factor $r^{\text{QED}}$ calculated in 
$\mathcal{O}(\alpha_{\text{QED}})$ perturbation theory and 
given in Table~\ref{tab:Zm-qed}. These numbers are also 
very close to 1. The $\alpha_s\alpha_{\text{QED}}$ term could in 
principle have some impact here but it is very small and we 
neglect it~\cite{Bednyakov:2016onn}.  

Multiplying $R_{\text{QED}}[am_c]$ and $R_{\text{QED}}[Z_m^{\overline{\text{MS}}}]$ together allows us to determine the 
ratio of the $c$ quark mass in the $\overline{\text{MS}}$ scheme 
at 3 GeV from QCD+QED to that in pure QCD, 
i.e.\ $R_{\mathrm{QED}}[\overline{m}_c(3\,\mathrm{GeV})]$. 
The values that we have for this ratio come from results at multiple 
values of $\mu$ and multiple values of the lattice spacing. 
To determine the physical ratio in the continuum limit with condensate 
contamination removed (in this case QED corrections to QCD condensates) 
we need to fit the results to a similar form to that used in~\cite{Lytle:2018evc}.   
We use
\begin{eqnarray}
\label{eq:fitmc}
  R_{\mathrm{QED}}[\overline{m}_c(3\ \mathrm{GeV},\mu,a)] &=& R_{\mathrm{QED}}[\overline{m}_c(3\ \mathrm{GeV})] \times  \\
  && \hspace{-7.0em}\left[ 1 + \alpha_{\mathrm{QED}} Q^2 \sum_{i=1} c_{a^2}^{(i)} (a (1\ \mathrm{GeV}))^{2i} \right] \times \nonumber \\
  && \hspace{-5.0em}\bigg[ 1 + \alpha_{\mathrm{QED}} Q^2 \bigg(\sum_{j=1} c_{\mu^2 a^2}^{(j)} (a\mu)^{2j} \nonumber \\
  && + \sum_{n=1} \alpha_s(\mu) c_{\mathrm{cond}}^{(n)} \frac{(1\ \mathrm{GeV})^{2n}} {\mu^{2n}} \bigg)\bigg] . \nonumber
\end{eqnarray}

The first term on the second line of Eq.~(\ref{eq:fitmc}) accounts for 
discretisation effects in
$R_{\mathrm{QED}}[am_c]$; a scale of 1 GeV is chosen in these effects 
as this is close to the $c$ quark mass. The
term multiplying this (on the bottom two lines) 
models the $a$ and $\mu$ dependence of the QED correction
to $Z_m$. This includes discretisation effects of the form $(a\mu)^{2i}$ and
terms to model condensate contributions, starting at $1/\mu^2$.
The priors on all coefficients are taken as $0.0\pm 1.0$, except for 
the physical result, $R_{\mathrm{QED}}[\overline{m}_c(3\,\mathrm{GeV})]$, for which  
we take prior 1.00(1). 

The lattice QCD results for $R_{\mathrm{QED}}[\overline{m}_c(3\,\mathrm{GeV})]$ and 
the fit output are shown in Figure~\ref{fig:Zm}. The fit has a
$\chi^2/\mathrm{dof}$ of 0.87 and returns a physical value of 
$R_{\mathrm{QED}}[\overline{m}_c(3\,\mathrm{GeV})]$ of 0.99823(17). 
We conclude that the impact of quenched QED is to lower the $c$ quark 
mass, $\overline{m}_c$(3 GeV) by a tiny amount: 0.18(2)\%. 

We obtain our final answer for the $c$ quark mass in QCD+QED by multiplying 
$R_{\mathrm{QED}}[\overline{m}_c(3\,\mathrm{GeV})]$ by our pure QCD result for 
$\overline{m}_c(3\,\mathrm{GeV})$. 
This gives the QCD+QED result of  
\begin{equation}
\label{eq:mcval-qcdqed}
\overline{m}_c(3\,\mathrm{GeV})_{\mathrm{QCD+QED}} = 0.9841(51) \, \mathrm{GeV} . 
\end{equation} 
Running down to the scale of the mass with QCD+QED gives:
\begin{equation}
\label{eq:mcmc-qcdqed}
\overline{m}_c(\overline{m}_c)_{\mathrm{QCD+QED}} = 1.2719(78) \, \mathrm{GeV} , 
\end{equation} 
very close to the pure QCD value at this scale. 
This is the first determination of the $c$ quark mass to include QED effects 
explicitly, rather than estimate them phenomenologically as has been done in 
the past. 
The uncertainty achieved here of 0.5\% is smaller than the 0.6\% from
\cite{Lytle:2018evc} because we have reduced several sources of uncertainty, 
mainly those from the extrapolation to the continuum limit and from missing 
higher order terms in the SMOM to $\overline{\text{MS}}$ matching. 

\subsection{Discussion: $m_c$}
\label{sec:discuss-mc}

\begin{figure}
  \includegraphics[width=0.47\textwidth]{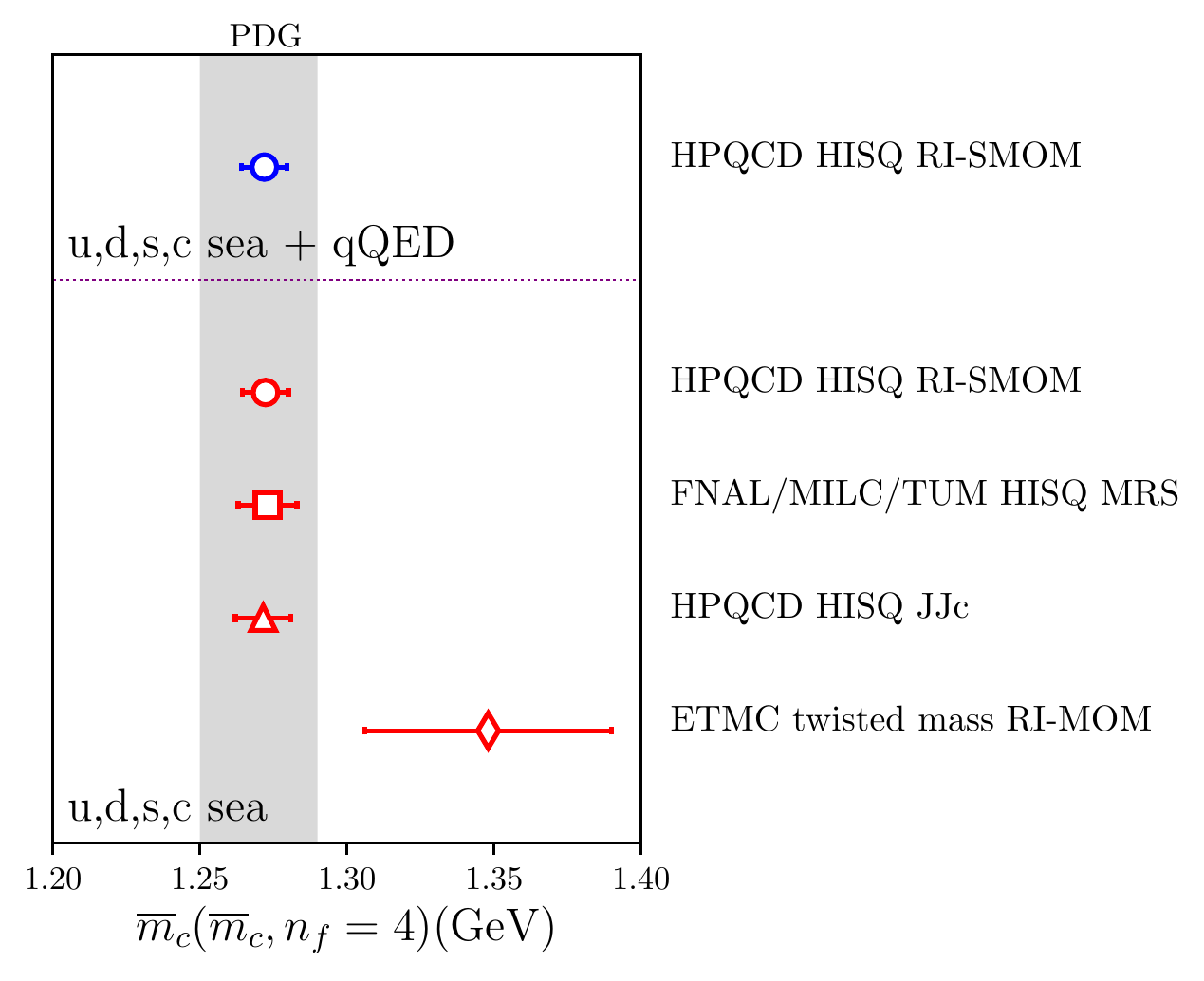}
  \caption{Comparison of lattice QCD results for $m_c$ that include 
$u$, $d$, $s$ and $c$ quarks in the sea. The top two results are the ones 
from this paper. Our QCD + quenched QED result is given in 
Eq.~(\ref{eq:mcval-qcdqed}). Our pure QCD result, Eq.~(\ref{eq:mcval-qcd}), supersedes 
our earlier result in~\cite{Lytle:2018evc}. The Fermilab/MILC/TUM result 
is from~\cite{Bazavov:2018omf} and uses a method based on charm-light meson masses.   
The `HPQCD HISQ JJc' result is from~\cite{Chakraborty:2014aca} and uses current-current 
correlator techniques. These three results agree to better than 1\%. 
The ETMC result is from~\cite{Carrasco:2014cwa} and uses the RI-MOM intermediate 
scheme.  The grey band gives the $\pm 1 \sigma$ uncertainty band from the 
Particle Data Group~\cite{Tanabashi:2018oca}. 
  }
  \label{fig:mccomp}
\end{figure}

Figure~\ref{fig:mccomp} gives a comparison of lattice QCD results for $\overline{m}_c$.
We plot the masses at the scale of the mass, as is conventional even though 
this is a rather low scale. 
We restrict the comparison to results that were obtained on gluon field 
configurations including $u$, $d$, $s$ and $c$ quarks in the sea. 
The top result is our value from Eq.~(\ref{eq:mcval-qcdqed}) 
that explicitly includes a calculation of 
the impact of quenched QED on the determination of the quark mass.

The top three results in the pure QCD section of the figure all 
include an estimate of, and correction for, QED effects. 
These corrections are made, however, by allowing for `physical' 
QED effects such as those arising from the Coulomb interaction between 
quark and antiquark in a meson. They do not allow for the QED 
self-energy contribution which is substantial. Although a large part 
of this is cancelled by the impact of QED on the mass renormalisation, 
a consistent calculation has to include both effects, as we have done here.  

An important point about Figure~\ref{fig:mccomp} is that the top 
three pure QCD results all have uncertainties of less than 1\% and 
agree to better than 1\%, using completely 
different methods. This implies a smaller uncertainty on $\overline{m}_c$ 
than the 1.5\% allowed for by the 
Particle Data Group~\cite{Tanabashi:2018oca}. 
This impressive agreement is not changed by our new result including 
quenched QED because, as we have shown, the impact of this is at the 0.2\% 
level. 

\section{$J/\psi$ and $\eta_c$ decay constants}
\label{sec:jpsidecay}

The decay constant of the $J/\psi$, $f_{J/\psi}$, is defined from 
the matrix element between the vacuum and a $J/\psi$ meson at rest by
\begin{equation}
\label{eq:fjpsidef}
  \langle 0 \vert \overline{\psi} \gamma_{\mu} \psi \vert J/\psi \rangle =
  f_{J/\psi} M_{J/\psi} \epsilon_{\mu} ,
\end{equation}
where $\epsilon_{\mu}$ is the component of the 
polarisation of the $J/\psi$ in the direction of the vector current. In terms of the
ground state amplitude, $A_0^V$, and mass, $M_0^V$ ($\equiv E_0^V$), 
obtained from the fit of 
Eq.~(\ref{eq:vfit}) to the 
charmonium vector correlator it is (in lattice
units)
\begin{equation}
\label{eq:fvamp}
  f_{J/\psi} = Z_V \sqrt{\frac{2A_0^V}{M_0^V}} .
\end{equation}
$Z_V$ is the renormalisation factor required to match the lattice vector current 
to that in continuum QCD if a nonconserved lattice vector current is used (as here). 
We discuss the renormalisation of vector currents using intermediate momentum-subtraction 
schemes in~\cite{Hatton:2019gha} and we will make use of the results based 
on the RI-SMOM scheme here (see Section~\ref{sec:setup}). Note that there is no 
additional renormalisation required to get from the RI-SMOM scheme to $\overline{\text{MS}}$
because the RI-SMOM scheme satisfies the Ward-Takahashi identity~\cite{Hatton:2019gha}. 

The partial decay width of the $J/\psi$ to an $\ell^+\ell^-$ pair ($\ell = e, \mu$) 
is directly 
related to the decay constant. At leading order in $\alpha_{\mathrm{QED}}$ 
and ignoring $(m_{\ell}/M_{J/\psi})^4$ correction terms, the
relation is
\begin{equation} \label{eq:width-decayconst}
  \Gamma(J/\psi \to \ell^+\ell^-) = \frac{4\pi}{3} \alpha^2_{\mathrm{QED, eff}}(M^2_{J/\psi}) Q_c^2 \frac{f_{J/\psi}^2}{M_{J/\psi}} ,
\end{equation}
where $Q_c$ is the electric charge of the charm quark in units of the charge of the
proton. Note that the formula contains the effective coupling, 
$\alpha_{\mathrm{QED, eff}}$ evaluated at the 
scale of $M_{J/\psi}$ but without including the effect of the $J/\psi$ resonance in the 
running of $\alpha_{\mathrm{QED}}$ to avoid double-counting~\cite{Keshavarzi:2018mgv}.  

Experimental values of $\Gamma(J/\psi \to e^+e^-)$ are obtained by 
mapping out the cross-section for $e^+e^- \to e^+e^-$ and $e^+e^- \to$ hadrons 
through the resonance region~\cite{Anashin:2018ldj} or by using initial-state 
radiation to map out this region via 
$e^+e^- \rightarrow \mu^+\mu^-\gamma$~\cite{Ablikim:2016xbg}. 
In either case initial-state radiation and non-resonant background 
must be taken care 
of~\cite{Anashin:2011ku, Alexander:1988em}. A cross-section fully inclusive of 
final-state radiation is obtained; interference between initial and final-state 
radiation is heavily suppressed~\cite{Fadin:1993kt}. 
The resonance parameter determined by the experiment is then the `full' partial 
width~\cite{Alexander:1988em,Anashin:2011yd}, 
\begin{equation}
\label{eq:gamma0}
\Gamma_{\ell\ell} = \frac{\Gamma_{\ell\ell}^{(0)}}{|1-\Pi_0|^2} 
\end{equation}  
where $\Gamma^{(0)}$ is the partial width to lowest order in QED 
and $\Pi_0$ is the photon vacuum polarisation.  
The effect of the vacuum polarisation is simply to replace 
$\alpha_{\mathrm{QED}}$ in the lowest-order QED formula for 
the width with 
$\alpha_{\mathrm{QED, eff}}(M^2)$, 
as we have done in Eq.~(\ref{eq:width-decayconst}).  

The experimental determination of $\Gamma_{\ell\ell}$ is accurate to 2\% for 
the $J/\psi$~\cite{Tanabashi:2018oca}. This allows us to infer a decay constant 
value from experiment, accurate to 1\%, using Eq.~(\ref{eq:width-decayconst}). 
\begin{eqnarray}
f_{J/\psi}^{\mathrm{expt}} &=& \left(\frac{3M_{J/\psi}}{4\pi Q_c^2}\right)^{1/2}\frac{\Gamma^{1/2}_{e^+e^-}}{\alpha_{\mathrm{QED, eff}}} \\
&=& 40.786 (\mathrm{MeV})^{1/2} \frac{\Gamma^{1/2}_{e^+e^-}}{\alpha_{\mathrm{QED, eff}}} \, .\nonumber
\end{eqnarray}
Using the experimental average of $\Gamma_{e^+e^-}=$ 5.53(10) keV~\cite{Tanabashi:2018oca}, and 
$\alpha_{\mathrm{QED, eff}}(M^2_{J/\psi})=1/134.02(3)$~\cite{alex-priv} gives 
\begin{equation}
\label{eq:fjpsiexpt}
f_{J/\psi}^{\mathrm{expt}} = 406.5(3.7)(0.5) \, \mathrm{MeV} . 
\end{equation}
The first uncertainty comes from the experimental uncertainty in $\Gamma$ and 
the second is an $\mathcal{O}(\alpha_{\mathrm{QED}}/\pi)$ uncertainty for higher-order 
in QED terms, for example from final-state radiation, in the connection between $\Gamma$ and $f$ in Eq.~(\ref{eq:width-decayconst}). 
Note that using $\alpha_{\mathrm{QED}}$ of 1/137 would increase this number by 
2.3\% (9 MeV). 

This experimental value can then be compared to our 
lattice QCD results for a precision test of QCD. 
Here we improve on HPQCD's earlier calculation~\cite{Donald:2012ga} by working 
on gluon field configurations that cover a wider range of lattice spacing values 
and with sea $u/d$ quark masses now down to their physical values. In addition 
we now include $c$ quarks in the sea and have a more accurate determination 
of the vector renormalisation factor $Z_V$~\cite{Hatton:2019gha}. 
We will also test the impact on $f_{J/\psi}$ of the $c$ quark's electric charge.  

The decay constant of the pseudoscalar $\eta_c$ meson is determined 
from our 
pseudoscalar correlators (of spin-taste $\gamma_5 \otimes \gamma_5$) using 
the ground-state mass and amplitude parameters from the correlator 
fit, Eq.~(\ref{eq:psfit}): 
\begin{equation}
\label{eq:psdecay}
  f_{\eta_c} = 2m_c \sqrt{\frac{2A_0^P}{(M_0^P)^3}} .
\end{equation}
Note that this is absolutely normalised and 
no $Z$ factor is required.
Because the $\eta_c$ does not annihilate to a single particle 
there is no experimental process from which we can directly determine 
$f_{\eta_c}$. Nevertheless it is a useful quantity to calculate for 
comparison to $f_{J/\psi}$ and to fill out the picture of these hadronic 
parameters from lattice QCD~\cite{Colquhoun:2015oha}.  
Again we will improve on HPQCD's earlier calculation~\cite{Davies:2010ip} 
as discussed above for the $J/\psi$. 

\subsection{Pure QCD}
\label{sec:QCDdecay}

\begin{table}
  \caption{Columns 2 and 3 give the results in lattice units for the 
$J/\psi$ and $\eta_c$ decay constants respectively in pure QCD on 
the ensembles listed in Table~\ref{tab:ensembles}. 
The values of $am_c^{\mathrm{val}}$ are those given in Table~\ref{tab:ensembles} 
except for two cases with a deliberately mistuned $c$ quark mass: 
set 6 denoted by a * where $am_c=0.643$ and set 14 denoted by a $\dag$ where 
$am_c$=0.188. 
The results for $f_{J/\psi}$ do not include the multiplication by 
$Z_V$ needed to normalise them (Eq.~(\ref{eq:fvamp})).
Columns 4 and 5 give the 
 electromagnetic corrections for the $J/\psi$ and
  $\eta_c$ decay constants, as the ratio of the QCD+QED result to that 
in pure QCD. Again, the electromagnetic corrections for $f_{J/\psi}$ do
  not include the corrections to $Z_V$. 
$Z_V$ values are given in Table~\ref{tab:ZVs}. 
  }
  \label{tab:fjpsi}
\begin{ruledtabular}
\begin{tabular}{lllll}
Set & $af_{J/\psi}/Z_V$ & $af_{\eta_c}$ & $R^0_{\mathrm{QED}}[f_{J/\psi}/Z_V]$ & $R^0_{\mathrm{QED}}[f_{\eta_c}]$ \\
\hline
1 & 0.43370(55) & 0.37659(18) & - & - \\
2 & 0.42346(48) & 0.370332(91) & 1.00410(64) & 1.00294(50) \\
3 & 0.4163(11) & 0.366127(57) & - & - \\
\hline
4 & 0.29411(21) & 0.268331(61) & - & - \\
6 & 0.28835(15) & 0.263727(60) & 1.00341(37) & 1.00326(13) \\
$6^*$ & 0.28671(15) & 0.262077(48) & - & - \\
8 & 0.285592(88) & 0.261676(26) & - & - \\
\hline
9 & 0.19406(30) & 0.18191(12) & - & - \\
10 & 0.191341(79) & 0.179362(26) & 1.00295(12) & 1.002951(54) \\
11 & 0.18961(15) & 0.178039(24) & - & - \\
\hline
12 & 0.12334(10) & 0.117535(28) & 1.00283(33) & 1.00311(47) \\
13 & 0.119606(63) & 0.114151(26) & - & - \\
\hline
14 & 0.091380(85) & 0.087772(39) & - & - \\
$14^{\dag}$ & 0.09069(29) & 0.086774(59) & - & - \\
\end{tabular}
\end{ruledtabular}
\end{table}

\begin{table}
  \caption{Vector current renormalistion constants, $Z_V(\mu)$, using the RI-SMOM 
scheme at $\mu =$ 2 GeV (column 2) and
  $\mu =$ 3 GeV (column 3) in pure QCD for each $\beta$ value corresponding 
to a group of ensembles in Table~\ref{tab:ensembles}. 
Column 4 gives the QED correction to $Z_V$ at 2 GeV in the form 
of the ratio of the QCD+QED value to that of pure QCD. 
Most of these values are taken
  from~\cite{Hatton:2019gha} although the $Z_V$ value at $\beta=7$ 
  and the QED correction at $\beta=5.8$ are new here.}
  \label{tab:ZVs}
\begin{ruledtabular}
\begin{tabular}{llll}
$\beta$ & $Z_V$(2 GeV) & $Z_V$(3 GeV) & $R_{\mathrm{QED}}[Z_V(2 \, \mathrm{GeV})]$ \\
\hline
5.80 & 0.95932(18) & - & 0.999544(14) \\
6.00 & 0.97255(22) & 0.964328(75) & 0.999631(24) \\
6.30 & 0.98445(11) & 0.977214(35) & 0.999756(32) \\
6.72 & 0.99090(36) & 0.98702(11) & 0.999831(43) \\
7.00 & 0.99203(108) & 0.99023(56) & - \\
\end{tabular}
\end{ruledtabular}
\end{table}

\begin{figure}
  \includegraphics[width=0.48\textwidth]{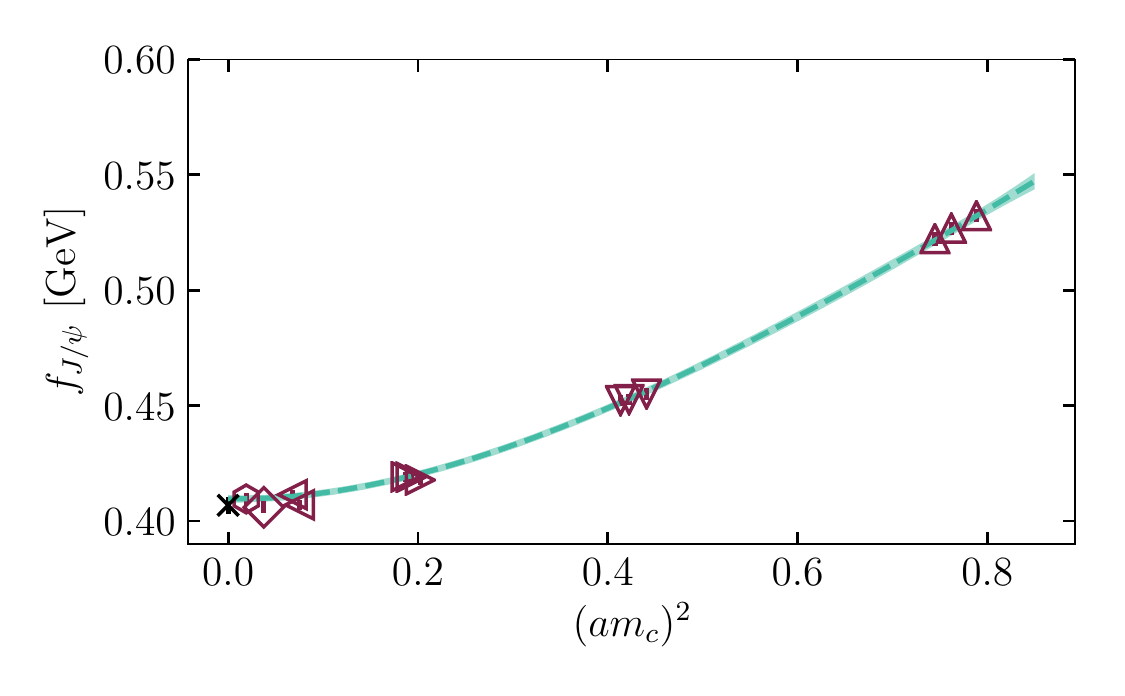}
  \caption{The $J/\psi$ decay constant calculated on the ensembles of
  Table~\ref{tab:ensembles} in pure QCD and plotted against the square of the 
bare $c$ quark mass in lattice units. 
The different red shapes correspond to different groups of ensembles with 
similar lattice spacing and the error bars shown include the full uncertainty 
on the points. Points at mistuned $m_c$ are not plotted but are 
included in the fit. The green curve marks our extrapolation to the physical 
point, where the black cross
shows the result determined from the 
experimental average for $\Gamma_{e^+e^-}$ from Eq.~(\ref{eq:fjpsiexpt}).}
  \label{fig:fjpsi}
\end{figure}

\begin{figure}
  \includegraphics[width=0.48\textwidth]{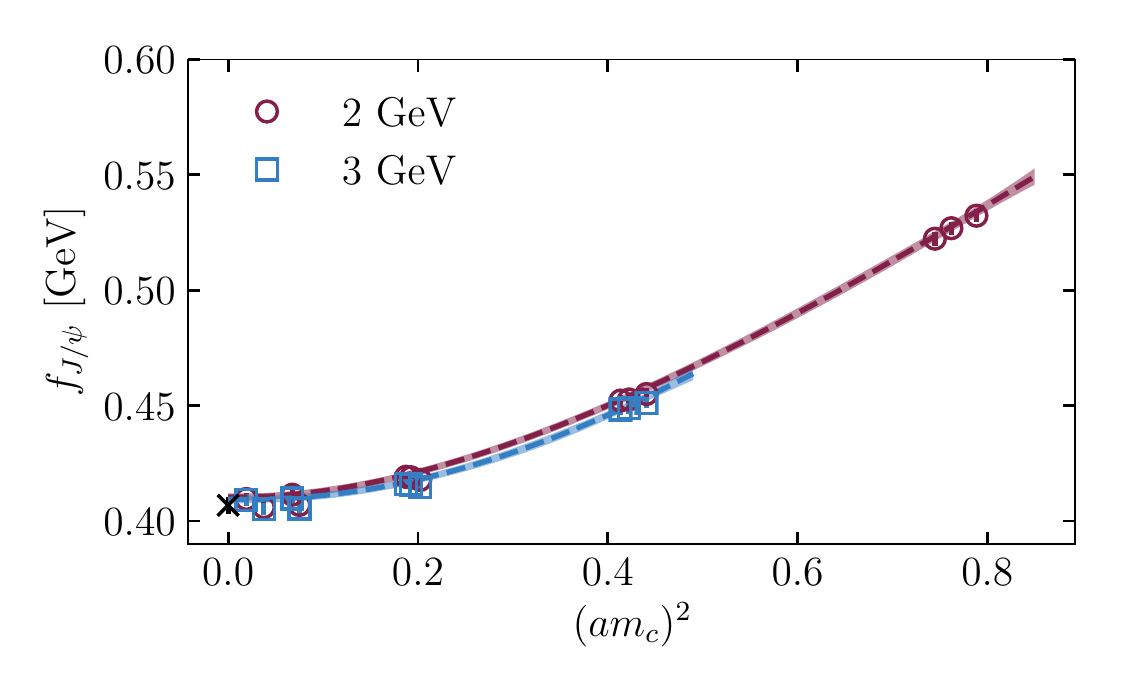}
  \caption{The continuum extrapolation of $f_{J/\psi}$ using 
  $Z_V$(2 GeV) (results and fit curve in red, same values 
as in Fig.~\ref{fig:fjpsi}) and $Z_V$(3 GeV) (results and fit 
curve in blue). 
  The continum extrapolated results agree in the two cases as they should. 
A black cross shows the experimental average result from Eq.~(\ref{eq:fjpsiexpt}).}
  \label{fig:2-3-comp}
\end{figure}

The second column of Table~\ref{tab:fjpsi} gives our results for 
the (unnormalised) values of $af_{J/\psi}$ in
pure QCD on 12 of the sets from Table~\ref{tab:ensembles}. 
We multiply $af_{J/\psi}/Z_V$ by the value of $Z_V$ and convert to physical units 
using the inverse lattice spacing. $Z_V$ values are taken 
from~\cite{Hatton:2019gha} except for a new value calculated here for
$\beta=7.0$ (ultrafine) set 14. We collect these values in Table~\ref{tab:ZVs}.
See Appendix~\ref{appendix:zv} for a discussion of the $Z_V$ results. 
The $Z_V$ values are very precise and so have little impact on the uncertainty 
in $f_{J/\psi}$.

Our results for $f_{J/\psi}$ for the pure QCD case 
are shown in Fig.~\ref{fig:fjpsi} plotted against the square 
of the lattice spacing (in units of the bare $c$ quark mass). 
Clear dependence on the lattice spacing is seen. This dependence comes 
from the amplitudes of the two-point correlators; the lattice spacing 
dependence of $Z_V$ contributes very little to it. 
We also plot in Fig.~\ref{fig:fjpsi} the results of the fit 
using Eq.~(\ref{eq:X-fit}). The priors for the fit are as given 
in Section~\ref{sec:fitstrategy} with the prior on the physical value 
of $f_{J/\psi}$ (i.e.\ $x$ in Eq.~(\ref{eq:X-fit})) of 0.4(1). 
 The
$\chi^2/\mathrm{dof}$ of the fit is 0.43. 
The
agreement with the result derived from experiment can clearly be seen. 
We obtain an $f_{J/\psi}$ value in pure QCD of
\begin{equation}
\label{eq:fjpsiqcd}
f_{J/\psi,\mathrm{QCD}}= 409.6(1.6) \, \mathrm{MeV}. 
\end{equation}
We will discuss this result further in Section~\ref{sec:discussdecay}. 

We have used vector current renormalisation factors, $Z_V$, 
in the RI-SMOM scheme at a scale of
2 GeV. The $\mu$ dependence of $Z_V$ should just be the result of discretisation
effects and results for the physical quantity, $f_{J/\psi}$, 
using different renormalisation scales $\mu$ should agree in the
continuum limit. Here we verify that this is the case using $Z_V(2\,\mathrm{GeV})$ 
 and $Z_V(3\,\mathrm{GeV})$ results from 
Table~\ref{tab:ZVs}~\cite{Hatton:2019gha}. 
There is no 3 GeV result on the very coarse lattices since $\mu a$ would be 
too large. 
The comparison for $f_{J/\psi}$ using $\mu$=2 and 3 GeV is shown in
Fig.~\ref{fig:2-3-comp} for the pure QCD case. 
The difference between the two values of $\mu$ is barely visible. 
The values at $\mu=$ 3 GeV give a continuum limit result of 
$f_{J/\psi}=$ 408.7(1.8) MeV, in good agreement with that 
at $\mu =$ 2 GeV in Eq.~(\ref{eq:fjpsiqcd}) but slightly less 
accurate. The $\chi^2/\mathrm{dof}$ of the fit was 0.45.  

\begin{figure}
 \includegraphics[width=0.48\textwidth]{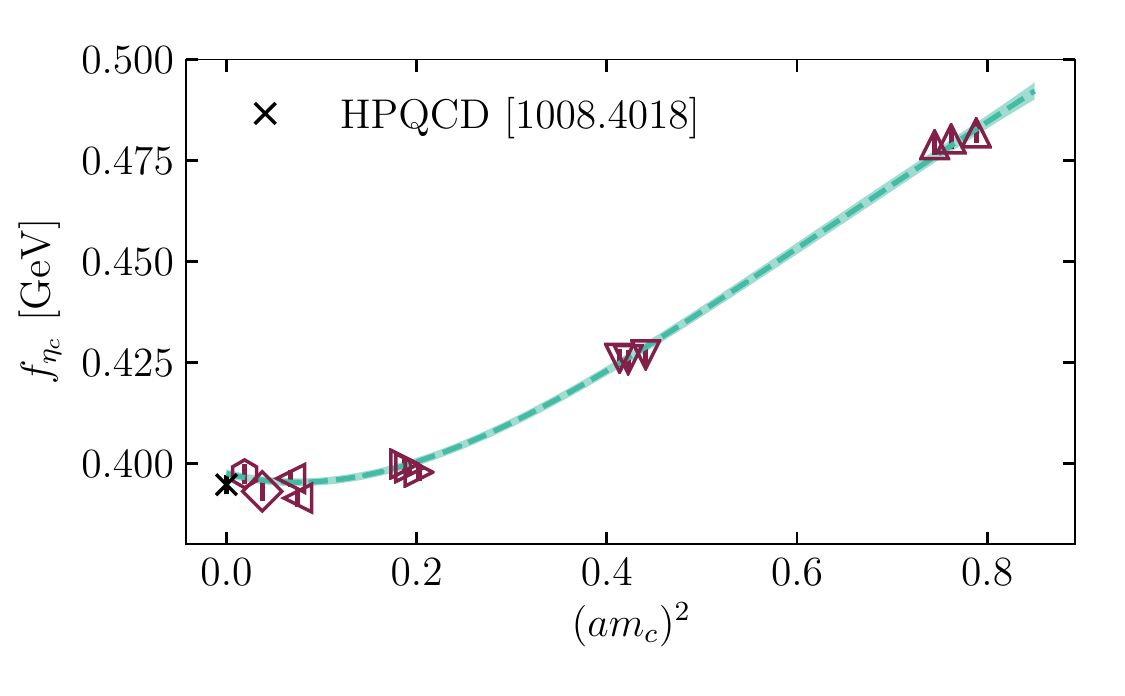}
  \caption{The $\eta_c$ decay constant calculated on the ensembles of
  Table~\ref{tab:ensembles} in pure QCD and plotted against the square of the 
bare $c$ quark mass in lattice units. 
The different red shapes correspond to different groups of gluon field 
ensembles with 
similar lattice spacing. The error bars on each point are the full uncertainty, 
including correlated uncertainties from, for example, the determination of 
the lattice spacing. The green curve shows our fit and 
extrapolation to the physical 
point. The black cross
gives the earlier HPQCD result on $n_f=2+1$ gluon field configurations 
from~\cite{Davies:2010ip}.}
  \label{fig:fetac-qcd}
\end{figure}

\begin{figure}
 \includegraphics[width=0.48\textwidth]{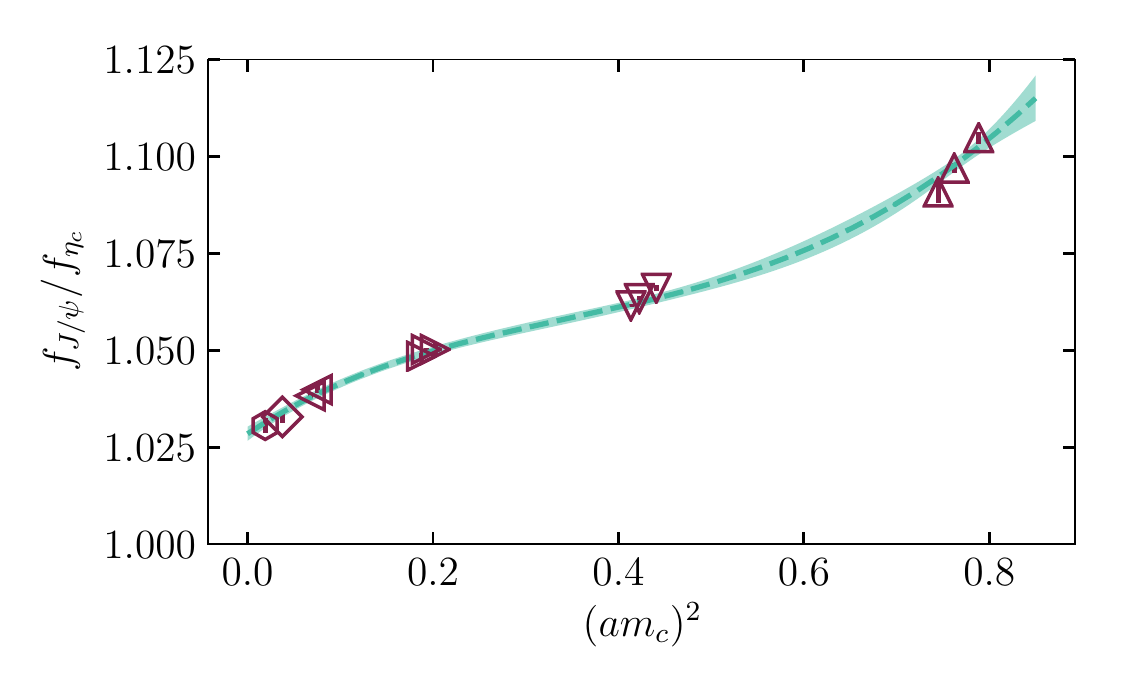}
  \caption{ The ratio of $J/\psi$ to $\eta_c$ decay constant determined in 
pure QCD, plotted against the square of the bare $c$ quark mass in lattice units. 
The different red shapes correspond to different groups of gluon field 
ensembles with 
similar lattice spacing. The error bars on each point are the full uncertainty, 
including correlated uncertainties from, for example, the determination of 
the lattice spacing. The green curve shows our fit and 
extrapolation to the physical 
point. 
}
  \label{fig:fratio-qcd}
\end{figure}

Our results for $af_{\eta_c}$, the $\eta_c$ decay constant in 
lattice units, are given in the third column of 
Table~\ref{tab:fjpsi} for the pure QCD case. 
After conversion to physical units, they are plotted in 
Figure~\ref{fig:fetac-qcd}. The curve is similar to that 
for $f_{J/\psi}$ but with somewhat smaller discretisation effects. 
We also plot the results of performing the same fit as for $f_{J/\psi}$ 
using Eq.~(\ref{eq:X-fit}). 
 The
$\chi^2/\mathrm{dof}$ of the fit is 0.88 giving a result for 
the decay constant in pure QCD of 
\begin{equation}
\label{eq:fetacqcd}
f_{\eta_c,\mathrm{QCD}}= 397.5(1.0) \, \mathrm{MeV}. 
\end{equation}
This agrees well with the earlier HPQCD value on $n_f=2+1$ 
gluon field configurations~\cite{Davies:2010ip} of 0.3947(24) GeV but has half 
the uncertainty. 
In \cite{Davies:2010ip} the effects
from neglecting the charm quark in the sea are estimated to be
$\mathcal{O}(0.01\%)$ which is negligible and means that the two calculations 
should give the same result.

Figure~\ref{fig:fratio-qcd} shows our results for the ratio of $f_{J/\psi}$ 
to $f_{\eta_c}$ in pure QCD. A lot of the discretisation effects cancel in
the ratio, as is evident in comparing this figure to Figures~\ref{fig:fjpsi} 
and~\ref{fig:fetac-qcd}. Systematic uncertainties, for example from the determination 
of the lattice spacing, are also reduced. The shape of the curve again, as in the hyperfine 
splitting case, reflects the fact that we have successfully reduced sources of 
$a^2$ error to the point where $a^4$ and $a^6$ are visible. 

We fit the ratio to the same fit 
as before (Eq.~(\ref{eq:X-fit})) with a prior on the physical value of 1.0(1). 
The fit has a $\chi^2/\mathrm{dof}$ of 0.62 and returns a physical value for 
the decay constant ratio in pure QCD of 
\begin{equation}
\label{eq:fratqcd}
\frac{f_{J/\psi,\mathrm{QCD}}}{f_{\eta_c,\mathrm{QCD}}}= 1.0285(18) .
\end{equation}
Thus we see that the $J/\psi$ decay constant is nearly 3\% larger than 
that of the $\eta_c$ with an uncertainty of 0.2\%. 

\begin{table}
  \caption{Error budget for the $J/\psi$ and $\eta_c$ decay constants as a 
percentage of the final answer. 
}
  \label{tab:f-errbudg}
\begin{ruledtabular}
\begin{tabular}{lll}
 & $f_{J/\psi}$ & $f_{\eta_c}$ \\
\hline
$a^2 \to 0$ & 0.09 & 0.03 \\
$Z_V$ & 0.05 & - \\
Pure QCD Statistics & 0.12 & 0.05 \\
QCD+QED Statistics & 0.05 & 0.02 \\
$w_0/a$ & 0.11 & 0.08 \\
$w_0$ & 0.34 & 0.24 \\
Valence mistuning & 0.05 & 0.01 \\
Sea mistuning & 0.01 & 0.00 \\
\hline
Total & 0.40\% & 0.26\% \\
\end{tabular}
\end{ruledtabular}
\end{table}

Table~\ref{tab:f-errbudg} gives the error budget 
for our final values of $f_{J/\psi}$ and
$f_{\eta_c}$, both for the pure QCD case and the QCD+QED case to be 
discussed in Section~\ref{sec:f-qed}. 
The contributions from different sources are very similar between the two
decay constants. 
It is clear from this that the
dominant sources of error are related to the determination of the lattice
spacing, as for the hyperfine splitting. 

The error budget presented here for the $J/\psi$ decay constant is markedly
different from that of \cite{Donald:2012ga}. There the dominant contribution to
the error was from the vector renormalisation constant, $Z_V$, 
obtained using a matching
between lattice time moments and high order perturbative QCD~\cite{Kuhn:2007vp}. Here that error is
substantially reduced by using the $Z_V$ values obtained in lattice QCD 
fully nonperturbatively in the RI-SMOM 
scheme~\cite{Hatton:2019gha}.  
Note that the uncertainty from scale-setting in the decay 
constant is much smaller than that for the hyperfine splitting 
(Table~\ref{tab:hf-errbudg}). This is because the decay constant has opposite
behaviour as a function of quark mass, increasing as the quark mass increases 
rather than decreasing. This then offsets, rather than augments (as in the 
hyperfine splitting case), its  
sensitivity to changes in the scale-setting parameter, $w_0$. 

\subsection{Impact of Quenched QED}
\label{sec:f-qed}

\begin{figure}
  \includegraphics[width=0.48\textwidth]{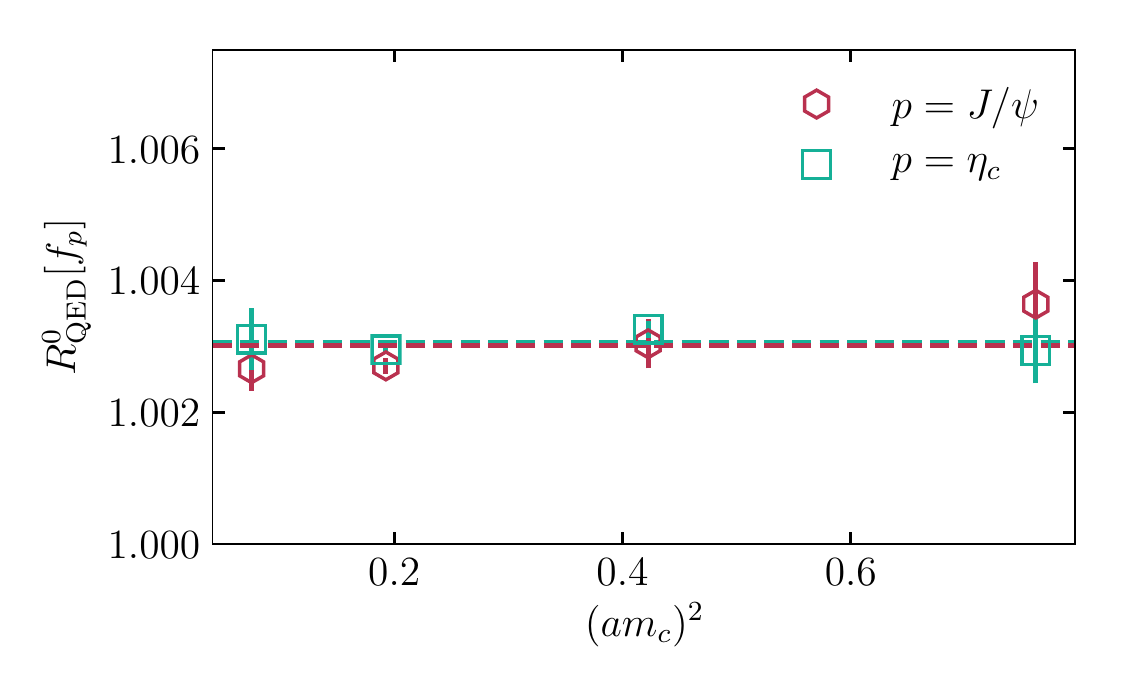}
  \caption{The fractional QED correction to the $J/\psi$ and $\eta_c$ decay constants as 
a function of lattice spacing. The horizontal dashed lines mark the weighted average 
of the points. 
  }
  \label{fig:fjpsi-qed}
\end{figure}

\begin{figure}
  \includegraphics[width=0.47\textwidth]{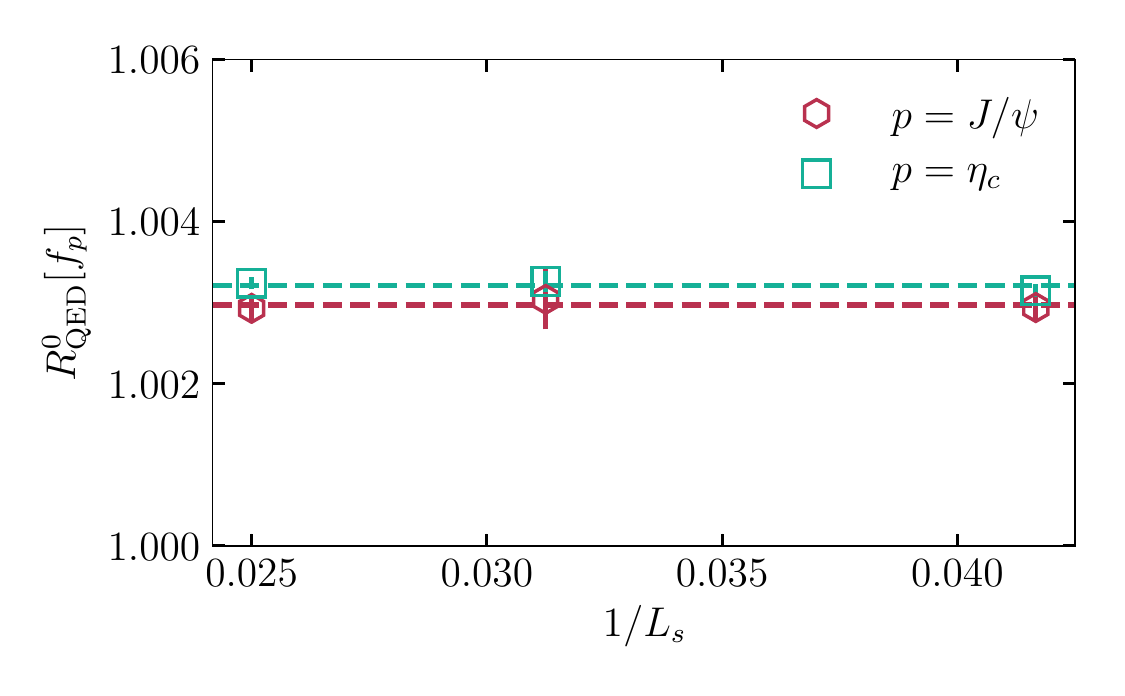}
  \caption{The volume dependence of the fractional QED effect on the $J/\psi$
  and $\eta_c$ decay constants measured on sets 5-7. 
The dashed lines are horizontal and indicate the weighted average 
of the points. 
  There is no observable finite volume
  effect at the level of our statistical uncertainties.}
  \label{fig:fjpsi-vol}
\end{figure}

\begin{figure}
  \includegraphics[width=0.48\textwidth]{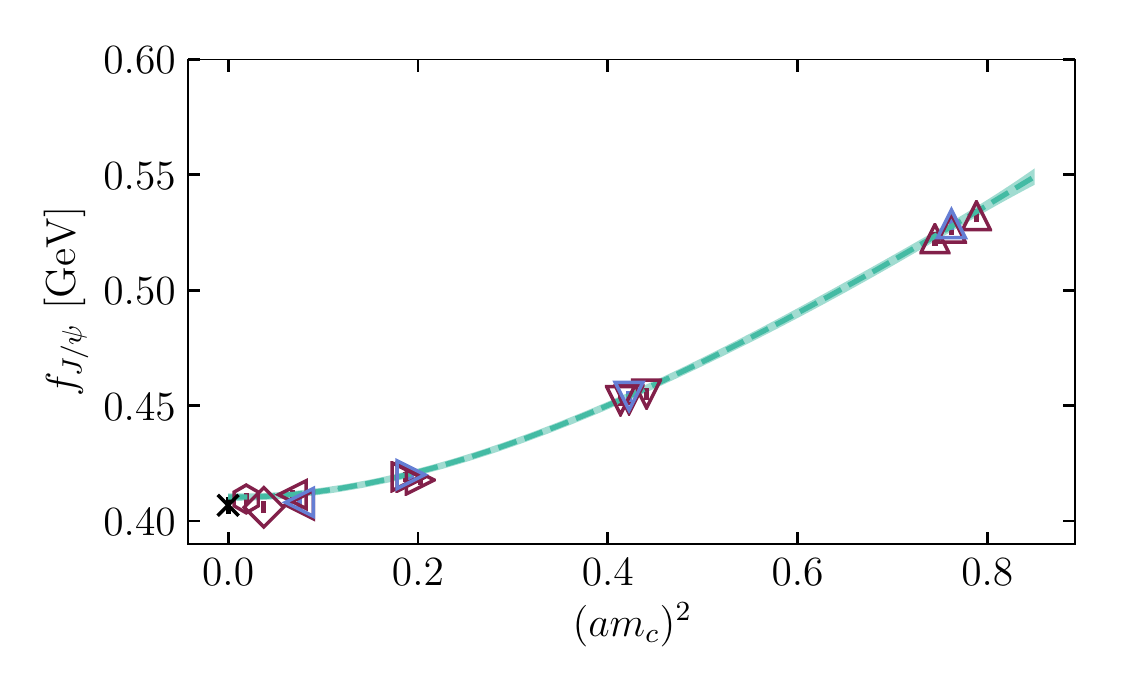}
  \caption{The $J/\psi$ decay constant calculated on the ensembles of
  Table~\ref{tab:ensembles} in pure QCD (red points) and including also quenched 
QED (blue points) plotted against the square of the 
bare $c$ quark mass in lattice units. 
The green curve marks our extrapolation to the physical 
point, where the black cross
shows the experimental average result from Eq.~(\ref{eq:fjpsiexpt}).}
  \label{fig:fjpsi-full}
\end{figure}

\begin{figure}
  \includegraphics[width=0.48\textwidth]{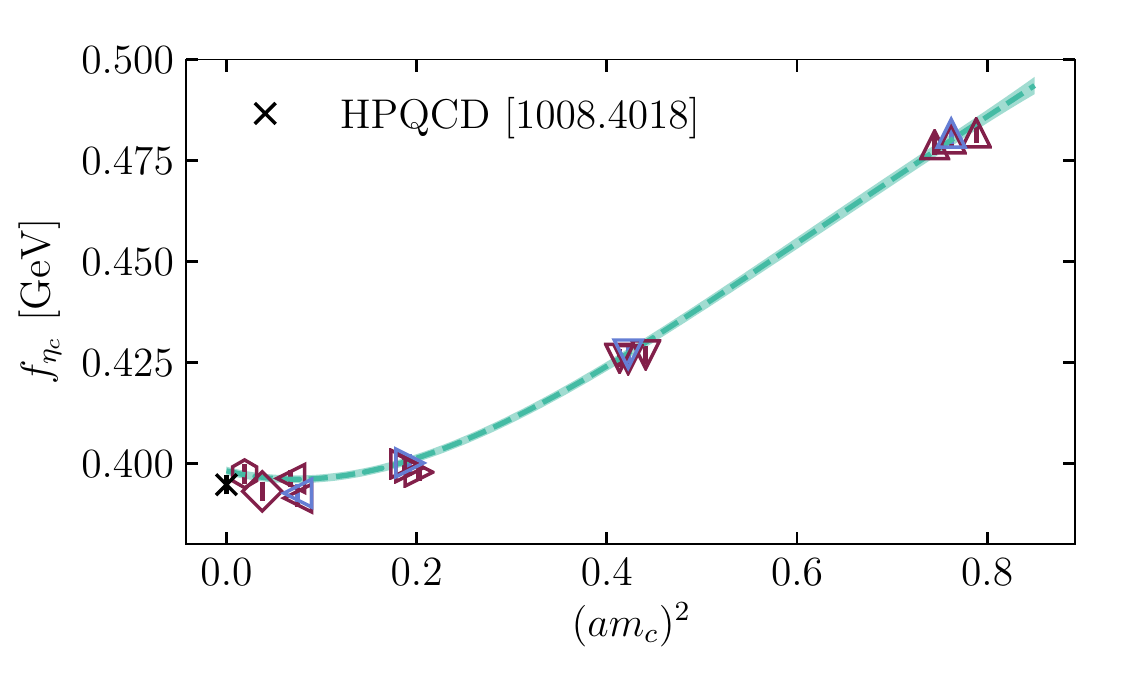}
  \caption{The $\eta_c$ decay constant calculated on the ensembles of
  Table~\ref{tab:ensembles} in pure QCD (red points) and including also quenched 
QED (blue points) plotted against the square of the 
bare $c$ quark mass in lattice units. 
The green curve marks our extrapolation to the physical 
point. The black cross
gives the earlier HPQCD result on $n_f=2+1$ gluon field configurations 
from~\cite{Davies:2010ip}.}
  \label{fig:fetac-full}
\end{figure}

Including quenched QED effects into our calculations allows us to determine the 
effect on the $J/\psi$ and $\eta_c$ decay constants of the electric charge of 
the valence $c$ quarks. Because the $J/\psi$ and $\eta_c$ are electrically 
neutral particles, there is no long-distance infrared component to cause problems 
(as there is for $f_{\pi^+}$) 
and we can simply proceed to determine the decay constants after the addition 
of the QED field as we did in the pure QCD case. 

The fractional QED effect on the $J/\psi$ and $\eta_c$ decay constants 
at fixed bare $c$ quark mass in lattice units is given in Table~\ref{tab:fjpsi}. 
We see a 0.3\% increase, offset slightly by the change in $Z_V$ in the $J/\psi$ 
case. The fractional QED effect on $Z_V$ is given in Table~\ref{tab:ZVs}. 
The fractional QED effect at fixed bare mass is plotted in Figure~\ref{fig:fjpsi-qed}. 
We see that the effect is similar for the $J/\psi$ and $\eta_c$ in the 
continuum limit and shows very little dependence on the lattice spacing. 

The volume dependence of the fractional QED effect is shown in Figure~\ref{fig:fjpsi-vol} 
on sets 5--7. We find that
the effect is negligible well below the 0.1\% level.

We now combine our QCD+QED results with our pure QCD results and the full fit of 
Eq.~(\ref{eq:X-fit}), which takes into account the retuning of the quark mass 
needed when quenched QED is included. 
Figure~\ref{fig:fjpsi-full} shows our pure QCD results, QCD+QED results and fit 
curve for the $J/\psi$ decay constant. We obtain 
\begin{equation}
\label{eq:fjpsifull}
f_{J/\psi,\mathrm{QCD+QED}}= 410.4(1.7) \, \mathrm{MeV}. 
\end{equation}
This is a 0.2\% increase over the value in pure QCD (Eq.~(\ref{eq:fjpsiqcd})) 
because retuning reduces the quark mass and offsets some of the impact 
of quenched QED seen in Table~\ref{tab:fjpsi}. 
A more accurate statement is that the final fractional effect 
from quenched QED is $R_{\mathrm{QED}}[f_{J/\psi}]=$1.00188(36). 

A very similar picture is seen for $f_{\eta_c}$ in Figure~\ref{fig:fetac-full}. 
We obtain
\begin{equation}
\label{eq:fetacfull}
f_{\eta_c,\mathrm{QCD+QED}}= 398.1(1.0) \, \mathrm{MeV}. 
\end{equation}
The final fractional effect from quenched QED is then $R_{\mathrm{QED}}[f_{\eta_c}]=$1.00166(25). 

\begin{figure}
  \includegraphics[width=0.48\textwidth]{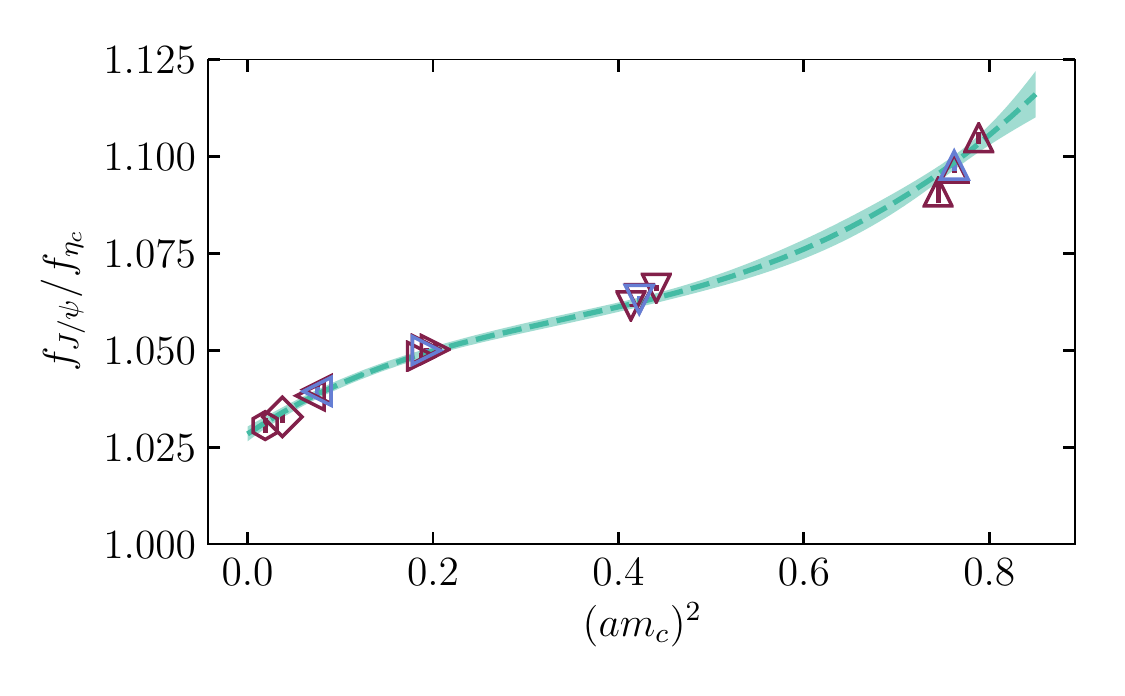}
  \caption{The ratio of $J/\psi$ to $\eta_c$ decay constants calculated on the ensembles of
  Table~\ref{tab:ensembles} in pure QCD (red points) and including also quenched 
QED (blue points) plotted against the square of the 
bare $c$ quark mass in lattice units. 
The green curve marks our extrapolation to the physical 
point. 
}
  \label{fig:fratio-full}
\end{figure}

Finally, in Figure~\ref{fig:fratio-full} we plot results for the ratio of 
$J/\psi$ to $\eta_c$ decay constants and show the fit curve extrapolated 
to the continuum limit. This gives   
\begin{equation}
\label{eq:fratfull}
\frac{f_{J/\psi,\mathrm{QCD+QED}}}{f_{\eta_c,\mathrm{QCD+QED}}}= 1.0284(19) .
\end{equation}
This is almost the same as the pure QCD result. 

\subsection{Discussion : $f_{J/\psi}$ and $f_{\eta_c}$}
\label{sec:discussdecay}
\begin{figure}
  \includegraphics[width=0.47\textwidth]{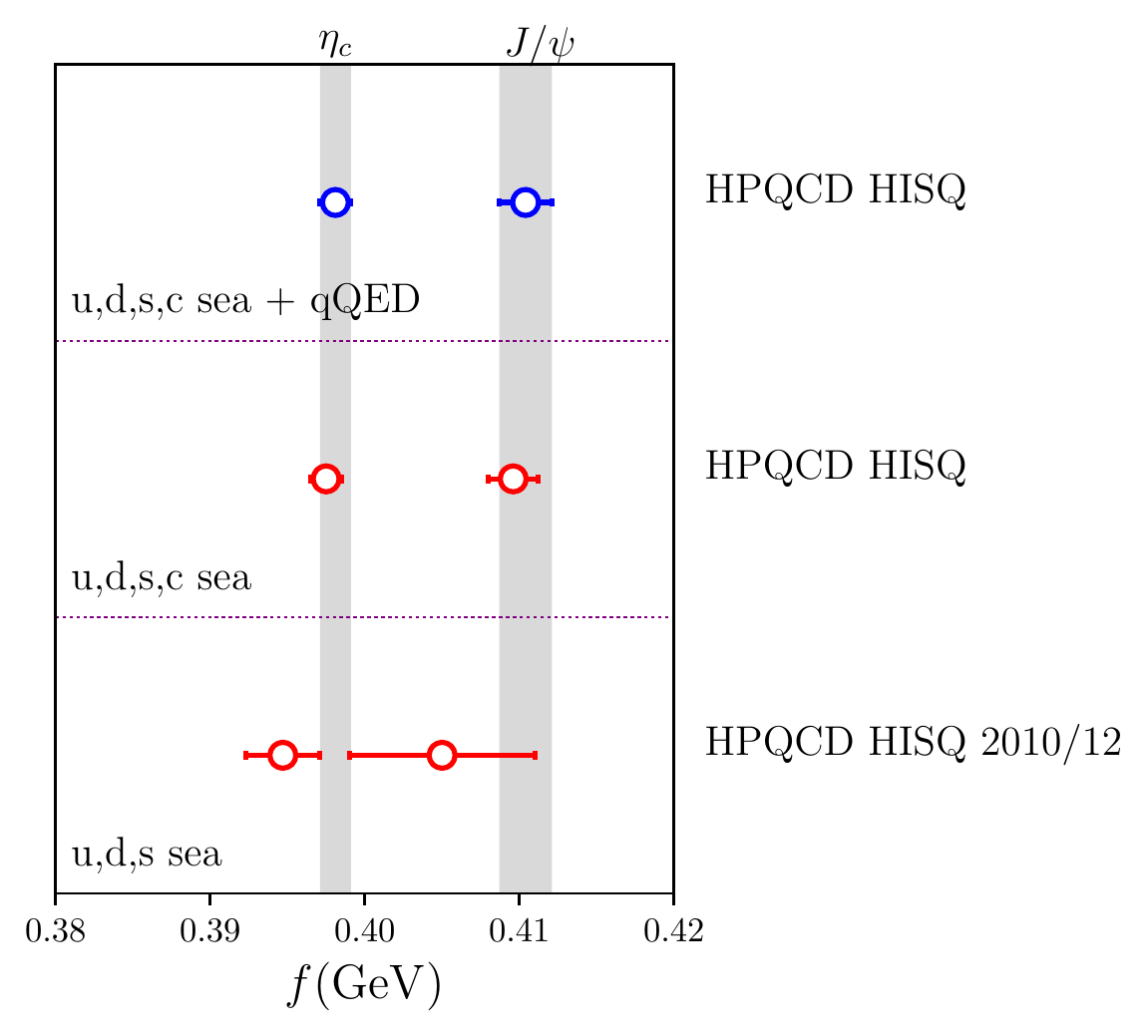}
  \caption{A comparison of our new results for $f_{\eta_c}$ and $f_{J/\psi}$ 
with earlier lattice QCD results, also by HPQCD, on gluon field configurations that 
include $n_f=2+1$ flavours of quarks in the sea. The results labelled `HPQCD HISQ' 
are from this paper and the results labelled `HPQCD HISQ 2010/12' are from~\cite{Davies:2010ip, Donald:2012ga}. The grey bands are the $\pm 1\sigma$ bands from our new QCD+QED results.  
  }
  \label{fig:fcomp}
\end{figure}

\begin{figure}
  \includegraphics[width=0.47\textwidth]{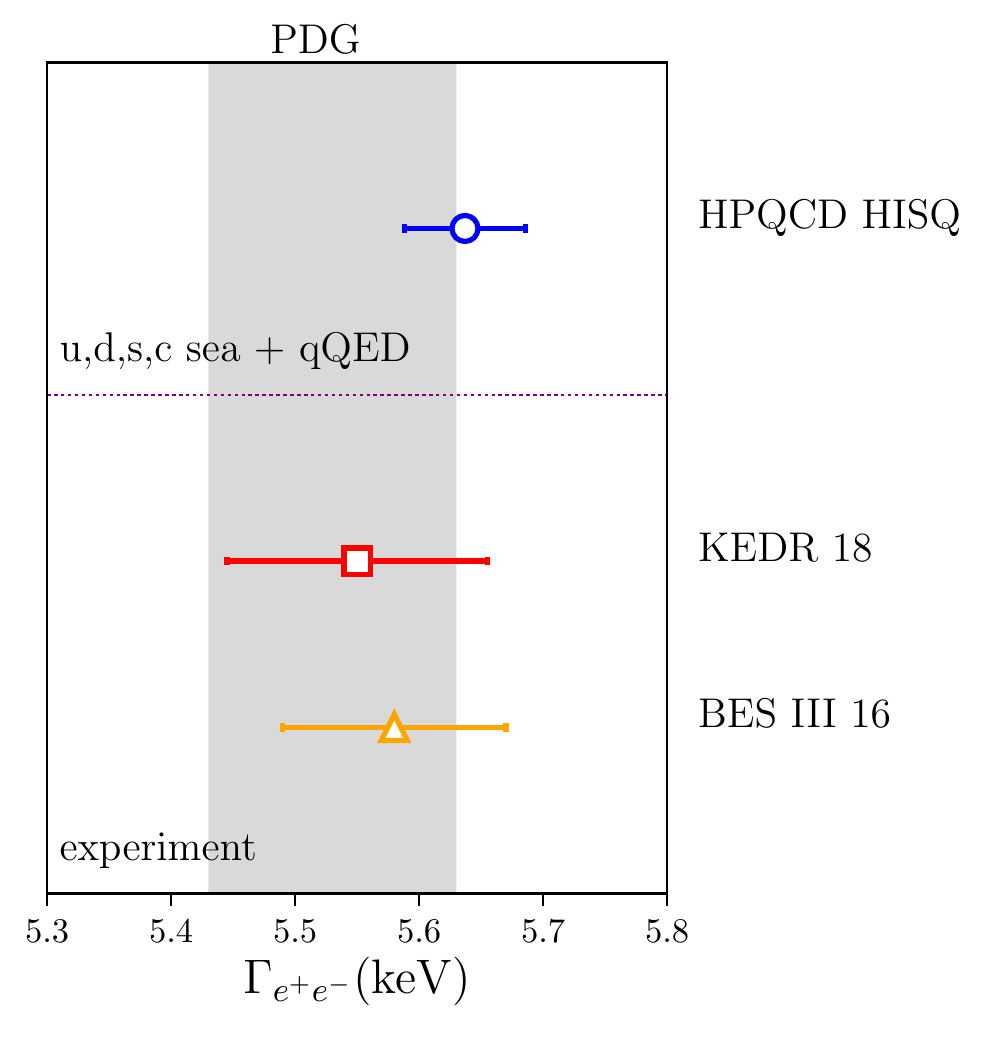}
  \caption{A comparison of the width for $J/\psi$ decay to $e^+e^-$ implied by 
our new results for $f_{J/\psi}$ to that obtained from two recent experiments. 
The point labelled `KEDR 18' is from~\cite{Anashin:2018ldj} and the point 
labelled `BES III 16' is from~\cite{Ablikim:2016xbg}. 
The grey bands are the $\pm 1\sigma$ bands from the Particle Data Group 
average~\cite{Tanabashi:2018oca}.  
  }
  \label{fig:gamcomp}
\end{figure}

Figure~\ref{fig:fcomp} compares our new pure QCD and QCD+QED results to 
previous results including $n_f=2+1$ flavours of sea quarks for 
$f_{\eta_c}$~\cite{Davies:2010ip} and $f_{J/\psi}$~\cite{Donald:2012ga}. There is good agreement. 
These earlier calculations also 
used HISQ quarks but our new results are more accurate, 
particularly for $f_{J/\psi}$ because of the use of more accurate values of $Z_V$~\cite{Hatton:2019gha}.  

There have also been calculations that use $n_f=2$ flavours of sea quarks. 
It is harder to make a comparison to these results because it is not clear 
what the systematic error is from not including at least the $s$ quarks in 
the sea, and no uncertainty is included for this. 
The calculation of~\cite{Becirevic:2013bsa} uses twisted-mass 
quarks on $n_f=2$ gluon field configurations and obtains $f_{\eta_c}=$387(7)(2) MeV
and $f_{J/\psi}=$418(8)(5) MeV. 
The calculation of~\cite{Bailas:2018car} uses
clover quarks on $n_f=2$ CLS gluon-field configurations to give 
$f_{\eta_c}=$387(3)(3) MeV and $f_{J/\psi}=$399(4)(2) MeV. The $n_f=2$ results 
for $f_{\eta_c}$ agree with each other and have a central value about 
2$\sigma$ below ours. The $\sigma$ here is that from the $n_f=2$ results 
since our uncertainty is much smaller.  
The $n_f=2$ results for $f_{J/\psi}$ are compatible with each other 
and with our result, again at $2\sigma$.  

As discussed in Section~\ref{sec:jpsidecay}, the $J/\psi$ decay 
constant is the hadronic quantity that is needed to determine the 
rate of $J/\psi$ annihilation to leptons via Eq.~(\ref{eq:width-decayconst}). 
Our result for $\Gamma$ using Eq.~(\ref{eq:width-decayconst}) with 
$f_{J/\psi}$ from Eq.~(\ref{eq:fjpsifull}) is 
\begin{equation}
\label{eq:ourgam}
\Gamma(J/\psi \rightarrow e^+e^-) = 5.637(47)(13) \, \mathrm{keV} . 
\end{equation}
The first uncertainty is from our lattice QCD+QED result
for $f_{J/\psi}$ and the second uncertainty allows for
a relative $\mathcal{O}(\alpha_{\mathrm{QED}}/\pi)$  
correction to Eq.~(\ref{eq:width-decayconst}) from higher-order effects. 

Figure~\ref{fig:gamcomp} compares this width $\Gamma(J/\psi \rightarrow e^+e^-)$ 
to results from experiment. Recent experimental results from 
KEDR~\cite{Anashin:2018ldj} and 
BES III~\cite{Ablikim:2016xbg} are shown along with the 
Particle Data Group average~\cite{Tanabashi:2018oca} (as a grey band).  
Figure~\ref{fig:gamcomp} shows good agreement between our result and 
the experimental values shown, as well as the experimental average. 

The lattice QCD result is now more accurate than the experimental values. 

\section{Vector correlator moments and $a_{\mu}^c$}
\label{sec:amuc}

\begin{table*}
  \caption{Time-moments of charmonium vector current-current correlators calculated on the ensembles in Table~\ref{tab:ensembles}.
The results tabulated are values of $(G_n/Z_V^2)^{1/(n-2)}$ in lattice units 
along with their statistical uncertainties, 
given for $n=4$, 6, 8 and 10 in columns 2, 3, 4 and 5. 
Uncertainties are statistical only. Note that the results for different moments are 
correlated because they are determined from the same correlation functions. 
The results marked with a $*$ and $\dag$ are for deliberately mistuned $am_c$ values as 
detailed in the caption to Table~\ref{tab:masses}. 
  Also included are the ratios at fixed $am_c$, denoted $R^0_{\mathrm{QED}}$, of QCD+QED to 
QCD results for each rooted moment on a subset of ensembles.
}
\label{tab:moments-data}
\begin{ruledtabular}
\begin{tabular}{lllllllll}
Set & $n=4$ & $n=6$ & $n=8$ & $n=10$ & $R_{\mathrm{QED}}^{0}[n=4]$ & $R_{\mathrm{QED}}^{0}[n=6]$ & $R_{\mathrm{QED}}^{0}[n=8]$ & $R_{\mathrm{QED}}^{0}[n=10]$ \\
\hline
1 & 0.389670(40) & 0.949791(62) & 1.410524(75) & 1.815497(88) & - & - & - & - \\
2 & 0.396283(22) & 0.961260(35) & 1.425498(42) & 1.833868(49) & 0.999954(26) & 0.999910(17) & 0.999858(15) & 0.999810(15) \\
3 & 0.400779(15) & 0.969045(24) & 1.435671(28) & 1.846369(33) & - & - & - & - \\
\hline
4 & 0.511194(12) & 1.164351(19) & 1.701040(26) & 2.184698(34) & - & - & - & - \\
6 & 0.5206344(85) & 1.181180(14) & 1.724311(19) & 2.214708(24) & 0.9998455(15) & 0.9997169(11) & 0.9995987(10) & 0.9994908(11) \\
$6^*$ & 0.5254224(87) & 1.189687(14) & 1.736041(19) & 2.229780(25) & - & - & - & - \\
8 & 0.5254560(47) & 1.1897785(76) & 1.736217(10) &  2.230069(13) & - & - & - & - \\
\hline
9 & 0.70981(13) & 1.53941(21) & 2.24688(27) & 2.90799(32) & - & - & - & - \\
10 & 0.723760(11) & 1.566115(20) & 2.285959(27) & 2.959283(36) & 0.999554(24) & 0.999312(20) & 0.999124(20) & 0.998995(22) \\
11 & 0.731489(11) & 1.580936(18) & 2.307649(25) & 2.987715(32) & - & - & - & - \\
\hline
12 & 1.070736(33) & 2.276543(58) & 3.355470(80) & 4.37418(10) & 0.999096(59) & 0.998767(49) & 0.998584(48) & 0.998489(48) \\
13 & 1.114660(44) & 2.366266(78) & 3.48827(11) & 4.54699(14) & - & - & - & - \\
\hline
14 & 1.431378(91) & 3.03675(16) & 4.49434(22) & 5.86769(29) & - & - & - & - \\
$14^{\dagger}$ & 1.46556(17) & 3.10710(31) & 4.59734(43) & 6.00058(56) & - & - & - & - \\
\hline
15 & 1.91475(23) & 4.06357(42) & 6.02429(55) & 7.86806(66) & - & - & - & - \\
\end{tabular}
\end{ruledtabular}
\end{table*}

With new results expected from the Fermilab $g-2$ experiment soon there has been
a concerted effort by the lattice community to understand and control systematic
effects in the lattice QCD calculation of the hadronic vacuum polarisation (HVP)
contribution to the anomalous magnetic moment of the muon. 
Since the most accurate values for the HVP currently come from experimental results 
on $R(e^+e^- \rightarrow \mathrm{hadrons})$, it is also important to compare 
lattice QCD results to these, disaggregated by flavour where possible. 
 
The first calculation
of the quark-line connected $c$-quark contribution to the HVP, $a_{\mu}^c$, 
was given in~\cite{Chakraborty:2014mwa} using results for the time-moments 
of vector charmonium current-current correlators calculated in~\cite{Donald:2012ga}. 
The time moments are defined by
\begin{equation} \label{eq:time-moments-latt}
  G_{n} = Z_V^2 \sum_t t^{n} C_V(t) ,
\end{equation}
where $C_V(t)$ is the vector current-current correlator and
$Z_V$ is the vector current renormalisation factor, discussed in 
Section~\ref{sec:jpsidecay}. Note that $t \in \{-L/2+1,-L/2+2,\ldots,L/2\}$. 

The even-in-$n$ time-moments for $n\ge 4$ can be related to the derivatives at $q^2=0$ of the renormalised vacuum 
polarisation function~\cite{Allison:2008xk}, $\hat{\Pi}(q^2) \equiv \Pi(q^2)-\Pi(0)$, by 
\begin{equation}
G_{n} = \left. (-1)^{n/2} \frac{\partial^{n}}{\partial q^{n}} q^2\hat{\Pi}(q^2)\right|_{q^2=0} .
\end{equation}
This means that $\hat{\Pi}(q^2)$ can be reconstructed, using Pad\'{e} approximants, 
from the $G_{n}$~\cite{Chakraborty:2014mwa} and 
fed into the integral over $q^2$ that yields the quark-line 
connected HVP contribution
to $a_{\mu}$~\cite{Blum:2002ii}. Only 
time-moments of low moment number 
are needed to give an accurate result for $a_{\mu}^c$ because the integrand is 
dominated by small $q^2$. We will give results for 
the four lowest moments, $n={4,6,8,10}$, improving on the values given 
in~\cite{Donald:2012ga}. The improvement comes mainly 
through the use of a more accurate vector current renormalisation    
as well as an improved method for reducing lattice spacing uncertainties
but we also use second-generation gluon field configurations that include $c$ quarks 
in the sea and calculate, rather than estimate, the impact of 
the leading QED effects. 

$\hat{\Pi}(q^2)$ and hence $a_{\mu}^c$ can also be determined from experimental results for 
$R_c \equiv R(e^+e^-\rightarrow c\overline{c} \rightarrow \mathrm{hadrons})$ 
as a function of squared 
centre-of-mass energy, $s$. 
This can be done using inverse-$s$ moments  
\begin{equation}
\label{eq:Mdef}
  \mathcal{M}_k \equiv \int \frac{ds}{s^{k+1}}R_c(s) \, .
\end{equation}
$R_c$ is obtained from the full $e^+e^-$ rate from 
just below the $c$ threshold upwards by subtracting the background 
contribution from $u$, $d$, and $s$ quarks perturbatively, see e.g.~\cite{Kuhn:2007vp}.

The relationship between $G_{2k+2}$ and $\mathcal{M}_k$ is then  
\begin{equation}
\label{eq:GMmatch}
  G_{2k+2} = \frac{(2k+2)!\mathcal{M}_k}{12\pi^2 Q^2} .
\end{equation}
A comparison of our correlator time-moments calculated on the lattice and extrapolated to 
the continuum limit to the inverse-$s$ moments determined from 
experiment is equivalent to a test of the agreement of the results for $a_{\mu}^c$
in the two cases. 

\subsection{Vector correlator moments: Pure QCD and QCD+QED results}
\label{sec:momentsres}

Table~\ref{tab:moments-data} gives our raw results for the time-moments of 
the same vector current-current correlators from which we have 
determined the mass and decay constant of the $J/\psi$ meson in 
Sections~\ref{sec:hyperfine} and~\ref{sec:jpsidecay}. 
Notice that the statistical uncertainties are tiny. 
The correlators make use of a local vector current
that must be renormalised as discussed in Section~\ref{sec:jpsidecay}. 
The results in Table~\ref{tab:moments-data} are calculated {\it before} 
renormalisation and are given in lattice units. The quantity that 
is tabulated is 
\begin{equation}
\left(\frac{G_n}{Z_V^2}\right)^{1/(n-2)}\,.
\end{equation}
We take the $(n-2)$th root to reduce all results to the same dimensions~\cite{Donald:2012ga}. 
To normalise the time-moments we use the $Z_V$ values at $\mu$= 2 GeV given 
in Table~\ref{tab:ZVs} that were used for $f_{J/\psi}$ in Section~\ref{sec:jpsidecay}. 

\begin{figure}
  \includegraphics[width=0.47\textwidth]{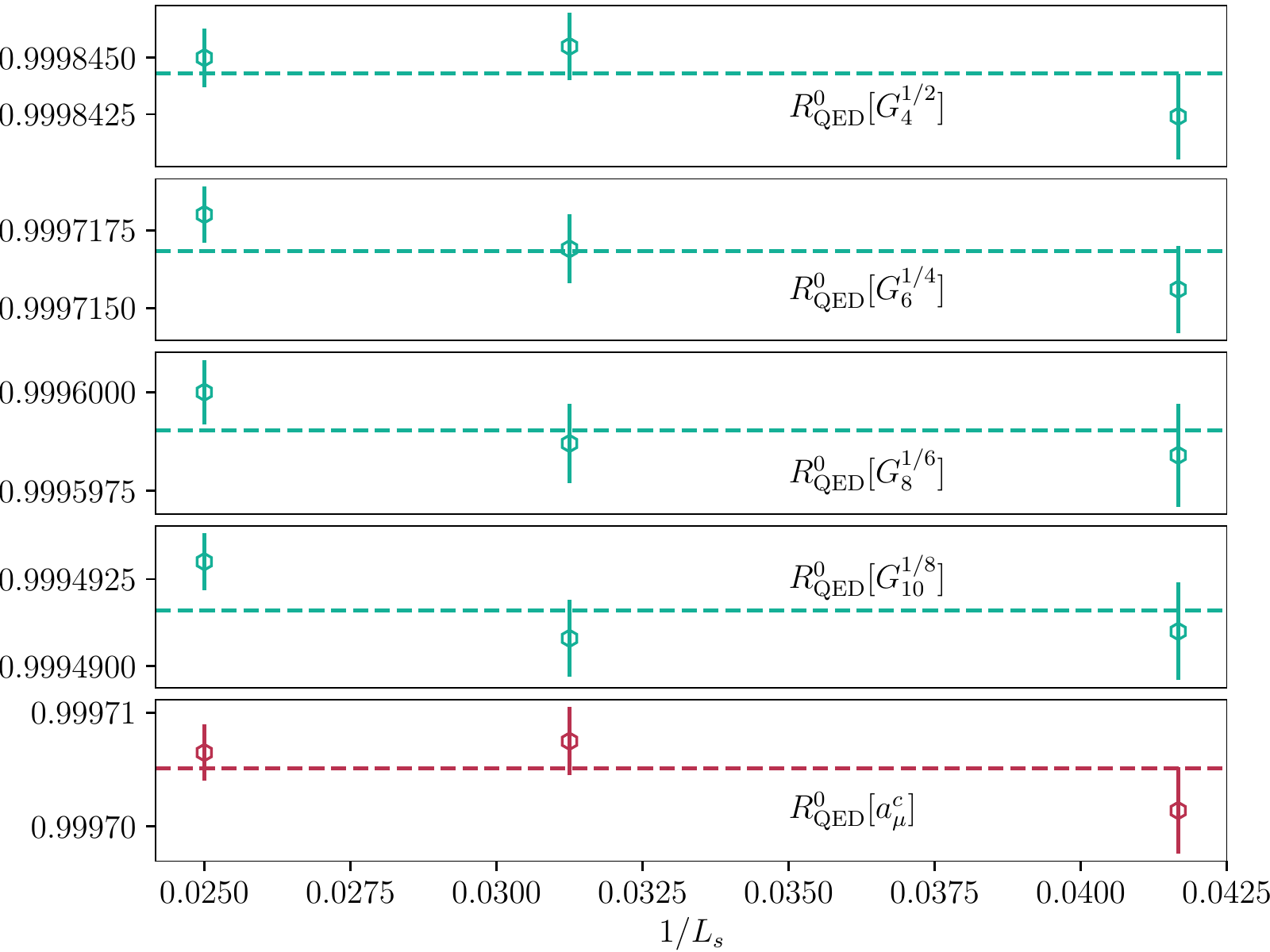}
  \caption{A study of the volume dependence of the electromagnetic correction 
  to the first four time moments of the 
vector current-current correlator and to $a_{\mu}^c$ on sets 5-7. There
  is no observable dependence on the lattice spatial extent, $L_s$, as can be 
judged by comparison to the dashed horizontal lines at the weighted averages 
of the points. 
}
  \label{fig:mom-vol-dep}
\end{figure}

Table~\ref{tab:moments-data} also gives the result of including 
quenched QED as the ratios $R^0_{\mathrm{QED}}$ for each rooted moment 
(at fixed $am_c$). These values are all very slightly 
less than 1, by up to 0.1\% for $n$=4, and 0.2\% for $n$=10.   
We can also test the finite-volume dependence of the quenched QED effect 
using sets 5, 7 and 8 and the results are shown in Fig.~\ref{fig:mom-vol-dep}. 
There is no visible volume dependence in the QED effect on time-moments at 
the level of our statistical uncertainties. This is to be expected, as seen for 
the $J/\psi$ mass and decay constant in Sections~\ref{sec:hyperfine} and~\ref{sec:jpsidecay}, 
since the vector current being used here is electrically neutral. 

To fit the results as a function of lattice spacing it is convenient to 
work with the dimensionless combination:
\begin{equation}
\label{eq:MG}
M_{J/\psi} \left(G_n\right)^{1/(n-2)} \, ,
\end{equation}
using our $M_{J/\psi}$ masses from Table~\ref{tab:masses}. 
The removal of dimensions reduces the uncertainty coming 
from the value of the lattice spacing. 
At the same time this quantity also has reduced sensitivity 
to mistuning of the valence 
$c$ quark mass because the $m_c$-dependence is largely cancelled by 
$M_{J/\psi}$. 
We fit the quantity defined in Eq.~(\ref{eq:MG}) as a function of lattice 
spacing, allowing for 
quark-mass mistuning 
effects of both valence and sea quarks to derive results in the physical continuum limit. 
We do this as before using the fit function given in Eq.~(\ref{eq:X-fit}). 
Values in the physical continuum limit are then divided by 
$M_{J/\psi}$ from experiment to obtain our final results for $G_n^{1/(n-2)}$ in 
$\mathrm{GeV}^{-1}$.  In doing this we allow an uncertainty of 0.7 MeV in 
$M_{J/\psi}$ from annihilation to a photon (see Section~\ref{sec:hyperfine})
since this effect is not included in our results. 

\begin{table}
  \caption{ QCD+QED results in the physical continuum limit for 
the first four time-moments (column 2) compared
  with the results extracted from experiment in column 3~\cite{Dehnadi:2011gc}.
  Agreement within 2$\sigma$ is seen for all except the 
4th moment, but the lattice 
QCD results are much more accurate. Column 4 gives the effect of 
quenched QED as a ratio of the physical results in QCD+QED to those in pure QCD. 
  }
  \label{tab:moments}
\begin{ruledtabular}
\begin{tabular}{llll}
$n$ & $G_{n}^{1/(n-2)}$ & $(G_{n}^{\mathrm{exp.}})^{1/(n-2)}$ & $R_{\mathrm{QED}}[G_n^{1/(n-2)}]$ \\
& $\mathrm{GeV}^{-1}$ & $\mathrm{GeV}^{-1}$ & $-$ \\
\hline
4 & 0.31715(49) & 0.3110(26) & 1.00106(13) \\
6 & 0.67547(84) & 0.6705(31) & 1.00069(11) \\
8 & 1.0041(11) & 0.9996(36) & 1.00047(10) \\
10 & 1.3117(13) & 1.3080(37) & 1.00037(10) \\
\end{tabular}
\end{ruledtabular}
\end{table}

\begin{figure}[h!]
  \includegraphics[width=0.47\textwidth]{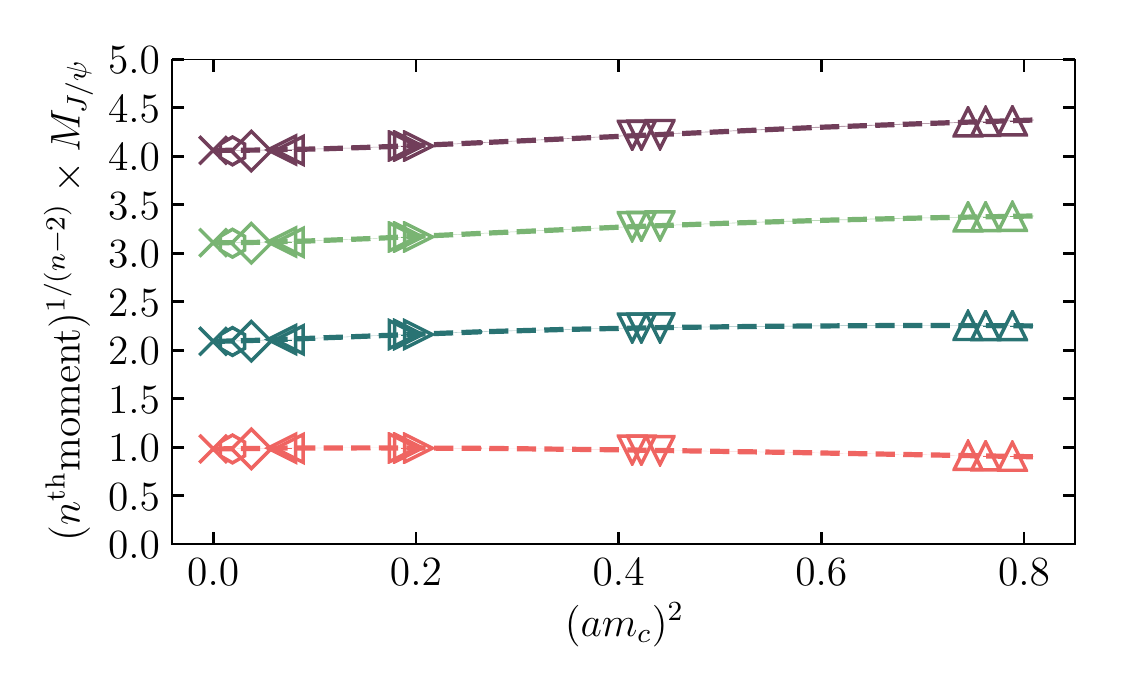}
  \caption{The four lowest time moments and their extrapolation to $a=0$. The symbols 
give results for the rooted moment multiplied by $M_{J/\psi}$ with different symbols 
denoting different groups of ensembles with similar lattice spacing. The diffferent 
colours pick out the different moments, from $n=4$ at the bottom to $n=10$ at the 
top of the plot. 
Only the results from the pure QCD calculation at well-tuned $c$ quark masses are shown 
for clarity. 
Uncertainties are too small to be visible. The
  extrapolation for all the moments are performed simultaneously including correlations 
between moments determined from the same current-current correlators. 
The dashed lines give the QCD+QED fit curve using Eq.~(\ref{eq:X-fit}) and 
the coloured crosses mark the result in the continuum limit at physical quark masses. }
  \label{fig:time-moments}
\end{figure}

Fig.~\ref{fig:time-moments} plots the results for the rooted moments 
multiplied by $M_{J/\psi}$
as a function of $(am_c)^2$. Also shown 
is the fit result from Eq.~(\ref{eq:X-fit}). 
Only the pure QCD lattice results are shown for clarity; those including the 
effect of quenched QED are very close to them. The fit result plotted is 
that for the QCD+QED case. 

Table~\ref{tab:moments} gives our QCD+QED results for the 
4th, 6th, 8th and 10th rooted 
moments in the physical continuum limit. These are obtained from a simultaneous 
fit to all of the moments, including the correlations between them (since they 
are derived from the same correlators). The fit has a $\chi^2/\mathrm{dof}$ of 0.62
for an svdcut of $1\times 10^{-3}$. 

Column 3 of Table~\ref{tab:moments} gives the results derived from experimental data for 
$R(e^+e^- \rightarrow \mathrm{hadrons})$ and $\Gamma_{\ell\ell}$ and mass
for the $J/\psi$ and $\psi^{\prime}$ by~\cite{Dehnadi:2011gc} for comparison 
(see Section~\ref{sec:amuc}). 
We have converted these results into the quantity that we calculate 
according to Eq.~(\ref{eq:GMmatch}). We see agreement within 2$\sigma$ 
for $n\ge 6$ 
with the values derived from experiment, but a 2.4$\sigma$ tension at 
$n=4$. The results given from~\cite{Dehnadi:2011gc} correspond to 
their standard selection of experimental datasets. Results shift 
by $\pm 1\sigma$ for other selections. Note that the lattice QCD results 
are now much more precise than those determined from experiment. 
The comparison between lattice QCD and experiment 
will be discussed further in Section~\ref{sec:momdiscuss}. 

Column 4 of Table~\ref{tab:moments} gives the final impact of quenched 
QED, including the quark mass retuning, on the moments. We see that 
the ratios, $R_{\mathrm{QED}}[G_n^{1/(n-2)}]$, are greater than 1, 
in contrast to the $R^0_{\mathrm{QED}}$ of Table~\ref{tab:moments-data} that 
are less than 1. As discussed in Section~\ref{sec:incqed}, the inclusion 
of QED means that the $c$ quark mass must be retuned downwards. The rooted time-moments 
are approximately inversely proportional to the quark mass and so they increase 
under this retuning, more than offsetting the direct effect of QED seen in 
$R^0_{\mathrm{QED}}$.   

\begin{table}
  \caption{Error budget for $G_n^{1/(n-2)}$ 
as a percentage of the final answer. 
}
  \label{tab:G4G10-errbudg}
\begin{ruledtabular}
\begin{tabular}{lllll}
 & $G_4^{1/2}$ & $G_6^{1/4}$ & $G_8^{1/6}$ & $G_{10}^{1/8}$ \\
\hline
$a^2 \to 0$ & 0.06 & 0.05 & 0.04 & 0.03 \\
$Z_V$ & 0.04 & 0.02 & 0.02 & 0.02 \\
Pure QCD Statistics & 0.03 & 0.02 & 0.02 & 0.02 \\
QCD+QED Statistics & 0.01 & 0.01 & 0.01 & 0.01\\
Sea mistunings & 0.06 & 0.03 & 0.03 & 0.03 \\
Valence mistunings & 0.01 & 0.00 & 0.00 & 0.00 \\
$M_{J/\psi}$ & 0.02 & 0.02 & 0.02 & 0.02 \\
$w_0$ & 0.10 & 0.08 & 0.06 & 0.05 \\
$w_0/a$ & 0.06 & 0.05 & 0.05 & 0.04\\
\hline
Total & 0.15 & 0.12 & 0.11 & 0.10 \\
\end{tabular}
\end{ruledtabular}
\end{table}

Table~\ref{tab:G4G10-errbudg} gives the error budget for the 
$G_n$. 
The total uncertainty in all cases is below 0.2\%. 

\subsection{Discussion: Vector correlator moments}
\label{sec:momdiscuss}

\begin{figure}
  \includegraphics[width=0.47\textwidth]{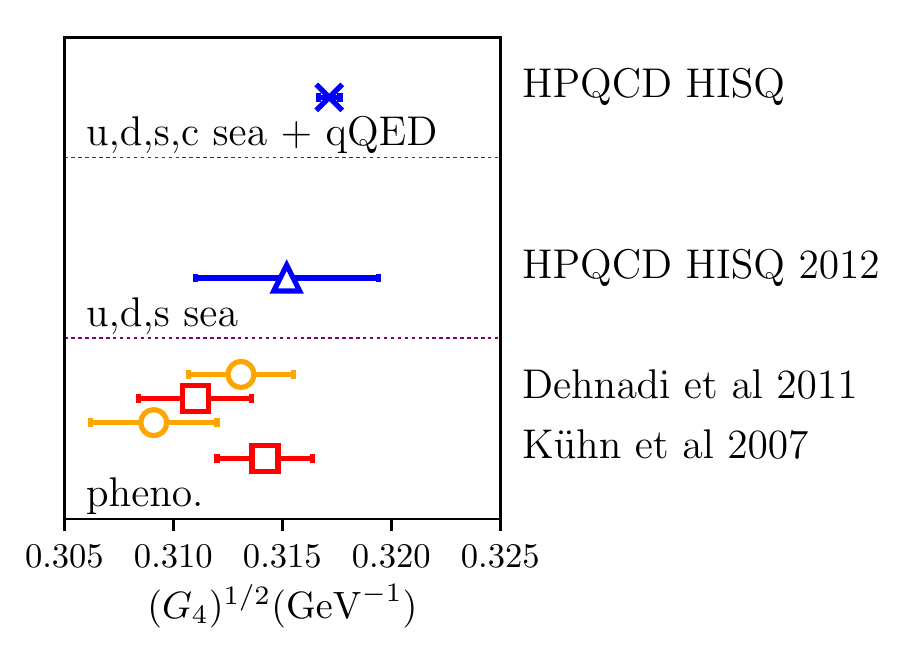}
  \includegraphics[width=0.47\textwidth]{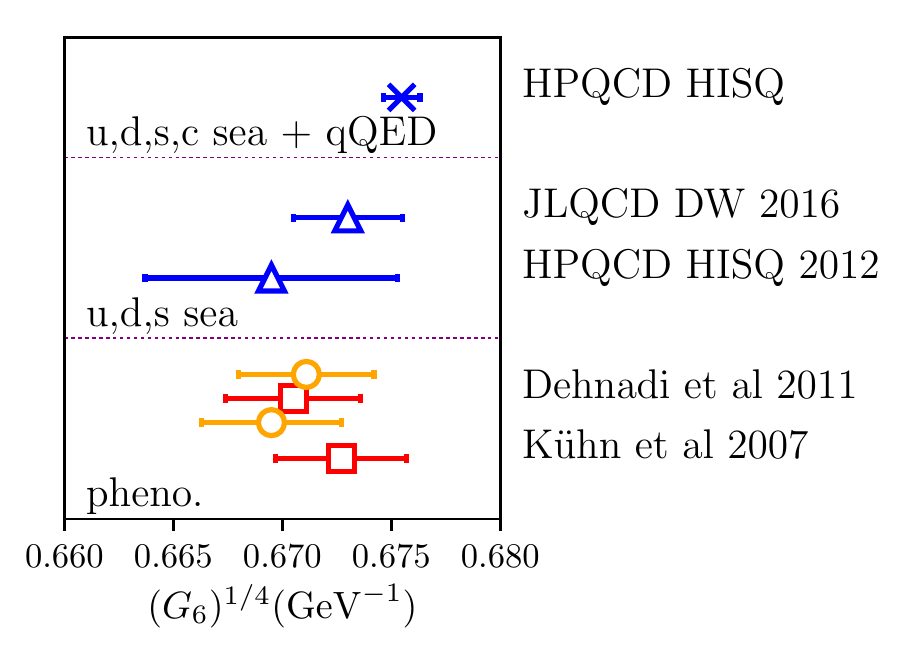}
  \caption{Comparison of our new result to those of 
previous lattice QCD calculations for the 4th and 6th time-moments (appropriately 
rooted) of the charmonium vector current-current correlator. 
Our new result obtained here on $n_f=2+1+1$ gluon field configurations
and including the effect of quenched QED 
is given at the top (blue cross). HPQCD's 2012 result on 
$n_f=2+1$ gluon field configurations with HISQ valence $c$ quarks 
is marked `HPQCD HISQ 2012'~\cite{Donald:2012ga}. JLQCD's 2016 
result using $n_f=2+1$ domain-wall quarks is marked `JLQCD DW 2016' for 
the 6th moment only. We also compare to results (open red squares) denoted `pheno.' 
that are derived from experimental data for the cross-section for $e^+e^-$ 
to hadrons as a function of centre-of-mass energy, by determining the $c\overline{c}$ 
component. The points plotted come from~\cite{Kuhn:2007vp} 
and~\cite{Dehnadi:2011gc}. Open orange circles show alternative 
selections of datasets from~\cite{Dehnadi:2011gc}; the upper value is 
for the `maximal' set and the lower value for the `minimal' set.     
  }
  \label{fig:momcomp}
\end{figure}

Our new results for the time-moments of vector current-current correlators 
improve significantly on earlier lattice QCD calculations and are now 
more accurate than results derived from experiment. 

Figure~\ref{fig:momcomp} compares our new results for the 4th and 6th 
moments to earlier lattice QCD results. Comparison for the 8th and 
10th moments gives a very similar picture and so is not shown. 
The first lattice QCD calculation of the time-moments of vector 
charmonium current-current correlators was given by HPQCD in~\cite{Donald:2012ga} 
using HISQ valence quarks on $n_f=2+1$ asqtad gluon field configurations. 
Our new results have an uncertainty almost ten times smaller than these. 
The error budget in the earlier results was dominated by the uncertainty 
in $Z_V$ from the use of continuum perturbation theory in the matching 
factors and there was also a sizeable uncertainty from the lattice spacing. 
These uncertainties have been enormously reduced here 
and in addition we no longer need an uncertainty from missing QED effects. 
We also show a comparison with the subsequent results from the 
JLQCD collaboration~\cite{Nakayama:2016atf} using domain-wall quarks on 
$n_f=2+1$ gluon field configurations. JLQCD do not give a value 
for the 4th moment because of discretisation effects in their 
formalism (tree-level $\mathcal{O}(a^2)$). The dominant uncertainties 
in their results are from statistics and from the value of $t_0^{1/2}$ used
to fix the lattice spacing, on which they have a 2\% uncertainty. 
Good agreement is seen for all of the lattice QCD results. 

Two of the most recent results from phenomenological determinations 
of the moments~\cite{Kuhn:2007vp, Dehnadi:2011gc} 
are also compared in Fig.~\ref{fig:momcomp}. The results 
from~\cite{Dehnadi:2011gc} include experimental datasets
for the inclusive cross-section 
that are both older and newer than those used in~\cite{Kuhn:2007vp}. 
Results from~\cite{Dehnadi:2011gc}'s  `standard' selection of datasets 
were given in Table~\ref{tab:moments} and are shown in Fig.~\ref{fig:momcomp} 
in red. We also show, in orange, the results from the `maximal' set (all 
experimental information available at that point) 
and the `minimal' set (datasets that are needed to cover the full $\sqrt{s}$ 
range from 2 GeV to 10.5 GeV without gaps, keeping the most accurate results).   
Note that the resonance parameters are the same for all selections. 
We see that the variation with dataset selection covers almost $1\sigma$ 
for the 4th moment, but much less for the 6th moment. This is also 
reflected in the differences between~\cite{Dehnadi:2011gc} and~\cite{Kuhn:2007vp}. 

These phenomenological analyses must subtract the `non-charm' background 
from experimental results for $R(e^+e^- \rightarrow \mathrm{hadrons})$ to 
leave $R_c$ for Eq.~(\ref{eq:Mdef}). $R_c$ is defined to be the 
result from diagrams with a charm quark loop
connected to a photon at both ends~\cite{Kuhn:2007vp} i.e.\ the quark-line 
connected vector current-current correlator that we study on the lattice. 
The subtracted background includes QED effects for the non-charm and singlet 
(quark-line disconnected) contributions. The remainder, $R_c$ 
then includes the QED effects associated with the $c\overline{c}$ loop. 
The dominant source 
of uncertainty in $R_c$ comes from the charmonium resonance ($J/\psi$ and 
$\psi^{\prime}$) region and is set by the uncertainty in 
$\Gamma_{ee}$ for these states. 
The fractional uncertainty is approximately the same for all 
moments~\cite{Kuhn:2007vp, Dehnadi:2011gc}. When the $(n-2)$th root is taken 
the fractional uncertainty then falls with increasing $n$.  

Good agreement is seen between the phenomenological results and our 
new lattice results for $n=6$, 8 and 10, although the lattice results 
are systematically at the upper end of the phenomenological range. 
The largest discrepancy is a 2.8$\sigma$ tension for the 
4th moment between us and the results of~\cite{Dehnadi:2011gc} for their 
minimal selection of datasets. The tension is 2.4$\sigma$ for the 
standard selection, and below 2$\sigma$ for the maximal selection 
and for the results of~\cite{Kuhn:2007vp}.
The $\sigma$ here is that for the phenomenological results since 
the lattice uncertainty is much smaller. 
Because the 4th moment dominates the determination of 
$a_{\mu}^c$, this tension between lattice QCD+QED 
and some of the phenomenological results carries 
over to $a_{\mu}^c$, to be discussed in the next section. 

The time-moments can also be used to determine a value for $\overline{m}_c$ 
by comparing to $\mathcal{O}(\alpha_s^3)$ continuum QCD perturbation theory 
and this was the focus of~\cite{Kuhn:2007vp,Dehnadi:2011gc}. 
We do not do this here because the scale of $\alpha_s$ is rather low 
in these determinations meaning that uncertainties from missing higher-order 
corrections can be substantial. We prefer instead the method 
of~\cite{Chakraborty:2014aca}, which enables a higher scale to be used in 
the perturbation theory. We have checked, however that the $m_c$ value 
that would be obtained from the time-moments 
is consistent with both~\cite{Chakraborty:2014aca} and the 
value given in Section~\ref{sec:mass}.  

\subsection{$a_{\mu}^c$: Pure QCD and QCD+QED results}
\label{sec:amucres}

\begin{table}
  \caption{Values of $a_{\mu}^c$ on the ensembles of Table~\ref{tab:ensembles} 
and the direct quenched QED correction on a subset of those ensembles. 
Those marked with a $*$ and $\dag$ are at deliberately mistuned $c$ masses 
(see caption to Table~\ref{tab:masses}). 
  The uncertainties quoted are correlated through the value of $M_{J/\psi}$ (for all ensembles, see text) and $Z_V$ 
(for ensembles at a given $\beta$).  
}
  \label{tab:amuc}
\begin{ruledtabular}
\begin{tabular}{lll}
Set & $a_{\mu}^c \times 10^9$ & $R_{\mathrm{QED}}^{0} \left[a_{\mu}^c \right]$ \\
\hline
1 & 1.23183(78) & - \\
2 & 1.24522(75) & 1.000478(80) \\
3 & 1.25431(77) & - \\
3A & 1.25518(49) & - \\
3B & 1.25485(48) & - \\
\hline
4 & 1.40782(91) & - \\
6 & 1.41738(91) & 1.001080(89) \\
$6^*$ & 1.42370(91) & - \\
8 & 1.42234(91) & - \\
\hline
9 & 1.47866(97) & - \\
10 & 1.48514(75) & 1.001416(83) \\
11 & 1.48853(75) & - \\
\hline
12 & 1.4725(13) & 1.00141(15) \\
13 & 1.4805(13) & - \\
\hline
14 & 1.4610(33) & - \\
$14^{\dagger}$ & 1.4702(33) & - \\
\hline
15 & 1.4572(10) & - \\
\end{tabular}
\end{ruledtabular}
\end{table}

\begin{figure}
  \includegraphics[width=0.47\textwidth]{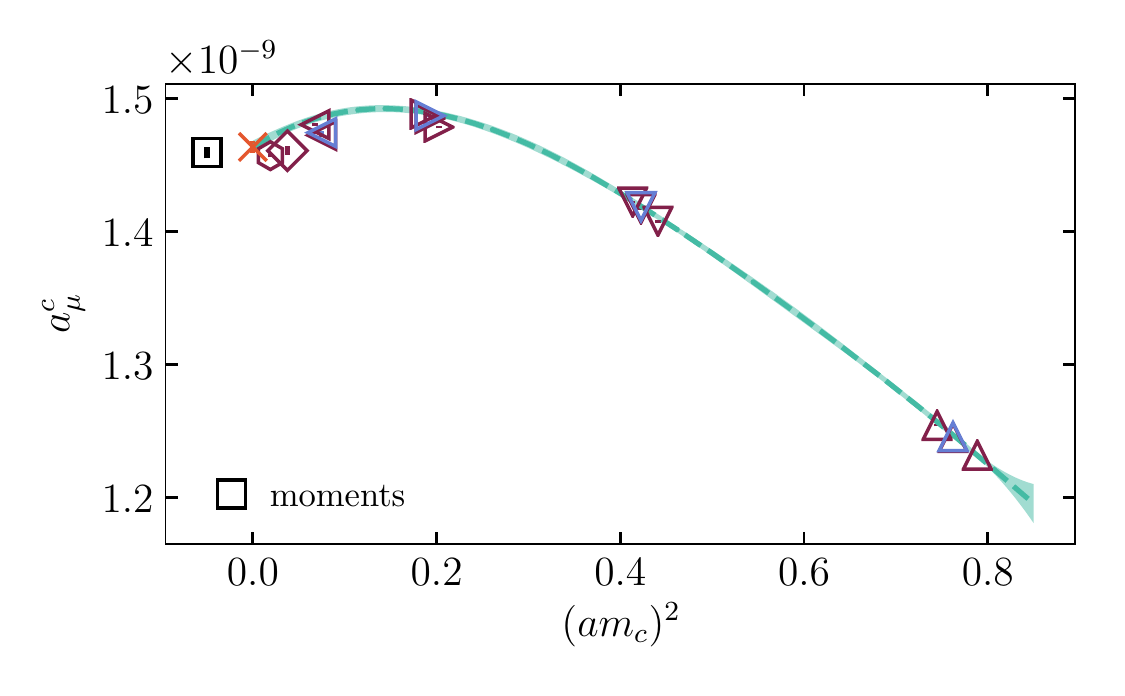}
  \caption{Extrapolation to the continuum physical point 
of the connected charm HVP contribution to the
  anomalous magnetic moment of the muon. Different symbols denote results 
on groups of ensembles with similar lattice spacing. Results at deliberately 
mistuned $c$ quark masses are not plotted but are included in the fit. 
The red points correspond to pure QCD, 
the light blue points to QCD+QED and the dashed green fit curve 
plotted is that for QCD+QED. The continuum result (red cross)
  is compared to the result (open black square) obtained by calculating $a_{\mu}^c$ from the
  individually extrapolated time-moments in Section~\ref{sec:momentsres}. 
}
  \label{fig:amuc}
\end{figure}

To calculate the quark-line connected HVP contribution to 
$a_{\mu}$ from $c$ quarks, $a_{\mu}^c$, we can either 
use the physical results for the vector current-current correlator 
time-moments discussed in the previous subsection or 
we can calculate $a_{\mu}^c$ on each lattice ensemble and 
perform a fit as a function of lattice spacing to extrapolate 
directly to the continuum limit. We will do the latter here.  

The values of $a_{\mu}^c$ on each lattice 
are given in Table~\ref{tab:amuc}. These are determined from 
the time-moments rescaled by the $J/\psi$ mass on each lattice :
\begin{equation}
\frac{M_{J/\psi}G_n^{1/(n-2)}}{M_{J/\psi}^{\mathrm{expt}}} \,. 
\end{equation}
As discussed in Section~\ref{sec:momentsres} rescaling by 
$M_{J/\psi}$ reduces the lattice spacing uncertainty and the 
impact of mistuning the $c$ quark mass. This was used for lighter 
quark masses in~\cite{Chakraborty:2016mwy} 
(see also~\cite{Burger:2013jya}). The reduced effect 
of mistuning is clear from comparing the mistuned 
results in Table~\ref{tab:amuc} to those in Table~\ref{tab:moments-data}. 

Table~\ref{tab:amuc} also includes results on 
the two ensembles, sets 3A and 3B, that allow direct comparison 
of the effects of strong-isospin breaking the sea. 
The ratio of the two results is 0.99974(14), so the results 
agree to within 0.03\% (and 2$\sigma$), consistent with 
these effects being negligible here (see Section~\ref{sec:seaeffects}). 

Table~\ref{tab:amuc} also gives the direct effect of quenched 
QED through the ratio $R^0_{\mathrm{QED}}[a_{\mu}^c]$. Because the 
rescaled moments have less sensitivity to the $c$ quark mass 
these numbers are larger than 1 (unlike the results 
in Table~\ref{tab:moments-data}) and reflect more closely 
the final impact of QED on $a_{\mu}^c$. 
We observe no finite-volume dependence for the quenched QED 
corrections to $a_{\mu}^c$, as for the correlator time
moments. This is shown in Fig.~\ref{fig:mom-vol-dep}.

The results from Table~\ref{tab:amuc} are shown in Fig.~\ref{fig:amuc} 
along with our standard fit of the form given in Eq.~(\ref{eq:X-fit}). 
The fit has a $\chi^2/\mathrm{dof}$ of 0.44. This fit obtains
the physical values 
\begin{eqnarray}
\label{eq:amudirect}
a_{\mu}^c &=& 14.606(47)\times 10^{-10} , \quad \mathrm{QCD}\\ 
  a_{\mu}^c &=& 14.638(47)\times 10^{-10} , \quad \mathrm{QCD+QED} \nonumber
\end{eqnarray}
along with $R_{\mathrm{QED}}[a_{\mu}^c] = 1.00214(19)$.
These results agree well (within $1\sigma$) with those obtained using the 
extrapolated values for the moments from Table~\ref{tab:moments} 
and calculating $a_{\mu}^c$ in 
the continuum, as is seen 
in Fig.~\ref{fig:amuc}. 

The error budget for our final value of $a_{\mu}^c$ is given in
Table~\ref{tab:amu-errbudg}. The largest uncertainties come from  
the determination of the lattice spacing, although this uncertainty 
is much reduced by our rescaling with $M_{J/\psi}$. 

\begin{table}
  \caption{Error budget for $a_{\mu}^c$ from our fit to 
   $a_{\mu}^c$ values on each ensemble.}
  \label{tab:amu-errbudg}
\begin{ruledtabular}
\begin{tabular}{ll}
 & $a_{\mu}^c$ \\
 \hline
 $a^2 \to 0$ & 0.15  \\
 $Z_V$ & 0.07 \\
 Pure QCD Statistics & 0.08 \\
 QCD+QED Statistics & 0.01  \\
 $w_0/a$ & 0.16 \\
 $w_0$ & 0.18 \\
 Sea mistunings & 0.09 \\
 Valence mistunings & 0.03 \\
 $M_{J/\psi}^{\mathrm{exp.}}$ & 0.05 \\
 \hline
 Total & 0.32 \\
 \end{tabular}
\end{ruledtabular}
\end{table}

\subsection{Discussion: $a_{\mu}^c$ }
\label{sec:amudiscuss}

\begin{figure}
  \includegraphics[width=0.47\textwidth]{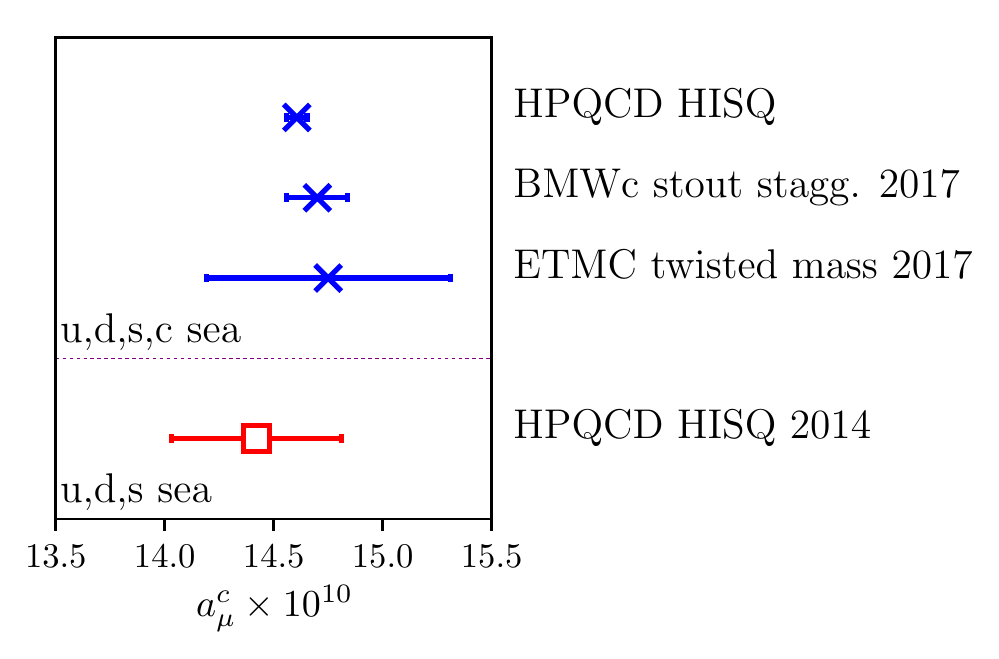}
  \caption{Comparison of lattice QCD results (not including QED) 
for the connected $c$ quark 
HVP contribution to $a_{\mu}$, $a_{\mu}^c$. 
Results are divided according to the number of sea quark flavours included 
in the gluon field configurations on which the calculation was done. The first 
result, labelled `HPQCD HISQ 2014' is from~\cite{Chakraborty:2014aca} using 
values of time-moments determined in~\cite{Donald:2012ga} using HISQ 
valence quarks on gluon field configurations including $n_f=2+1$ flavours of 
asqtad sea quarks. The other results all include $n_f=2+1+1$ flavours of sea 
quarks. The result labelled `ETMC twisted mass 2017' uses the twisted mass 
formalism~\cite{Giusti:2017jof} and that labelled `BMW stout stagg. 2017' 
a smeared staggered quark action~\cite{Borsanyi:2017zdw}. 
Our new result (from Eq.~(\ref{eq:amudirect})) 
labelled `HPQCD HISQ' agrees with, but is much more accurate
than, these earlier results. 
}
  \label{fig:amuc-lattcomp}
\end{figure}

\begin{figure}
  \includegraphics[width=0.47\textwidth]{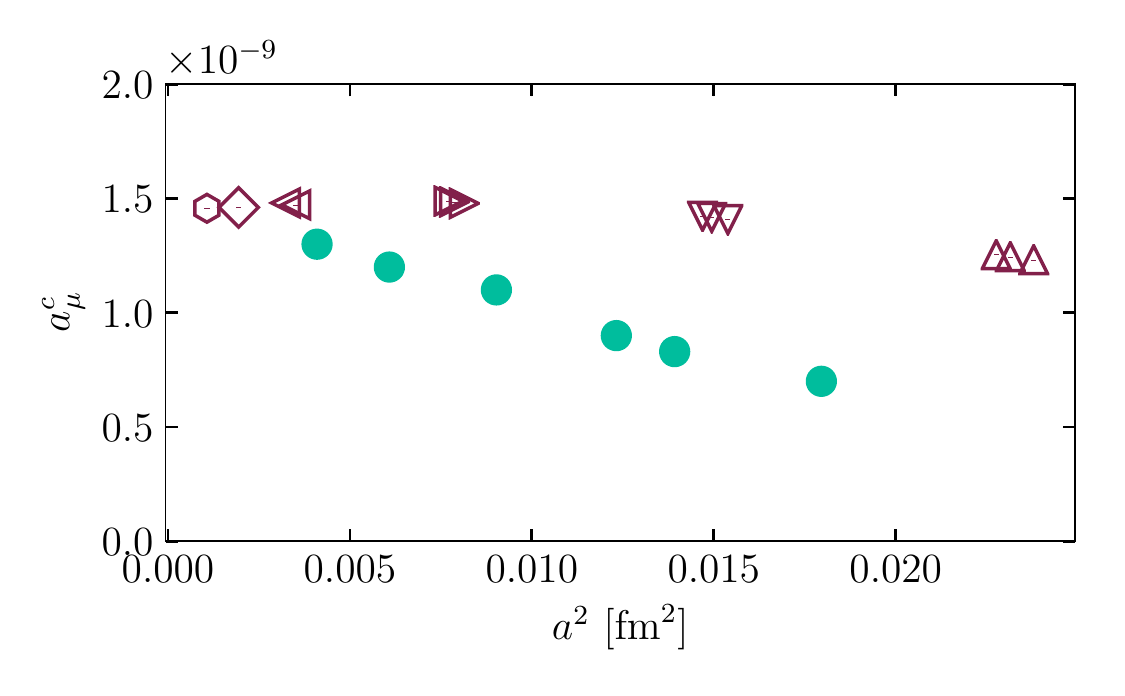}
  \caption{Comparison of the discretisation effects in $a_{\mu}^c$ in
our results from 
  Table~\ref{tab:amuc} (red open symbols) and those from the 
BMW collaboration in~\cite{Borsanyi:2017zdw} (filled blue circles) 
that use a less highly-improved staggered quark action. 
 The filled blue circles only give estimates of the position of 
the BMW data points and do not indicate the statistical uncertainties 
which are much smaller than the size of the points. 
}
  \label{fig:BMW-comp}
\end{figure}

Our new result for the pure QCD case 
(Eq.~(\ref{eq:amudirect})) is compared to earlier lattice QCD results 
with realistic sea quark content in Fig.~\ref{fig:amuc-lattcomp}. 
Our new result (the top point on the plot) is much more accurate 
than the earlier results. With respect to the first calculation of 
$a_{\mu}^c$ that also used HISQ quarks~\cite{Chakraborty:2014aca} we 
have greatly reduced the previous dominant sources of uncertainty 
from $Z_V$ and the determination of the lattice spacing. 

With respect to results using other formalisms, we give one figure 
to demonstrate the control of discretisation effects that is 
possible with the HISQ formalism. 
Figure~\ref{fig:BMW-comp} compares the approach to the continuum 
limit of our results (from Table~\ref{tab:amuc}) with results 
from BMWc~\cite{Borsanyi:2017zdw}, for which the continuum 
extrapolated result is shown in Fig.~\ref{fig:amuc-lattcomp}. 
The points plotted from~\cite{Borsanyi:2017zdw} are estimates of 
the positions read from Figure S4, and do not include any indication of 
statistical uncertainties. 
The BMWc stout staggered quark action has $\mathcal{O}(a^2)$ 
discretisation errors at tree-level since it uses an unimproved 
derivative in its version of the Dirac equation. 
Figure~\ref{fig:BMW-comp} shows that the price to be paid 
for not improving the discretisation is a very large discretisation 
effect. This is 
particularly evident when working with heavier quarks such as charm, 
since it means that the dominant $(\Lambda a)^2$ effects 
have $\Lambda \approx$ 1 GeV, as here. 
This means, for example, that the BMWc points at finest lattice spacing 
($\sim$0.06 fm) are about as far (20\% away) from the continuum limit as our points 
at our coarsest lattice spacing ($\sim$0.15 fm). 
Our results at $a\sim$ 0.06 fm are within 1\% of the continuum limit, 
allowing us to achieve a sub-1\% uncertainty in the final value. 

We find that the impact of quenched QED on the result for $a_{\mu}^c$ 
is +0.214(19)\% (Eq.~\ref{eq:amudirect}). This is a shift in 
$a_{\mu}^c$ in the presence of the dominant QED effect of
\begin{equation}
\label{eq:amuc-shift}
\delta a_{\mu}^c = +0.0313(28) \times 10^{-10} . 
\end{equation}
We can compare this to the result obtained by ETMC in~\cite{Giusti:2019xct}
following work on the QED effect on the renormalisation of quark 
bilinears in~\cite{DiCarlo:2019thl}. The ETMC value is 
$+0.0182(36) \times 10^{-10}$, 2.8$\sigma$ smaller than ours. 

The two calculations of the quenched QED effect 
are different. Ours is a direct calculation of 
the quenched QED effect, retuning the valence quark mass through determination 
of a meson mass in the usual way. The ETMC calculation is perturbative 
in quenched QED and fixes the valence quark masses so that they agree 
in the $\overline{\mathrm{MS}}$ scheme at 2 GeV in QCD+QED and pure QCD.  
It is likely that it is this difference in scheme for 
definining how QCD and QCD+QED are compared that is responsible for
the tension between the two values.
Our results in Section~\ref{sec:mass} show that the quark mass in the 
$\overline{\mathrm{MS}}$ scheme
is lower in QCD+QED than in QCD (at scales above $\overline{m}_c$) 
when the quark mass is tuned so that $M_{J/\psi}$ agrees with experiment
in the two cases.  
This leads us to expect a larger result for the 
impact of quenched QED on $a_{\mu}^c$ with our tuning. 
Once full lattice QCD+QED calculations are underway tuning of the 
quark masses will be done through matching of 
meson masses to experiment. 

\begin{figure}
  \includegraphics[width=0.47\textwidth]{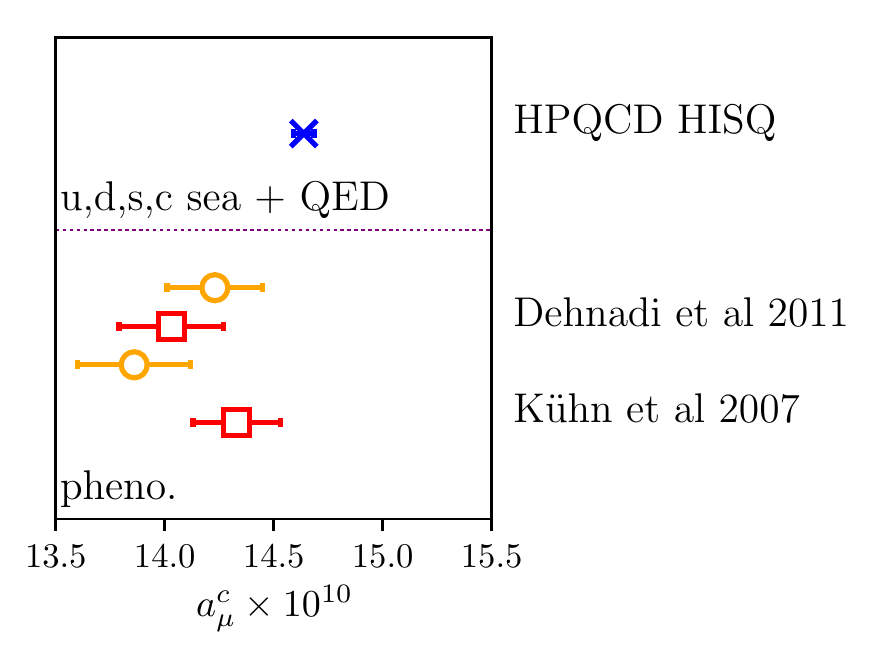}
  \caption{Comparison of our lattice QCD+QED result (blue cross)
for the connected $c$ quark 
HVP contribution to $a_{\mu}$, $a_{\mu}^c$, with results determined 
from experimental information. 
The red open squares, denoted `pheno.', 
use the charmonium moments from $R_{e^+e^-}$ given 
in~\cite{Kuhn:2007vp, Dehnadi:2011gc}
to determine $a_{\mu}^c$. The orange open circles give the alternative 
maximal (upper) and minimal (lower) inclusive cross-section datasets 
from~\cite{Dehnadi:2011gc}. 
}
  \label{fig:amuc-phencomp}
\end{figure}

Figure~\ref{fig:amuc-phencomp} includes a comparison of our 
result for $a_{\mu}^c$, now in QCD+QED, with 
values obtained using the moments determined from 
$J/\psi$ and $\psi^{\prime}$ properties and the inclusive 
cross-section for $e^+e^- \rightarrow {\mathrm{hadrons}}$, 
removing contributions from quarks other than $c$. 
The results for these moments from~\cite{Kuhn:2007vp, Dehnadi:2011gc} 
were discussed and compared to our results in Section~\ref{sec:momdiscuss}. 
In Fig.~\ref{fig:amuc-phencomp} we have converted the moments 
into a result for $a_{\mu}^c$ for comparison. 
As with the moments, we see a $2.5\sigma$ tension with the~\cite{Dehnadi:2011gc} 
result using the standard selection of 
datasets ($a_{\mu}^c = 14.03(24) \times 10^{-10}$), and a larger tension 
with the minimal selection of datasets. There is agreement within 2$\sigma$ 
for the maximal selection of datasets from~\cite{Dehnadi:2011gc} 
and with~\cite{Kuhn:2007vp}. Our result may then provide a pointer to the 
selection of $R_{e^+e^-}$ datasets for the phenomenological 
determination and/or indicate an issue with the perturbation 
theory used to subtract the $u/d/s$ contribution to obtain $R_c(s)$. 

More recent determinations of the complete HVP contribution to 
$a_{\mu}$ using results from $R_{e^+e^-}$ are given 
in~\cite{Keshavarzi:2018mgv, Davier:2019can}. 
Although the $c$ component is not separated out, the contribution 
from the $J/\psi$ resonance is given as $6.26(19)\times 10^{-10}$ 
by~\cite{Keshavarzi:2018mgv} and $6.20(11)\times 10^{-10}$ by~\cite{Davier:2019can}. 
We can also readily determine the contribution to $a_{\mu}^c$ from the $J/\psi$ 
resonance alone since in that case~\cite{Chakraborty:2015ugp} 
\begin{equation}
G_n = n! \frac{f_{J/\psi}^2}{M_{J/\psi}^n} \,.
\end{equation}
Using our value for $f_{J/\psi}$ from Eq.~(\ref{eq:fjpsifull}) and 
the experimental value for $M_{J/\psi}$~\cite{Tanabashi:2018oca} gives
\begin{equation}
\label{eq:amucjpsi}
a_{\mu}^c (J/\psi) = 6.345(53)\times 10^{-10}. 
\end{equation} 
This is in good 
agreement with the results determined above from the experimental $J/\psi$ 
parameters, but more accurate, reflecting the fact that our 
result for $\Gamma_{e^+e^-}$ in Fig.~\ref{fig:gamcomp} agrees with 
experiment but has smaller uncertainty.  

The $J/\psi$ contribution provides almost half of $a_{\mu}^c$ - 
we conclude that it is the rest of the contribution, from the inclusive 
cross-section above the resonance region, that causes the tension 
between our results for $a_{\mu}^c$ and that from $R_{e^+e^-}$ for 
some selections of experimental datasets. 
The tension then amounts to 7(3)\% of this non-resonant cross-section. 

\section{Conclusions}
\label{sec:conclusions}

We have performed the first $n_f = 2+1+1$ lattice QCD computations 
of the properties of ground-state charmonium mesons. 
These have been done using the HISQ action to calculate 
quark-line connected two-point 
correlation functions on gluon field configurations that include 
$u/d$ quark masses going down to the physical point. The 
small discretisation effects in the HISQ action and high 
statistics achievable have given us good 
control over both the continuum and chiral extrapolations 
and has enabled us to obtain smaller uncertainties than previous 
lattice QCD computations of these properties (including previous 
calculations by HPQCD). At the same time we have improved the tuning of the bare 
$c$ quark mass to update the value of $\overline{m}_c$. 
 From the same correlators that we use for 
the masses and decay constants we have also derived an improved 
result for the $c$ quark HVP contribution to the anomalous magnetic 
moment of the muon.  

The precision possible for $c$ quark correlators with the HISQ action 
makes it possible to determine the impact of the $c$ quark's 
electric charge. We do this directly and nonperturbatively in quenched 
QED by multiplying an appropriate U(1) field into 
our gluon field configurations. We tune the bare $c$ quark 
mass so that the $J/\psi$ mass agrees with experiment in both 
QCD+QED and QCD and we calculate mass and vector renormalisation 
constants in the RI-SMOM scheme in both cases, performing a full 
analysis as a function of $\mu$ to determine the $c$ quark mass 
in the $\overline{\mathrm{MS}}$ scheme.  

Here we collect our final QCD+QED results 
(from Eqs.~(\ref{eq:finalhyp}),~(\ref{eq:mcval-qcdqed}),~(\ref{eq:fjpsifull}),~(\ref{eq:fetacfull}),~(\ref{eq:amudirect})) 
before discussing each in turn:
\begin{eqnarray}
\label{eq:finalres}
    M_{J/\psi} - M_{\eta_c} &= 0.1203(11)\ \mathrm{GeV} \nonumber \\
    \overline{m}_c(3\ \mathrm{GeV}) &= 0.9841(51)\ \mathrm{GeV} \nonumber \\
    f_{J/\psi} &= 0.4104(17)\ \mathrm{GeV} \nonumber \\
    f_{\eta_c} &= 0.3981(10)\ \mathrm{GeV} \nonumber \\
    a_{\mu}^c &= 14.638(47) \times 10^{-10} .
\end{eqnarray}
Error budgets are given in Tables~\ref{tab:hf-errbudg}, \ref{tab:mc-errbudg}, \ref{tab:f-errbudg} and~\ref{tab:amu-errbudg} respectively. 
Our error budgets do not include an error for missing QED 
effects for sea quarks or from having $m_u=m_d$ in the sea. 
We expect the former to dominate (see Sections~\ref{sec:quenching} 
and~\ref{sec:seaeffects}) and estimate this at 10\% of the size of the valence 
quark QED effects. 
This is then a negligible error in every case. 

The precision of our result for the charmonium 
hyperfine splitting allows us to resolve, for the
first time, the sign and magnitude of the anticipated difference between the
lattice and experimental results arising from the fact that we 
do not include quark-line disconnected correlation functions. 
We take this difference to be the effect of the $\eta_c$
decay to two gluons which is prohibited in the lattice calculation, 
and conclude that $\Delta M_{\eta_c}^{\mathrm{annihln}}=+7.3(1.2)$ MeV. 

The effect of QED on the hyperfine splitting is fairly substantial (1.4\%) and 
the largest effect that we observe here. This has 3 components that all act 
in the same direction: a direct effect 
of 0.7\%, 0.1\% from retuning the $c$ quark mass and 0.6\% from $J/\psi$ 
annihilation to a photon that we add by hand.  

Our updated value of the charm quark mass (in the $\overline{\mathrm{MS}}$
scheme at a scale of 3 GeV) includes the effect of QED on the bare $c$ quark 
mass, tuned so that $M_{J/\psi}$ matches experiment, and on the 
mass renormalisation constant $Z_m$ determined on the lattice using the 
intermediate RI-SMOM scheme. The addition of results at a finer lattice 
spacing has improved the uncertainty slightly over HPQCD's 
earlier result~\cite{Lytle:2018evc}. 

The impact of QED on $\overline{m}_c$ is small since QED effects 
tend to cancel between the retuning needed and changes in $Z_m$. 
At a scale of 3 GeV we see a -0.18(2)\% effect, falling towards zero 
at a scale of $\overline{m}_c$. 

Our result for the the $J/\psi$ decay constant is the most 
precise to date and acts as a subpercent
test of QCD. The gain in precision from the 2012 HPQCD calculation 
of~\cite{Donald:2012ga}
is a result of the use of a more accurate renormalisation of the vector 
current~\cite{Hatton:2019gha} as well as gluon field configurations with 
a wider range of sea $u/d$ quark masses and lattice spacing values. 
We can use our result for $f_{J/\psi}$ to determine a value for 
the width for $J/\psi$ decay to a lepton-antilepton pair 
(repeating Eq.~(\ref{eq:ourgam})):
\begin{equation}
\label{eq:ourgam2}
\Gamma(J/\psi \rightarrow e^+e^-) = 5.637(47)(13) \, \mathrm{keV} . 
\end{equation}
This is more accurate than the current average of
experimental results~\cite{Tanabashi:2018oca}.  

The impact of quenched QED on $f_{J/\psi}$ is +0.2\%, since the retuning 
of the $c$ quark mass offsets some of the direct effect. 
The effect on $f_{\eta_c}$ is almost the same so that the 
ratio of $f_{J/\psi}$ to $f_{\eta_c}$ remains the same. This 
ratio is determined here to be 1.0289(19), 
so that it is definitely greater than 1; this was not completely
clear from earlier calculations. 

Our results for the time-moments of the charmonium vector 
current-current correlators also provide a new level of 
accuracy for these quantities, improving by a factor of 
10 over the first such calculations in~\cite{Donald:2012ga}. 
Our results are given in Table~\ref{tab:moments}. We see some 
tension for the lowest (4th) moment with phenomenological 
results derived from $R(e^+e^- \rightarrow \mathrm{hadrons})$ 
in~\cite{Dehnadi:2011gc} when particular selections of 
experimental datasets are made.

We use the time-moments to derive the connected 
$c$ quark HVP contribution to the anomalous magnetic moment of 
the muon, $a_{\mu}^c$. Although this is not a large part of 
the total HVP contribution and so improving its uncertainty 
has little impact on the full HVP contribution, nevertheless 
it is a piece that can be calculated very accurately in lattice 
QCD and provides a test case for comparison of lattice calculations 
and a comparison with phenomenology.   

Our result for $a_{\mu}^c$ improves the accuracy by a factor 
of 3 over earlier lattice QCD  
values~\cite{Chakraborty:2014aca, Giusti:2017jof, Borsanyi:2017zdw}.  
Comparison of our result for $a_{\mu}^c$ to that determined 
from phenomenology can be divided into contributions from narrow 
resonances and from the continuum 
$e^+e^- \rightarrow c\overline{c}$. The contribution 
from the $J/\psi$ 
agrees well between our result and phenomenology, 
with our result being more accurate (see Section~\ref{sec:amudiscuss}). This reflects 
the situation described above for $f_{J/\psi}$ and $\Gamma_{\ell\ell}$. 
The total $a_{\mu}^c$
derived from the time-moments determined from 
experimental data on $R(e^+e^- \rightarrow \mathrm{hadrons})$ 
when the component from a $c\overline{c}$ loop is separated out 
shows some tension with our results, depending on which experimental 
datasets are used above the resonance region. 
Our central value is higher, tending to reduce by a small amount the discrepancy 
in $a_{\mu}$ between existing experiment and the Standard Model. 
We should stress, however, that more complete determinations of 
$a_{\mu}$, for example in~\cite{Keshavarzi:2018mgv, Davier:2019can}, do not 
separate out $a_{\mu}^c$ and so we cannot make a direct comparison 
to them. 

We also determine the impact of quenched QED on $a_{\mu}^c$ in a 
scheme in which the $c$ quark mass is tuned from $M_{J/\psi}$ in 
both QCD+QED and pure QCD. We find (repeating Eq.~(\ref{eq:amuc-shift})) 
\begin{equation}
\label{eq:amuc-shift2}
\delta a_{\mu}^c = +0.0313(28) \times 10^{-10} . 
\end{equation}
This is a 0.2\% effect and dominated by the impact of retuning 
the $c$ quark mass. 

Our result for $a_{\mu}^c$ has an accuracy of 0.3\%. Sub-0.5\% 
uncertainty is the aim for lattice QCD calculations of the full 
HVP contribution to $a_{\mu}$. We have shown that this is possible, 
for a small piece of the HVP, at least.  

\subsection*{\bf{Acknowledgements}}

We are grateful to the MILC collaboration for the use of
their configurations. We are also grateful for the use of 
MILC's QCD code. We have modified 
it to generate quenched U(1) gauge fields and incorporate those 
into the quark propagator calculation as described here.
We thank P. Knecht for contributions at the start of the project, 
C. McNeile for the calculation
of $w_0/a$ and A. Keshavarzi, D. Miller, D. Nomura, D. Smaranda  
and T. Teubner for useful discussions.
Computing was done on the Darwin supercomputer at the University of
Cambridge High Performance Computing Service as part of the DiRAC facility,
jointly funded by the Science and Technology Facilities Council,
the Large Facilities Capital Fund of BIS and
the Universities of Cambridge and Glasgow.
We are grateful to the Darwin support staff for assistance.
Funding for this work came from the
Science and Technology Facilities Council
and the National Science Foundation.

\begin{appendix}

\section{$Z_V$ and $Z_m$ determination on the finest lattices}
\label{appendix:zv}

\begin{figure}
  \includegraphics[width=0.47\textwidth]{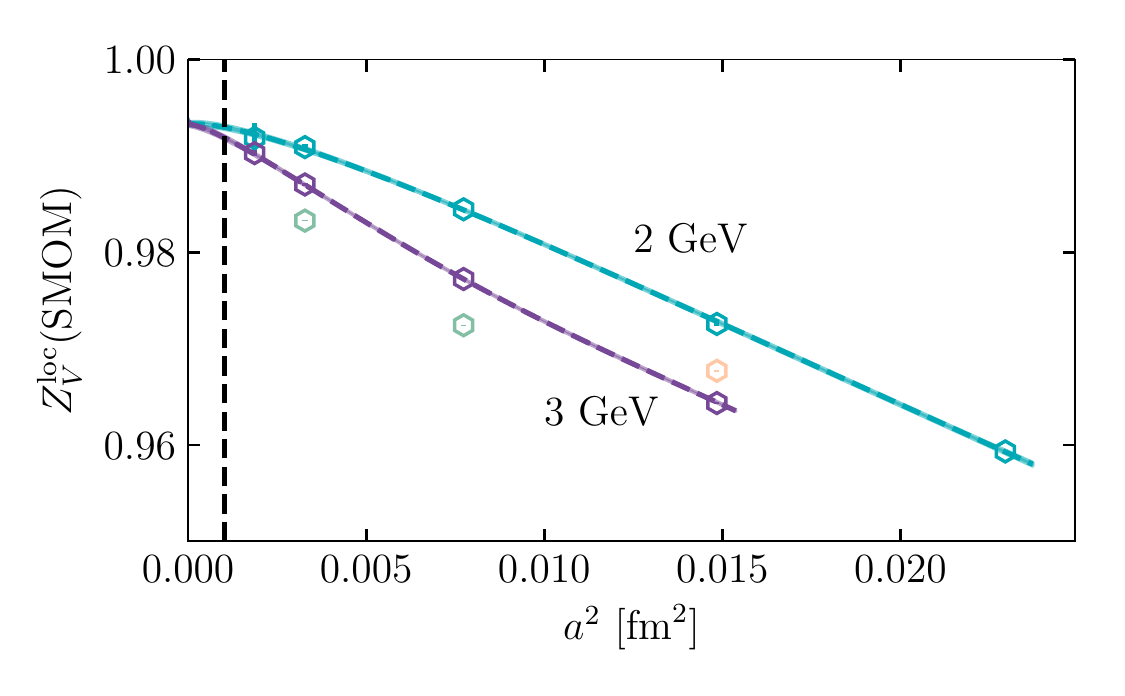}
  \caption{Results for the local vector current renormalisation factor 
in the RI-SMOM
  scheme compared to the fit form given in Eq.~\ref{eq:ZV-fit}. 
  All of the data
  points shown are included in the fit but only the 2 and 3 GeV fit lines are
  drawn. The data points not on the fit lines correspond to $\mu =$ 2.5 GeV (light orange) 
and 4 GeV (light green). The dashed line shows the value of $a$ on the exafine lattices that allows us to determine a $Z_V$ value there. }
  \label{fig:ZV-validation}
\end{figure}

We can use the results of \cite{Hatton:2019gha} to check the consistency of the
$Z_V$ value on set 14 that we present here and to obtain a value on set 15. 
We do this by fitting the RI-SMOM results for the local 
vector current in
Table III of \cite{Hatton:2019gha} to the expected functional form. 
This form is 
based on a power series in $\alpha_s$ evaluated in the $\overline{\mathrm{MS}}$
scheme at a scale of $1/a$, which is the perturbative 
lattice-to-$\overline{\text{MS}}$ 
renormalisation factor, remembering that the $\text{SMOM}$ to 
$\overline{\text{MS}}$ renormalisation factor is exactly 1 in this case. In 
addition we must allow for possible discretisation effects that 
depend on $a\mu$.
The form we use is:
\begin{equation} \label{eq:ZV-fit}
  Z_V^{\mathrm{loc}}(\mathrm{SMOM})(a,\mu) = 1 + \sum_{i,j=1}^{i=4,j=3} \left[ c_i + d_{ij} \left(\frac{a\mu}{\pi} \right)^{2j} \right] \alpha_s^i .
\end{equation}
This is very similar to the approach adopted in
Appendix B of~\cite{Chakraborty:2017hry} using results for $Z_V$ 
from the determination of form factors between two identical mesons 
at rest. As there, we fix
the $\alpha_s$ coefficient to its known perturbative value of -0.1164(3).

Fig.~\ref{fig:ZV-validation} shows the $Z_V$ data from \cite{Hatton:2019gha} as
hexagons, coloured according to their $\mu$ value. 
The fit lines for 2
and 3 GeV are also shown. 
The fit has a $\chi^2/\mathrm{dof}$ of 0.93. We note that the results from set 
14 are included in this fit. The results from set 14, however, 
also agree with the fit result when they are not included in the fit. 
As discussed in Section~\ref{sec:mass} $\mu$ was slightly mistuned 
on set 14; the true
values are 2.04 and 2.98 GeV rather than 2 and 3 GeV. At this small
lattice spacing the variation in $Z_V$ with $\mu$ (which is a discretisation 
effect) is small enough that this
small mistuning can be neglected. 

Note that the underlying perturbation theory 
of Eq.~(\ref{eq:ZV-fit}) should agree with that obtained from the fit 
in~\cite{Chakraborty:2017hry} 
and it does. The discretisation effects are different in the two cases, of course. 
For RI-SMOM there is a momentum scale $\mu$ which we take as 2 GeV (although 
it can be taken as much smaller for $Z_V$~\cite{Hatton:2019gha}) and this sets the 
size of discretisation effects as is clear from Figure~\ref{fig:ZV-validation}. 

Using the fit of Eq.~(\ref{eq:ZV-fit}) we may also extract $Z_V$ values at finer lattice spacings where
it may not be practical for us to perform direct calculations 
due to the computational
cost of Landau gauge fixing. This includes the exafine lattices of set 15 
in Table~\ref{tab:ensembles} 
with a lattice spacing adjusted for sea-mass mistuning of 0.032 fm. The
value of $Z_V$ from the fit for these lattices is 0.99296(21) at 
$\mu$ = 2 GeV and 0.99186(18) at 3 GeV. 

\begin{table}
  \caption{Values of $Z_m^{\mathrm{SMOM}}$ obtained in the RI-SMOM 
scheme on set 14 in pure QCD. 
Results are given at two values of $\mu$, listed in lattice units 
in column 1. They correspond approximately to 2 GeV and 3 GeV. 
The correlation between the two numbers is 0.122. 
  }
  \label{tab:Zm-uf}
\begin{ruledtabular}
\begin{tabular}{ll}
  $a\mu$ & $Z_m^{\mathrm{SMOM}}$ \\
\hline
0.4466  & 1.2925(39) \\
0.6525  & 1.1785(11)  \\
\end{tabular}
\end{ruledtabular}
\end{table}

Results for the mass renormalisation factor from the lattice 
to the RI-SMOM scheme, $Z_m^{\mathrm{SMOM}}(\mu)$, determined on ultrafine set 14 
are given in Table~\ref{tab:Zm-uf} at $\mu = $ 2 GeV and 3 GeV.  
These are calculated in the same way as that discussed in~\cite{Lytle:2018evc}. 
\end{appendix}

\bibliography{charmonium}

\end{document}